\documentclass[11pt,a4paper]{article}
\usepackage[margin=1in]{geometry}
\usepackage{booktabs}
\usepackage{indentfirst}
\usepackage{amssymb,epsfig,amsmath,accents,amsthm}
\usepackage{xcolor,xspace}
\usepackage{multirow}
\usepackage{lscape}
\usepackage{epstopdf}
\usepackage{enumitem}
\usepackage[title]{appendix}
\usepackage{mathtools}
\mathtoolsset{showonlyrefs=true}
\usepackage[labelfont=bf]{caption}
\usepackage[colorlinks=true,linkcolor=red, urlcolor=black, citecolor=blue,breaklinks]{hyperref}
\newtheorem{theorem}{Theorem}[section]
\newtheorem{proposition}{Proposition}[section]
\newtheorem{lemma}{Lemma}[section]

\newtheorem{assumption}{Assumption}[section]
\newtheorem{definition}{Definition}[section]

\newcommand*\dd{\mathop{}\!\mathrm{d}}

\usepackage[sort&compress]{natbib}
\usepackage{setspace}
\allowdisplaybreaks
\newcommand{\prob}{\mathbb{P}}

\newcommand{\expect}{\mathbb{E}}

\newcommand{\argmax}{\operatornamewithlimits{argmax}}

\newcommand{\halmos}{\qed}

\bibpunct{(}{)}{,}{a}{,}{,}
\newenvironment{pfof}[1]{\vspace{1ex}\noindent{\bf Proof of #1}\hspace{0.5em}}
{\vspace{1ex}}

\usepackage{fancyhdr}

\date{September 9, 2026}

\begin{document}

\title{Arbitrage on Decentralized Exchanges}

\author{Xue Dong He\thanks{Department of Systems Engineering and Engineering Management, The Chinese University of Hong Kong. Email: xdhe@se.cuhk.edu.hk} \and Chen Yang\thanks{Department of Systems Engineering and Engineering Management, The Chinese University of Hong Kong. Email: cyang@se.cuhk.edu.hk} \and Yutian Zhou\thanks{Department of Systems Engineering and Engineering Management, The Chinese University of Hong Kong. Email: yzhou@se.cuhk.edu.hk}}

\maketitle

\begin{abstract}
Decentralized exchanges using automated market makers create arbitrage opportunities with centralized exchanges, where gas fees and transaction ordering are critical. Existing models largely overlook competition among arbitrageurs, despite price discrepancies being public information. We develop the first equilibrium model of gas fee competition between two arbitrageurs under three transaction reversion settings: no-revert, auto-revert, and selectable-revert. We show that pure symmetric equilibria do not exist, but unique mixed equilibria can be characterized. Using data from Binance and Uniswap V2, we empirically confirm that arbitrageurs face positive inventory risk and validate our model's implications: gas fees increase with price discrepancies and liquidity, while trading amounts rise with both price discrepancies and gas fees. Comparative analysis reveals that under low inventory risk, the no-revert setting favors arbitrageurs in terms of profit, while auto-revert and selectable-revert settings enhance market efficiency. Under high inventory risk, the no-revert and selectable-revert settings dominate the auto-revert setting in both profitability and efficiency.
\end{abstract}

	
\section{Introduction}\label{sec:intro}
In recent years, decentralized exchanges (DEXs) have become popular platforms for trading crypto assets, enabling transactions directly between users without a central authority. The dominant trading mechanism on these platforms is the Automated Market Maker (AMM), which stands in contrast to the limit order books typically used by centralized exchanges (CEXs).\footnote{See for instance \url{https://www.theblock.co/data/decentralized-finance/dex-non-custodial}.} Consequently, we will assume all DEXs discussed hereafter utilize AMMs.

In this mechanism, an algorithmic pricing function determines the exchange rate based on the asset quantities within a liquidity pool. This is a key distinction from CEXs, where professional market makers actively set prices. Furthermore, on an AMM-based DEX, the exchange rate is only altered by trades; the provision or withdrawal of liquidity by users does not, by itself, impact the price.

Trading crypto assets across multiple DEXs and CEXs often leads to price discrepancies, creating arbitrage opportunities. Arbitrageurs capitalize on these differences by trading between platforms. A study by \cite{heimbach2024non} found that arbitrage between Ethereum-based DEXs and external exchanges constitutes over 25\% of the trading volume on Ethereum's five largest DEXs. While this activity enhances market efficiency by synchronizing exchange rates, it also exposes liquidity providers on DEXs to substantial losses. Therefore, a thorough understanding of cross-exchange arbitrage is critical for designing more robust and stable decentralized exchanges.

A further distinction between DEXs and CEXs lies in their trading models. CEXs operate on a continuous-time basis, whereas DEXs feature discrete trading, with transactions bundled into blocks. On the Ethereum network, for instance, a new block is generated approximately every twelve seconds. When submitting a transaction on a DEX, an investor must pay a ``gas fee" to the network validators who assemble these blocks. Validators typically prioritize transactions that offer higher gas fees \citep{buterin2013ethereum}. This ordering is critical because the AMM's pricing algorithm creates price impact (slippage), meaning the first executed order in a block receives a more favorable price than subsequent ones. Consequently, when two arbitrageurs compete for the same opportunity, the one who pays a lower gas fee will have their order executed later, potentially facing a reduced profit or even a loss due to the price impact from the first transaction.

With a few exceptions, the existing literature has largely assumed a single arbitrageur in their models, thus neglecting competition and the role of gas fees; see the literature review in Section \ref{sec:litRev}. However, since price discrepancies are public information, multiple arbitrageurs are likely to compete for the same arbitrage opportunity. This creates a trade-off: pay a high gas fee to increase the chance of being first, or pay a low gas fee to increase net profit.

This paper introduces the first equilibrium model for gas fee competition between two arbitrageurs. We consider a pair of crypto assets traded on both a CEX and a DEX. When a price discrepancy arises, two arbitrageurs decide whether to pursue the arbitrage and, if so, the optimal gas fee and trading amount to maximize their expected profit. They first submit an order on the CEX, which is executed immediately to lock in the price. They then submit a reverse order with a gas fee on the DEX to complete the arbitrage. The DEX order is executed at the end of the block, with the execution price depending on its position in the block. To avoid front-running attacks in public mempools, arbitrageurs often wait until the last moment to submit their orders. The recent emergence of private pools, like Flashbots, which use a sealed-bid auction mechanism, mitigates these attacks by hiding transactions from public view. Given this, we model the gas-fee competition as a symmetric Nash game in which each arbitrageur is unaware of the other's actions.

An alternative arbitrage strategy involves trading on the DEX first and then on the CEX. This approach, however, exposes the arbitrageur to price risk, as the reverse trade on the CEX cannot be executed until the trade on the DEX is confirmed at the end of the block, and the CEX price may change before that time. To mitigate this risk, the arbitrageur could choose to trade only at the end of each block, but this would mean missing opportunities that arise earlier. Our model, where the arbitrageur trades on the CEX first, allows for the exploitation of every price discrepancy, since the CEX price is locked and the DEX price remains unchanged within the block.

In many private pools, orders that are not executed first are automatically reverted (cancelled), and investors pay a fraction of the gas fee for these reverted orders.\footnote{While some private pools do not charge a gas fee for reverted orders, other costs may be incurred. For an investor, these additional costs are functionally equivalent to a gas fee.} In this scenario, the arbitrageur is not exposed to the risk of an unfavorable execution price if they lose the gas fee competition. However, they are exposed to inventory risk. Having executed the order on the CEX, a reverted DEX order leaves the arbitrageur with an unbalanced position, exposing them to the volatility of crypto asset prices. The amount of the inventory risk experienced by the arbitrageur depends both on the size of the unbalanced position and on the arbitrageur's degree of aversion to the inventory risk.

Some platforms, like Uniswap v2, allow investors to choose whether to revert orders that are not executed first by setting a slippage tolerance. A very low tolerance will cause the order to be reverted if it is not executed first, while a very high tolerance will ensure it is not reverted.

We categorize these three scenarios as {\em no-revert}, {\em auto-revert}, and {\em selectable-revert} settings. The most relevant setting in practice is selectable-revert. As previously mentioned, investors who trade on the Uniswap v2 platform can choose slippage tolerance which effectively specifies whether they want to revert transactions that are not executed first. Investors can also choose to trade in public pools, such as on the Uniswap v2 trading platform, or in private pools. In the former, the investor can decide not to revert transactions that are not executed first and in the latter, those transactions are automatically reverted. Then, the choice of the trading venue becomes the choice of whether to revert those transactions. Our goal is to analyze the arbitrageur's strategy in the gas fee competition game not only in the selectable-revert setting but also in the no-revert and auto-revert settings. There are two reasons for us to study the latter two settings as well, although they are less popular in practice. First, as we will show later, the selectable-revert setting reduces to the no-revert and auto-revert settings in extreme cases. Second, we will show that the revert and auto-revert settings can be superior to the selectable-revert setting, suggesting that the former two settings can be viable options in the design of trading mechanisms on DEX in practice. Our contributions are threefold:

First, we demonstrate that pure symmetric equilibrium strategies do not exist in any of the three settings and proceed to characterize the mixed equilibrium strategies. In the no-revert and auto-revert settings, we prove the existence of a unique mixed symmetric equilibrium strategy, which can be computed by solving a non-standard, highly nonlinear ordinary differential equation. In this equilibrium, the arbitrageur chooses a probability of taking the arbitrage opportunity and, conditional on this, randomly selects a gas fee and a corresponding trading amount. In the selectable-revert setting, we also characterize the mixed symmetric equilibrium strategy through a set of equations and numerically verify its uniqueness.

Second, our model provides several insights into arbitrage strategy, which we test through an empirical study. We prove that in the no-revert setting, the arbitrageur trades less than the maximum amount they would if there were no competition. In the other two settings, with zero inventory cost, they will trade the maximum amount. We also find that the gas fee density function is strictly decreasing in the no-revert setting, and the opposite is true in the auto-revert and selectable settings with zero inventory cost. Our empirical analysis of data from Binance and Uniswap v2, where investors can choose whether to revert their orders, supports our model. We find that in most arbitrage trades, the trading amount is less than the maximum amount, and high priority gas fees are less frequently selected than low priority fees, suggesting a non-zero inventory risk in the market. Our empirical study also confirms other implications of our model: the gas fee paid by the arbitrageur increases with the price discrepancy and liquidity level, and the trading amount, per unit of reserve, increases with the price discrepancy and the gas fee.

Third, we compare the no-revert, auto-revert, and selectable-revert settings from the perspectives of the arbitrageurs and the DEX. With low or zero inventory cost, the arbitrageur's expected profit is lower in the auto-revert and selectable-revert settings than in the no-revert setting. In the former two, arbitrageurs are forced or choose to revert orders that are not executed first, leading to aggressive bidding that drives expected profits to zero. In contrast, in the no-revert setting, arbitrageurs are more conservative in their bidding, allowing for positive expected profits. With a moderate to high inventory cost, however, the expected profits in the no-revert and selectable-revert settings are similar and positive, in contrast to the zero-expected profit in the auto-revert setting. Therefore, from the arbitrageurs' perspective, the no-revert setting is generally preferable. In terms of market efficiency, measured by the probability of the arbitrage opportunity disappearing, the auto-revert and selectable-revert settings are superior with low inventory costs. However, with moderate to high inventory costs, the no-revert and selectable-revert settings are more efficient.

The remainder of the paper is organized as follows. Section \ref{sec:litRev} reviews the related literature. Section \ref{sec:model} describes the trading mechanism on DEXs with AMMs and introduces our model of gas fee competition. Section \ref{sec:MainResults} derives the equilibrium strategies and analyzes their properties. Section \ref{sec:empirical} presents the empirical study. Section \ref{sec:compareSettings} compares the three transaction reversion settings. Section \ref{sec:conclude} concludes. The Online Appendix provides detailed theoretical and numerical results in Section \ref{appendix:theoreticalResults}, as well as proofs in Section \ref{sec:proofs}.

\subsection{Literature Review}\label{sec:litRev}

Most existing studies assume a single arbitrageur exploiting price discrepancies between a DEX and a CEX, thereby neglecting the role of gas fees; see, for instance, \citet{AngerisEtal2019:AnAnalysisOfUniswap}, \citet{AngerisChitraEvans2020:WhenDoesTheTailWag}, \citet{EvansAngerisChitra2021:OptimalFees}, \citet{AngerisEvansChitra2023:Replicating}, \citet{AngerisEtal2022:ConstantFunctionMarketMakers}, \citet{MilionisEtal2023:AMMFees}, \citet{hasbrouck2026need}, \citet{LeharParlour2023:DecentralizedExchange}, \citet{Aoyagi2022:LiquidityProvision}, \citet{AoyagiIto2024:CoexistingExchangePlatforms}, \citet{HeYangZhou2024:LiquidityPoolDesign}, \citet{fabi2025economics}, \citet{campbell2025optimal}, and \citet{capponi2026optimal}.

\citet{CapponiJia2021:TheAdoptionOfBlockchainBased} assume that arbitrageurs revert transactions not executed first without incurring gas fees and that there is no inventory risk. Consequently, each arbitrageur chooses the maximum trading amount as if there were no competition, and pays a gas fee that drives expected profit to zero. Our model differs from \citet{CapponiJia2021:TheAdoptionOfBlockchainBased} in three respects. First, we consider three transaction reversion settings, no-revert, auto-revert, and selectable-revert, whereas they focus only on auto-revert. Second, consistent with practice, in the auto-revert setting we assume a small cost (e.g., gas fee) for transactions that are submitted but reverted. Third, we incorporate inventory risk, consistent with our empirical findings.

\citet{grivoloptimal} discuss in Section 7.2 therein the case of two arbitrageurs, assuming collaboration to exploit price discrepancies and profit-sharing according to a fixed proportion. By contrast, we assume arbitrageurs compete with each other, as in reality.

\citet{capponi2025longer} study the effect of block time on priority fees and expected losses from adverse selection under competition between informed traders on a DEX. They assume that orders not executed first cannot be reverted and that price impact on the DEX is linear.

Several papers study maximum extractable value (MEV) on DEXs; see, for instance, \citet{zhu2024quantifying}, \citet{capponi2022allocative,capponi2023private}, \citet{Park2023:TheConceptualFlaws}, \citet{SchwarzSchillingEtal2023:TimeIsMoney}, \citet{LeharParlour2023:BattleoftheBots}, and the summary by \citet{john2025cryptoeconomics}. The profit of an arbitrageur exploiting a price discrepancy between a CEX and a DEX can be regarded as MEV. However, our model differs from these works. In those studies, MEV does not depend on competitors' actions. By contrast, with non-revertible transactions, arbitrageur profit in our model depends on competitors' trading amounts. For revertible transactions, inventory risk arises naturally in our model, but this risk is not accounted for in the MEV literature. Moreover, we analyze and compare three transaction reversion settings, a problem entirely unexplored in prior work.

In their empirical study, \citet{gogol2024quantifying} define maximum arbitrage value (MAV) to quantify the size of an arbitrage opportunity. They find that MAV is negatively correlated with the decay time of the opportunity and positively correlated with gas fees. These findings align with our model: (i) the larger the price discrepancy, the more likely arbitrageurs are to trade, thereby eliminating the discrepancy; and (ii) the larger the discrepancy, the higher the gas fee arbitrageurs are willing to pay.

\section{A Model of Arbitrageurs' Competition}\label{sec:model}

\subsection{Trading Mechanism on AMMs}\label{sec:AMM}

Consider a liquidity pool on a DEX consisting of two assets, $A$ and $B$, governed by a Constant Function Market Maker (CFMM). For $i \in {A, B}$, let $y_i$ denote the reserve of asset $i$ in the pool. An investor executing a trade deposits an amount $d_i$ of asset $i$ into the pool, where a negative $d_i$ represents a withdrawal. The relationship between the amounts traded is determined by the rule:
\begin{align}\label{eq:CFMM}
	F(y_A+d_A, y_B+d_B) = F(y_A, y_B),
\end{align}
where $F$ is a pricing function specified by the CFMM. This rule stipulates that the value of the pricing function remains invariant after a trade (excluding fees).

In this paper, we focus on the Constant Product Market Maker (CPMM) model, where $F(x,y)=xy$. This model is notably used by Uniswap, one of the largest AMM-based DEXs. Under this model, the pricing rule \eqref{eq:CFMM} becomes:
\begin{align}\label{eq:CPMM}
	(y_A+d_A)(y_B+d_B) = y_Ay_B.
\end{align}
From \eqref{eq:CPMM}, we can derive the {\em marginal exchange rate} of asset $B$ in terms of asset $A$, which is the limit of $-d_A/d_B$ for an infinitesimally small trade, yielding $y_A/y_B$.

Trading on a DEX incurs a proportional fee. Let $f$ be the unit trading fee. An investor wishing to deposit $d_A > 0$ of asset $A$ to receive $-d_B > 0$ of asset $B$ must pay a total of $(1+f)d_A$ of asset $A$. Following \citet{LeharParlour2023:DecentralizedExchange}, \citet{MilionisMoallemiRoughgardenZhang2024:AutomatedMarketMaking}, and \cite{hasbrouck2026need}, and consistent with the practice in Uniswap v3, we assume the trading fee $fd_A$ is paid directly to liquidity providers and is not added to the pool's reserves. Consequently, the reserves of asset $i$ in the pool after the trade remain $y_i+d_i$.

\subsection{Arbitrage Opportunity}\label{subse:ArbitrageurOpportunity}

Now, suppose assets $A$ and $B$ are also traded on a CEX, where the current price of asset $i$ is $p_i$ in US dollars, for $i\in\{A,B\}$. A price discrepancy between the CEX and the DEX creates an arbitrage opportunity. Consider an arbitrageur who identifies such an opportunity. The arbitrageur's strategy can be represented by a pair $(d_A, d_B)$, where a positive (negative, respectively) $d_i$ represents the amount of asset $i$ deposited into (withdrawn from, respectively) the DEX pool. To execute the arbitrage, the arbitrageur also incurs a gas fee $g$ for the DEX transaction. The reverse trade is done on the CEX to complete the arbitrage. The arbitrageur's profit, assuming their trades on both platforms are confirmed, is the net cash flow:
\begin{align}\label{eq:FMNetProfit}
	-(1+f \cdot \mathbf{1}_{d_{A}>0})d_{A}p_A - (1+f \cdot \mathbf{1}_{d_{B}>0})d_{B}p_B - g.
\end{align}
Here, the indicator function $\mathbf{1}_{d_{i}>0}$ ensures that the fee is applied only to the asset being paid into the DEX pool.

The arbitrageur's transaction on the CEX can be executed immediately. Their transaction on the DEX, however, cannot be confirmed until the end of the block. In the literature, it is commonly assumed that the arbitrageur's transaction on the DEX is confirmed prior to any other transactions in the block. As a result, the reserve of asset $i$ in the pool upon the confirmation of the arbitrageur's transaction is still $y_i$, so their trading amount $(d_A,d_B)$ satisfies \eqref{eq:CPMM}. The arbitrageur's optimization problem is thus to choose $(d_A, d_B)$ satisfying \eqref{eq:CPMM} to maximize this profit. This static arbitrage problem has been studied in the literature; see, for instance, \citet{EvansAngerisChitra2021:OptimalFees}, \citet{CapponiJia2021:TheAdoptionOfBlockchainBased}, and \citet{HeYangZhou2024:LiquidityPoolDesign}. Note that in \eqref{eq:FMNetProfit}, we assume trading on the CEX has no price impact due to the much higher market depth on CEXs than on DEXs. Trading on a DEX, however, has a price impact because the exchange rate $-d_A/d_B$ depends on the trading amount, as seen from \eqref{eq:CPMM}. As a result, an optimal trading amount exists.

To avoid triviality, we henceforth assume that $(p_B/p_A)/(y_A/y_B)$, the ratio of the exchange rates of asset $B$ in terms of asset $A$ on the CEX and DEX, is not in the interval $\big[(1+f)^{-1},1+f\big]$. Otherwise, the arbitrageur will not take the arbitrage opportunity even if there is no gas fee; see \citet[Proposition 3.1]{HeYangZhou2024:LiquidityPoolDesign}. Without loss of generality, we assume that $(p_B/p_A)/(y_A/y_B)>1+f$, i.e., that asset $B$, relative to asset $A$, is overpriced on the CEX compared to the price on the DEX. In this case, if the arbitrageur trades, they would buy asset $A$ and sell asset $B$ on the CEX and perform the reverse trade on the DEX; i.e., they set $d_A>0$ and $d_B<0$.

For notational simplicity, denote
\begin{align}
	O:=\frac{y_Bp_B}{y_Ap_A(1+f)},\quad L_B:=y_Bp_B\label{eq:LiquidityB_Mispricing}.
\end{align}
Note that $O$ is the exchange rate ratio on the CEX and DEX, adjusted by the trading fee on the DEX, thus representing the {\em price discrepancy}. $L_B$ stands for the \emph{liquidity level} on the DEX, as measured by the value of asset $B$ in the pool.

\subsection{Competition between Two Arbitrageurs}\label{subse:Competition}

We consider two arbitrageurs competing for the same arbitrage opportunity. In an arbitrageur's strategy of exploiting a price discrepancy between the CEX and DEX, their transaction on the CEX is executed immediately, while their transaction on the DEX incurs a gas fee and must wait until the end of the block to be confirmed. The gas fee consists of a {\em base gas fee} and a {\em priority gas fee}. The former is the minimal fee required, while the latter is chosen strategically by the arbitrageur. The execution order of transactions is determined by the magnitude of the gas fee, consistent with practice.\footnote{The base gas fee and priority gas fee are computed as the gas amount consumed in the transaction multiplied by the base fee per gas and the priority fee per gas, respectively. Validators typically prioritize transactions with higher priority fees per gas. In our model, we assume that the gas amount used by every arbitrageur is the same, so ranking transactions by the priority fee is equivalent to ranking them by the total gas fee.}

\begin{assumption}[Transaction Execution Order Rule]\label{assume}
	Validators prioritize transactions with higher gas fees. For transactions with the same gas fee, execution order is determined randomly with equal probability.
\end{assumption}

Let $\hat g_L>0$ denote the base gas fee. An arbitrageur can then choose a gas fee $g\ge \hat g_L$ when submitting their transaction on the DEX. If their transaction is executed before their competitor's, their profit is given by \eqref{eq:FMNetProfit} with $d_A>0$ and $d_B<0$ satisfying \eqref{eq:CPMM}:
\begin{align}
	-(1+f)d_{A}p_A-d_{B}p_B-g
	= d_A\left(\frac{y_B}{y_A+d_A}p_B-(1+f)p_A\right)-g
	=: R_F(g, d_A).\label{eq:FirstMoverReward}
\end{align}
Thus, if the arbitrageur's transaction is guaranteed to be executed first, they choose $d_A$ to maximize $R_F(g,d_A)$. Straightforward calculation yields:
\begin{align}
	\hat D_A:=\argmax_{d_A>0} R_F(g, d_A)
	= \sqrt{\frac{y_Ay_Bp_B}{p_A(1 + f)}} - y_A
	= (\sqrt{O}-1)y_A.\label{eq:MaximumTradingAmount}
\end{align}
The arbitrageur will not exploit the opportunity if the resulting profit is negative. Therefore, the maximum gas fee, denoted $\hat g_H$, that the arbitrageur is willing to pay when guaranteed first execution is:
\begin{align}
	\hat g_H =\hat D_A\left(\frac{y_B}{y_A+\hat D_A}p_B-(1+f)p_A\right)
	=L_B(1-O^{-1/2})^2.\label{eq:MaximumGasFee}
\end{align}

To avoid triviality, we impose the following assumption, ensuring that the arbitrage opportunity remains profitable if the arbitrageur pays the base gas fee $\hat g_L$ and is guaranteed first execution.

\begin{assumption}[Profitable Arbitrage Opportunity]\label{assumption:parameters}
	$L_B>0$, $\hat g_L>0$, and $O>\big(1-(\hat g_L/L_B)^{1/2}\big)^{-2}$.
\end{assumption}

If the arbitrageur's transaction is executed later than their competitor's, their profit depends not only on their trading amount and gas fee but also on (i) whether their trade can be reverted and (ii) their competitor's decision. We use the parameter $\ell$ to denote the three settings discussed in Section \ref{sec:intro} regarding reverted transactions: $\ell=\infty$ refers to the no-revert setting, $\ell=0$ to the auto-revert setting, and $\ell\in\{0,\infty\}$ to the selectable-revert setting.

Suppose the competitor submits an order to sell $\bar d_A>0$ units of asset $A$ on the DEX, and their order is executed first. They then withdraw $\frac{y_B}{y_A+\bar d_A}\bar d_A$ units of asset $B$ from the pool according to \eqref{eq:CPMM}. At the moment the arbitrageur's order is executed, the reserves of assets $A$ and $B$ in the pool become $y_A+\bar d_A$ and $\frac{y_Ay_B}{y_A+\bar d_A}$, respectively. Consequently, in the no-revert setting, the arbitrageur's order will be executed and their profit is given by \eqref{eq:FMNetProfit} with $d_A>0$ and $d_B<0$ satisfying the constant product rule:
\begin{align*}
	\big(y_A+\bar d_A+d_A\big)\left(\frac{y_Ay_B}{y_A+\bar d_A} +d_B\right)=y_Ay_B.
\end{align*}
Straightforward calculation yields that the profit is
\begin{align}
	-(1+f)d_{A}p_A-d_{B}p_B-g= d_A\left(\frac{\frac{y_Ay_B}{y_A+\bar d_A}}{y_A+\bar d_A+d_A}p_B-(1+f)p_A\right)-g=:R_S^\infty(g,d_A;\bar d_A).\label{eq:SecondMoverReward}
\end{align}

In the auto-revert setting, when the arbitrageur's trade is not executed first and is reverted, they still pay a proportion $r$ of the gas fee. In addition, because the trade on the CEX has already been executed, reverting the trade on the DEX creates inventory risk. The arbitrageur's profit in this case is:
\begin{align}
	-\Gamma(d_A)-rg=:R_S^0(g,d_A;\bar d_A), \label{eq:SecondMoverReward_0}
\end{align}
where $\Gamma(d_A)$ represents the inventory risk given trading amount $d_A$, and $rg$ is the gas fee paid for the reverted transaction. Note that $R_S^0(g,d_A;\bar d_A)$ does not depend on $\bar d_A$, but for notational consistency we retain it as an argument.

\begin{assumption}
	(i) $r>0$; (ii) $\Gamma$ is twice continuously differentiable on $[0,\infty)$, weakly increasing, convex, and satisfies $\Gamma(0)=0$.
\end{assumption}

We refer to the arbitrageur as the {\em first mover} if their transaction is executed first, and as the {\em second mover} otherwise. $R_F$ and $R_S$ denote the net profits of the first and second mover, respectively. The \emph{first-mover advantage} is defined as the difference between the first mover's and second mover's net profit:
\begin{align*}
	V^\ell (g,d_A;\bar d_A):=R_F(g,d_A)- R_S^\ell(g, d_A;\bar d_A).
\end{align*}
Straightforward calculation yields:
\begin{align}
	&V^\infty(g,d_A;\bar d_A)=\left[\frac{ d_A}{y_A+ d_A}-\frac{ d_A y_A}{(y_A+\bar d_A)(y_A+\bar d_A+d_A)}\right]y_Bp_B,\label{eq:FMAInfS}\\
	&V^0(g,d_A;\bar d_A) = d_A\left(\frac{y_B}{y_A+d_A}p_B-(1+f)p_A\right)+\Gamma(d_A)+(r-1)g.\label{eq:FMA0S}
\end{align}
Note that $V^\infty(g,d_A;\bar d_A)$ does not depend on $g$, and $V^0(g,d_A;\bar d_A)$ does not depend on $\bar d_A$.

For the moment, we focus on the no-revert and auto-revert settings; the selectable-revert setting will be discussed in Section \ref{subsect:TCR}. Note that the arbitrageur can also choose not to trade. Therefore, we can summarize an arbitrageur's action by the triplet $(\xi,g,d_A)$, where the binary variable $\xi$ takes the value 1 if the arbitrageur decides to trade and 0 otherwise. Given the competitor's action, denoted as $(\bar \xi,\bar g,\bar d_A)$, the arbitrageur's payoff is
\begin{align*}
	R^\ell(\xi,g,d_A;\bar \xi,\bar g,\bar d_A):=\begin{cases}
		0, & \text{if }\xi=0,\\
		R_F(g, d_A), & \text{if }\xi=1 \text{ and } (\bar \xi =0 \text{ or } \bar g<g),\\
		\tfrac{1}{2}R_F(g, d_A)+ \tfrac{1}{2}R_S^\ell(g, d_A;\bar d_A), & \text{if }\xi=1,\; \bar \xi =1 \text{ and } \bar g=g,\\
		R_S^\ell(g, d_A;\bar d_A), & \text{if }\xi=1,\;\bar \xi=1 \text{ and } \bar g>g.
	\end{cases}
\end{align*}

If the arbitrageur is guaranteed to be the first mover, they will not pay a gas fee higher than $\hat g_H$ as in \eqref{eq:MaximumGasFee}, nor will they choose a trading amount $d_A$ larger than $\hat D_A$ as in \eqref{eq:MaximumTradingAmount}. In the auto-revert setting, the arbitrageur prefers a smaller trading amount if they are the second mover, since the second mover's net profit decreases with the trading amount. Thus, they would not select $d_A>\hat D_A$. In the no-revert setting, the arbitrageur's optimal trading amount as the second mover is also less than $\hat D_A$, because the execution price on the DEX is less favorable than for the first mover. Consequently, the feasible range of the trading amount is $[0,\hat D_A]$. By similar reasoning, the feasible range of the gas fee is $[\hat g_L,\hat g_H]$. Therefore, the action space of each arbitrageur is
\begin{align*}
	U:=\{0,1\}\times [\hat g_L,\hat g_H]\times [0,\hat D_A].
\end{align*}

The competitor's payoff $\bar R^\ell(\bar \xi,\bar g,\bar d_A;\xi,g,d_A)$ for their action $(\bar \xi,\bar g,\bar d_A)$, given the arbitrageur's action $(\xi,g,d_A)$, is defined symmetrically:
$
	\bar R^\ell(\bar \xi,\bar g,\bar d_A;\xi,g,d_A) = R^\ell(\xi,g,d_A;\bar \xi,\bar g,\bar d_A).
$
Thus, the gas-fee competition is symmetric, and we seek Nash equilibria for this game.

Because of the first-mover advantage, the payoff function $R^\ell(\xi,g,d_A;\bar \xi,\bar g,\bar d_A)$ is discontinuous at $g=\bar g$. Therefore, our game does not satisfy the better-reply secure property in \citet{Reny1999:OnTheExistence}, which is a common condition for the existence of pure-strategy equilibria. We therefore search for a mixed-strategy Nash equilibrium. It turns out that the unique mixed-strategy Nash equilibrium is not a pure-strategy equilibrium.

Since the payoff function does not depend on the choice of gas fee $g$ or trading amount $d_A$ when $\xi=0$, the arbitrageur's randomized strategy can be represented by a pair $(\alpha,\mu)$, where $\alpha\in[0,1]$ is the probability of trading, and $\mu$, a probability measure on $[\hat g_L,\hat g_H]\times [0,\hat D_A]$, is the distribution of the gas fee and trading amount pair $(g,d_A)$ conditional on trading. In other words, the arbitrageur's choice of $\xi$ follows a Bernoulli distribution with $\prob(\xi=1)=\alpha$. For completeness, we define the gas fee and trading amount to be zero when $\xi=0$, i.e., when the arbitrageur chooses not to trade.

The payoff of the mixed game is
\begin{align}\label{eq:GamePayoffMixed}
	\begin{split}
		\Pi^\ell(\alpha,\mu;\bar \alpha,\bar \mu):&= \expect[R^\ell]=\alpha (1-\bar \alpha)\int R_F(g, d_A)\mu(\dd g\dd d_A)\\
		&\quad+\alpha \bar \alpha \bigg[ \int R_F(g, d_A)\mathbf 1_{g>\bar g}\mu(\dd g\dd d_A)\bar \mu(\dd \bar g\dd\bar d_A)\\
		&\quad+ \int \tfrac{1}{2}\left(R_F(g, d_A)+R_S^\ell(g, d_A;\bar d_A)\right)\mathbf 1_{g=\bar g}\mu(\dd g\dd d_A)\bar \mu(\dd\bar g \dd\bar d_A)\\
		&\quad+ \int R_S^\ell(g, d_A;\bar d_A)\mathbf 1_{g<\bar g}\mu(\dd g\dd d_A)\bar \mu(\dd\bar g\dd\bar d_A)\bigg],
	\end{split}
\end{align}
where the expectation is taken with respect to the random action $(\xi,g,d_A)$ chosen by the arbitrageur and $(\bar \xi,\bar g,\bar d_A)$ chosen by the competitor. On the right hand side, the first term corresponds to the case where the competitor does not trade, so the arbitrageur automatically becomes the first mover. The second term corresponds to the case where both trade, and whether the arbitrageur is the first or second mover follows from the analysis above. Finally, the payoff is zero if the arbitrageur does not trade.

Because the pure-strategy game is symmetric, the mixed-strategy game is also symmetric. Therefore, we focus on symmetric equilibria.

\begin{definition}[Symmetric Mixed Nash Equilibrium]
	Fix $\ell\in \{0,\infty\}$. A mixed strategy $(\alpha^*,\mu^*)\in[0,1]\times {\cal U}$ is a {\em symmetric mixed Nash equilibrium} of the gas-fee competition game if
	\begin{align*}
		(\alpha^*,\mu^*)\in \underset{(\alpha,\mu)\in[0,1]\times {\cal U}}{\mathrm{argmax}} \Pi^\ell(\alpha,\mu;\bar \alpha^*,\bar \mu^*),
	\end{align*}
	where ${\cal U}$ denotes the set of probability measures on the action space $[\hat g_L,\hat g_H]\times [0,\hat D_A]$.
\end{definition}

\section{Equilibrium Strategies}\label{sec:MainResults}

There is extensive literature on symmetric Nash games, and various sufficient conditions have been established for the existence of symmetric mixed Nash equilibria; see, for instance, \citet{Reny1999:OnTheExistence}. However, these conditions do not hold in our game under certain settings.\footnote{We verified that the conditions in \citet{Reny1999:OnTheExistence} hold in the no-revert setting but fail in the auto-revert and selectable-revert settings. Thus, we can prove the existence of a symmetric mixed Nash equilibrium in the former setting using their results. A formal proof is available from the authors upon request.} Moreover, these conditions neither yield uniqueness nor facilitate computation of the equilibrium. In what follows, we derive several properties of the mixed Nash equilibrium in our game, prove existence and uniqueness, and compute the equilibrium by solving an ordinary differential equation (ODE).

\subsection{Properties of Mixed Equilibria}\label{subse:MixedEquilibrium}

We define the {\em reward function} $h^\ell(g,d_A)$ as the arbitrageur's expected profit from trading an amount $d_A$ with gas fee $g$, given that the competitor follows the equilibrium strategy $(\alpha^*,\mu^*)$:
\begin{align}\label{eq:ResponseFunction}
	\begin{split}
		h^\ell(g,d_A):&=(1-\alpha^*)R_F(g, d_A)+\alpha^* \bigg[ \int R_F(g, d_A)\mathbf 1_{g>\bar g}\mu^*(\dd\bar g\dd\bar d_A)\\
		&\quad+ \int \tfrac{1}{2}\left(R_F(g, d_A)+R_S^\ell(g, d_A;\bar d_A)\right)\mathbf 1_{g=\bar g}\mu^*(\dd\bar g\dd\bar d_A)\\
		&\quad+ \int R_S^\ell(g, d_A;\bar d_A)\mathbf 1_{g<\bar g}\mu^*(\dd\bar g\dd\bar d_A)\bigg].
	\end{split}
\end{align}

\begin{proposition}\label{prop:propertyEq}
	Fix $\ell\in \{0,\infty\}$. Suppose $(\alpha^*,\mu^*)$ is a symmetric mixed Nash equilibrium. Then:
	\begin{enumerate}[leftmargin=2em]
		\item[(i)] $\alpha^*>0$.
		\item[(ii)] For almost every gas fee $g$ under $\mu^*$, the reward function $h^\ell(g,d_A)$ is continuous and strictly concave in $d_A\ge 0$, admitting a unique maximizer in $d_A\in[0,\hat D_A]$. Denoting
		\begin{align}\label{eq:DAs}
			D_A^*(g)=\argmax_{d_A\in[0,\hat D_A]} h^\ell(g,d_A),
		\end{align}
		we have $\mu^*\big(\{(g,d):d=D_A^*(g),g\in[\hat g_L,\hat g_H]\}\big)=1$.
	\end{enumerate}
\end{proposition}

Proposition \ref{prop:propertyEq} implies that the probability of trading is positive, since profits as the first mover are strictly positive under arbitrage. Furthermore, the arbitrageur's trading amount, given a selected gas fee $g$, is deterministic rather than random. Consequently, symmetric mixed Nash equilibria can be represented by the triplet $(\alpha,\Phi,D_A)$, where $\alpha\in(0,1]$ is the probability of trading, $\Phi$ is a probability measure on $[\hat g_L,\hat g_H]$ representing the randomized gas fee strategy, and $D_A$ is a function mapping each gas fee to a trading amount. With slight abuse of notation, we continue to denote the payoff function by $\Pi^\ell$:
\begin{align}
	\Pi^\ell(\alpha,\Phi,D_A;\bar \alpha,\bar \Phi,\bar D_A)&= \alpha (1-\bar \alpha)\int R_F(g, D_A(g))\Phi(\dd g)\notag\\
	&\quad+\alpha \bar \alpha \bigg[ \int R_F(g, D_A(g))\mathbf 1_{g>\bar g}\Phi(\dd g)\bar \Phi(\dd \bar g)\notag\\
	&\quad+ \int \tfrac{1}{2}\left(R_F(g, D_A(g))+R_S^\ell(g, D_A(g);\bar D_A(\bar g))\right)\mathbf 1_{g=\bar g}\Phi(\dd g)\bar \Phi(\dd \bar g)\notag\\
	&\quad+ \int R_S^\ell(g, D_A(g);\bar D_A(\bar g))\mathbf 1_{g<\bar g}\Phi(\dd g)\bar \Phi(\dd \bar g)\bigg].\label{eq:GamePayoffMixed2}
\end{align}

This structure $(\alpha,\Phi,D_A)$ allows us to further investigate equilibrium gas fee and trading amount strategies.

\begin{proposition}\label{le:RandomizedGasDensity}
	Fix $\ell\in \{0,\infty\}$. Suppose $(\alpha^*,\Phi^*,D_A^*)$ is a symmetric mixed Nash equilibrium. Then:
	\begin{enumerate}[leftmargin=2em]
		\item[(i)] $\Phi^*\big(\{g\in[\hat g_L,\hat g_H]:D_A^*(g)=0\}\big)=0$ and $\Phi^*\big(\{g\in[\hat g_L,\hat g_H]:V^\ell(g,D_A^*(g);D_A^*(g))\le0\}\big)=0$.
		\item[(ii)] $\Phi^*$ has no atoms on $[\hat g_L,\hat g_H]$.
		\item[(iii)] The reward function as defined in \eqref{eq:ResponseFunction} takes the form:
		\begin{align}
			h^\ell(g,d_A)&=R_F(g, d_A)-\alpha^* \int_{g}^{\hat g_H}V^\ell(g,d_A;D_A^*(\bar g))\Phi^*(\dd \bar g).\label{eq:ResponseFunction2}
		\end{align}
		Furthermore, $h^\ell(g,D_A^*(g))$ is constant for $g\in[\hat g_L,\hat g_H]$. More precisely, let $M$ and $m$ denote the essential supremum and infimum of $h(g,D_A^*(g))$ under $\Phi^*$, i.e.,
		\begin{align*}
			M:=\sup\{a\in\mathbb{R}: \Phi^*(\{g\in[\hat g_L,\hat g_H]: h^\ell(g,D_A^*(g))>a\})>0\},\\
			m:=\inf\{a\in\mathbb{R}: \Phi^*(\{g\in[\hat g_L,\hat g_H]: h^\ell(g,D_A^*(g))<a\})>0\}.
		\end{align*}
		Then $M=m$. Moreover, $M\ge 0$, with equality if $\alpha^*<1$.
		\item[(iv)] The support of $\Phi^*$ is $[\hat g_L,g_H]$ for some $g_H\in (\hat g_L,\hat g_H]$.
	\end{enumerate}
\end{proposition}

Proposition \ref{le:RandomizedGasDensity} establishes several key properties. First, the trading amount and the first-mover advantage must be positive if the arbitrageur chooses to trade. Second, in equilibrium, the probability of selecting any fixed gas fee cannot be positive, confirming that the equilibrium is not pure. Indeed, if her competitor chooses a certain gas fee $g_0$ with a positive probability, the arbitrageur would choose a gas fee slightly higher than $g_0$ in order to be the first mover. Third, the reward function at the optimal trading amount must be constant across feasible gas fees; otherwise, arbitrageurs would shift probability mass toward higher-reward fees. Finally, the support of $\Phi^*$ must be contiguous, ensuring that if an arbitrageur considers a fee $g_0$, they also consider all lower fees. The intuition is as follows: If the competitor does not choose gas fees in a certain range lower than $g_0$, the arbitrageur can lower her gas fee without changing the priority of trading; consequently, the game is not in equilibrium.

Based on these properties, we restrict attention to the strategy set:
\begin{align}\label{eq:ACGasFee}
	\begin{split}
		{\cal A}:=\big\{&(\alpha, \Phi,D_A): \alpha\in(0,1],\;\Phi \text{ admits a continuous density on its support},\\
		&\text{and } D_A \text{ is continuous on the support of }\Phi\big\}.
	\end{split}
\end{align}
We now show existence and uniqueness of the symmetric mixed Nash equilibrium in ${\cal A}$.

Before presenting the main result, we provide a heuristic derivation to illustrate the idea of finding the symmetric mixed Nash equilibrium. Suppose $(\alpha^*, \Phi^*,D_A^*)\in {\cal A}$ is a symmetric mixed Nash equilibrium, and let $\phi^*$ denote the density of $\Phi^*$. Recall the response function $h^\ell(g,d_A)$ defined in \eqref{eq:ResponseFunction}. By Proposition \ref{le:RandomizedGasDensity}-(iii), for $g\in[\hat g_L,g_H]$,
\begin{align*}
	0&=\frac{\partial}{\partial g}h^\ell(g,D_A^*(g))=\frac{\partial h^\ell}{\partial g}(g,d_A)\big|_{d_A=D_A^*(g)}\\
	&=\frac{\partial R_F}{\partial g}(g, d_A)\big|_{d_A=D_A^*(g)} + \alpha^* V^\ell(g,D_A^*(g);D_A^*( g))\phi^*(g)-\alpha^*\int_{g}^{\hat g_H}\frac{\partial V^\ell}{\partial g}(g,d_A;D_A^*(\bar g))\Phi^*(\dd \bar g),
\end{align*}
where we use $\frac{\partial h}{\partial d_A}(g,d_A)\big|_{d_A=D_A^*(g)}= 0$ from \eqref{eq:DAs}.
Since $\frac{\partial R_F}{\partial g}(g, d_A)=-1$, we immediately obtain
\begin{align}
	\phi^*(g) = \frac{1+\alpha^*\int_{g}^{\hat g_H}\frac{\partial V^\ell}{\partial g}(g,d_A;D_A^*(\bar g))\Phi^*(\dd \bar g)}{\alpha^* V^\ell(D_A^*(g),D_A^*( g))},\quad  g\in[\hat g_L,g_H].\label{eq:GasFeeDensityProp}
\end{align}
Note that $\frac{\partial V^\ell}{\partial g}$ equals $0$ in the setting $\ell=\infty$ and $r-1$ in the setting $\ell=0$. Thus, \eqref{eq:GasFeeDensityProp} implies that the density of the gas fee is {\em inversely proportional to} the first-mover advantage. Substituting \eqref{eq:GasFeeDensityProp} into the first-order condition $\frac{\partial h^\ell}{\partial d_A}(g,d_A)\big|_{d_A=D_A^*(g)}= 0$, we derive an equation satisfied by the trading amount $D_A^*$.

\subsection{Mixed Equilibrium in the No-Revert Setting}\label{subse:Slippageinfinity}

We first consider the no-revert setting, i.e., the case $\ell=\infty$. In this setting, the first-mover advantage $V^\infty(g,d_A;\bar d_A)$ does not depend on $g$, so we simply denote it as $V^\infty(d_A;\bar d_A)$.

Recalling \eqref{eq:GasFeeDensityProp} and noting that $\frac{\partial V^\infty}{\partial g}\equiv 0$, we obtain $\phi^*(g)=\big(\alpha^* V^\infty(D_A^*(g),D_A^*(g))\big)^{-1}$.
Substituting this into the first-order condition $\frac{\partial h^\infty}{\partial d_A}(g,d_A)\big|_{d_A=D_A^*(g)}=0$ and applying the change of variable $\hat x(z):=D_A^*(g_H-L_B z)/y_A$, we derive
\begin{align}
	Q(\hat x(z)) = \int_0^z K^\infty(\hat x(z),\hat x(\bar z))\,d\bar z,\quad z\ge 0,\label{eq:KeyODENormalized}
\end{align}
where
\begin{align}
	Q(x)&:=\frac{y_A}{L_B}\frac{\partial R_F}{\partial d_A}(g,y_Ax)=\frac{1}{(1+x)^2}-O^{-1}, \quad x>0,\label{eq:QFun}\\
	K^\infty(x,\bar x)&:=\frac{y_A\,\partial V^\infty/\partial d_A(y_Ax,y_A\bar x)}{V^\infty(y_A\bar x,y_A\bar x)}\notag\\
	&=\frac{2(1+x)(1+\bar x)(1+2\bar x)+\bar x(1+\bar x)(1+2\bar x)}{2\bar x(1+x)^2(1+\bar x+x)^2},\quad x,\bar x>0.\label{eq:KFun}
\end{align}

Thus, the search for equilibrium reduces to solving \eqref{eq:KeyODENormalized}.
Note that \eqref{eq:KeyODENormalized} is not a standard ODE, since the integrand on the right-hand side depends nonlinearly on $\hat x(z)$. In Lemma \ref{le:ODE} in the Appendix, we prove the existence and uniqueness of the solution under the following technical assumption on the price discrepancy $O$, which is satisfied empirically (see Section \ref{sec:empirical}):
\begin{assumption}
	The price discrepancy satisfies $O\le 3$.
\end{assumption}

We are now ready to state the main theorem on the existence and uniqueness of equilibrium.

\begin{theorem}\label{thm:Main}
	Consider the no-revert setting. There exists a unique symmetric mixed Nash equilibrium $(\alpha^*,\Phi^*,D_A^*)\in{\cal A}$, characterized as follows:
	\begin{enumerate}[leftmargin=2em]
		\item[(i)] If $L_B\hat z \le \hat g_H-\hat g_L$, then $\alpha^*=1$ and $g_H=\hat g_L+L_B\hat z$.
		\item[(ii)] If $L_B\hat z>\hat g_H-\hat g_L$, then $g_H=\hat g_H$ and
		\begin{align}
			\alpha^* =\int_0^{(\hat g_H-\hat g_L)/L_B}\frac{L_B}{V^\infty(y_A \hat x(\bar z),y_A \hat x(\bar z))}\,d\bar z<1.\label{eq:EquiAlpha}
		\end{align}
	\end{enumerate}
	Here, $\hat x(z)$ is the unique solution to \eqref{eq:KeyODENormalized}, and $\hat z$ is defined by
	\begin{align*}
		\int_0^{\hat z} \frac{L_B}{V^\infty(y_A \hat x(\bar z),y_A \hat x(\bar z))}\,d\bar z=1.
	\end{align*}
	In both cases (i) and (ii),
	\begin{align*}
		D_A^*(g)=y_A\hat x\big((g_H-g)/L_B\big),\quad
		\phi^*(g) = \frac{1}{\alpha^* V^\infty(D_A^*(g),D_A^*(g))},\quad g\in[\hat g_L,g_H],
	\end{align*}		
	where $\phi^*$ is the density of $\Phi^*$. Moreover, the arbitrageur's expected profit is $\hat g_H-g_H$, which equals zero if $L_B\hat z \ge \hat g_H-\hat g_L$ and is positive if $L_B\hat z< \hat g_H-\hat g_L$.
\end{theorem}

\begin{figure}
	\centering
	\includegraphics[width=0.5\textwidth]{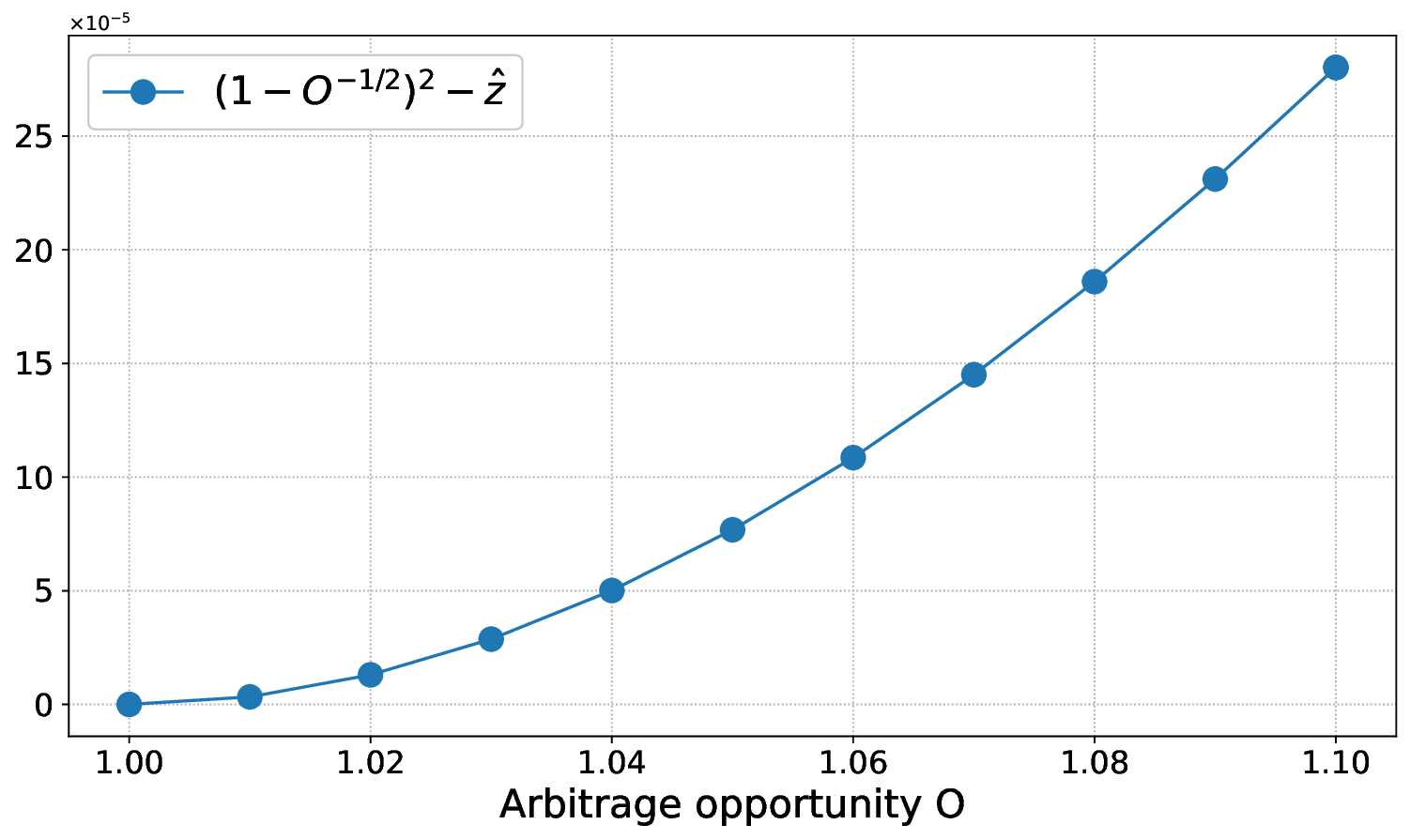}
	\caption{The quantity $(1-O^{-1/2})^2-\hat z$ with respect to the size of the price discrepancy $O$.}\label{quantity_o}
\end{figure}

The criterion for case (i) in Theorem \ref{thm:Main}, $L_B\hat z \le \hat g_H-\hat g_L$, can be rewritten as
\begin{align}\label{eq:conditionIneq}
	\big((1-O^{-1/2})^2-\hat z\big)\cdot L_B \ge \hat g_L.
\end{align}
Thus, case (i) occurs when the product of the price discrepancy and the liquidity level, as measured by the left-hand side of \eqref{eq:conditionIneq}, is large relative to the base gas fee.
Note that $(1-O^{-1/2})^2-\hat z$ depends only on $O$ and, as illustrated in Figure \ref{quantity_o}, it is positive and strictly increasing in the price discrepancy $O$. As a result, both cases in Theorem \ref{thm:Main} are possible.

Table \ref{tbl:property} summarizes several properties of the mixed Nash equilibrium strategy $(\alpha^*,\Phi^*,D_A^*)$ in Theorem \ref{thm:Main}. Formal statements and proofs are provided in Appendix \ref{appendix:theoreticalResults}. We highlight the key properties below.

First, the density function of the gas fee is strictly decreasing, implying that arbitrageurs are more likely to choose lower gas fees.

Second, as shown in Theorem \ref{thm:Main}, the arbitrageur's expected profit is positive if $\big((1-O^{-1/2})^2-\hat z\big)\cdot L_B > \hat g_L$, which occurs when the price discrepancy is large, pool liquidity is high, and the base gas fee is low. In Section \ref{sec:empirical}, we conduct an empirical study. Among all blocks with arbitrage opportunities in the data, approximately 10\% satisfy $\big((1-O^{-1/2})^2-\hat z\big)\cdot L_B > \hat g_L$, showing that the case of positive profit is empirically relevant.

Third, the arbitrageur's trading amount is strictly increasing in the chosen gas fee, with the trading amount at the highest possible gas fee equal to the maximal trading amount $\hat D_A$. This implies that arbitrageurs typically trade strictly less than $\hat D_A$, since they account for the possibility of being the second mover, in which case a large trading amount may incur losses.

Fourth, higher $L_B$ implies greater liquidity, and larger $O$ implies a larger price discrepancy, both of which are favorable to the arbitrageur. Accordingly, the arbitrageur's trading amount, the gas fee they are willing to pay, and their expected profit all increase with $L_B$ and $O$. Conversely, a higher base gas fee $\hat g_L$ is less favorable, leading to lower trading amounts and expected profits. When expected profit is positive, an increase in the base gas fee induces arbitrageurs to pay the same priority fee, effectively raising the total gas fee, to maintain the same probability of first execution. When expected profit is zero, a higher base gas fee reduces the arbitrageur's willingness to trade, resulting in a lower average gas fee.

\begin{table}
	\caption{Properties of the equilibrium strategy $(\alpha^*,\Phi^*,D_A^*)$ in the no-revert setting.}
	\label{tbl:property}
	\centering
	\footnotesize
	\begin{tabular}{c@{\hskip 2em}l@{\hskip 2em}l}
		\toprule
		& Case (i) & Case (ii)\\
		\midrule
		\textbf{Gas fee density $\phi^*(g)$} & \multicolumn{2}{c}{Strictly decreasing in $g\in[\hat g_L,g_H]$}\\ \midrule
		\textbf{Trading amount given gas fee $D_A^*(g)$} & \multicolumn{2}{c}{Strictly increasing in $g\in[\hat g_L,g_H]$ with $D_A^*(g_H)=\hat D_A$}\\\midrule
		\textbf{Effect of base gas fee $\hat g_L$} & $\hat g_L\le \hat g_H- L_B\hat z$ & $\hat g_L> \hat g_H- L_B\hat z$\\\midrule
		Trading probability $\alpha^*$ & $=1$ & $<1$, strictly decreasing in $\hat g_L$\\
		Arbitrageur's expected profit $\Pi^*$ & $>0$, strictly decreasing in $\hat g_L$ & $=0$\\
		Gas fee paid $\tilde{g}^*$ & Increasing in $\hat g_L$ & Decreasing in $\hat g_L$\\
		Relative trading amount $\tilde{D}^*/y_A$ & Independent of $\hat g_L$ & Decreasing in $\hat g_L$\\ \midrule
		\textbf{Effect of pool liquidity $L_B$} & $L_B\ge \hat g_L/\big((1-O^{-1/2})^2-\hat z\big)$ & $L_B< \hat g_L/\big((1-O^{-1/2})^2-\hat z\big)$\\\midrule
		Trading probability $\alpha^*$ & $=1$ & $<1$, strictly increasing in $L_B$\\
		Arbitrageur's expected profit $\Pi^*$ & $>0$, strictly increasing in $L_B$ & $=0$\\
		Gas fee paid $\tilde g^*$ & \multicolumn{2}{c}{Increasing in $L_B$}\\
		Relative trading amount $\tilde D^*/y_A$ & Independent of $L_B$ & Increasing in $L_B$\\ \midrule
		\textbf{Effect of price discrepancy $O$} & $O\ge \bar{O}$ & $O<\bar{O}$\\\midrule
		Trading probability $\alpha^*$ & $=1$ & $<1$, strictly increasing in $O$\\
		Arbitrageur's expected profit $\Pi^*$ & $>0$, strictly increasing in $O$ & $=0$\\
		Gas fee paid $\tilde g^*$ & \multicolumn{2}{c}{Increasing in $O$}\\
		Relative trading amount $\tilde D^*/y_A$ & \multicolumn{2}{c}{Increasing in $O$}\\
		\bottomrule
	\end{tabular}
	\vspace{0.5em}\par\flushleft\small  The effect of $O$ on $\alpha^*$, $\Phi^*$, $\tilde g^*$, and $\tilde D^*/y_A$ is verified numerically. All other properties are proved theoretically. See Theorem \ref{thm:Main} and Appendix \ref{subse:ComparativeStaticsNoRevert}.
\end{table}

\subsection{Mixed Equilibrium in the Auto-Revert Setting}\label{subse:Slippagezero}

We now consider the auto-revert setting, i.e., the case $\ell=0$. In this setting, the first-mover advantage $V^0(g,d_A;\bar d_A)$ does not depend on $\bar d_A$, so we simply denote it as $V^0(g,d_A)$. Consequently, the reward function in \eqref{eq:ResponseFunction2} takes the form
\begin{align}
	h^0(g,d_A)&=R_F(g, d_A)-\alpha^* V^0(g,d_A)\int_{g}^{\hat g_H}\Phi^*(\dd \bar g) = R_F(g, d_A)-\alpha^* V^0(g,d_A)\int_{g}^{g_H}\Phi^*(\dd \bar g),\label{eq:ResponseFunction2Slippage0}
\end{align}
where the second equality follows because $\Phi^*$ is supported on $[\hat g_L,g_H]$ for some $g_H\in (\hat g_L,\hat g_H]$.

\begin{proposition}\label{prop:Slippage0}
	Fix $\ell=0$. Suppose $(\alpha^*, \Phi^*,D_A^*)\in {\cal A}$ is a symmetric mixed equilibrium strategy. Then $\alpha^*\in (0,1)$, $g_H=\hat g_H$, and the arbitrageur's expected profit is zero.
\end{proposition}

Proposition \ref{prop:Slippage0} shows that $\alpha^*<1$. To see this, suppose $\alpha^*=1$. Then the competitor always trades, and if the arbitrageur chooses the minimum gas fee $\hat g_L$, they are guaranteed to be the second mover. The payoff is negative because they must pay $r \hat g_L$ for the reverted transaction. This is not optimal, since the arbitrageur could instead choose not to trade. Hence, $\alpha^*<1$, which implies that the arbitrageur is indifferent between trading and not trading, and the expected profit of trading is zero. When choosing the maximum gas fee $g_H$, the arbitrageur is guaranteed to be the first mover, so their expected payoff equals the first mover's payoff, maximized at $\hat D_A$ with value $\hat g_H-g_H$. Because the expected payoff of trading is zero, we conclude that $g_H=\hat g_H$.

In view of Proposition \ref{prop:Slippage0}, we restrict attention to $(\alpha^*, \Phi^*,D_A^*)\in {\cal A}$ with $g_H=\hat g_H$ in the search for the equilibrium strategy. Following the same approach as in Section \ref{subse:Slippageinfinity}, and recalling \eqref{eq:GasFeeDensityProp} with $\frac{\partial V^0}{\partial g}=r-1$, we obtain
\begin{align}
	\phi^*(g)=\frac{1-(1-r)\alpha^*\int_g^{\hat g_H}\phi^*(g)\dd g}{\alpha^* V^0(g,D_A^*(g))}.\label{eq:DensityEq1Slippage0}
\end{align}
On the other hand, recalling the first-order condition $\frac{\partial h^0}{\partial d_A}(g,d_A)\big|_{d_A=D_A^*(g)}=0$ and applying the change of variable $\hat x(z):=D_A^*(\hat g_H-L_B z)/y_A$, we derive
\begin{align}
	0 &=\frac{\partial R_F}{\partial d_A}(\hat g_H-L_Bz,y_A\hat x(z)) - \alpha^*\frac{\partial V^0}{\partial d_A}(\hat g_H-L_Bz,y_A\hat x(z))\int_{\hat g_H-L_Bz}^{\hat g_H}\Phi^*(\dd \bar g)\notag\\
	&=\frac{\partial R_F}{\partial d_A}(\hat g_H-L_Bz,y_A\hat x(z)) - \alpha^*L_B\frac{\partial V^0}{\partial d_A}(\hat g_H-L_Bz,y_A\hat x(z))\int_0^{z}\phi^*(\hat g_H-L_B\bar z)\dd\bar z,\label{eq:FOCEq1Slippage0}
\end{align}
for every $z\in[0,(\hat g_H-\hat g_L)/L_B]$. Define
\begin{align*}
	\hat \theta(z):=\alpha^*L_B\phi^*(\hat g_H-L_B z),\quad z\in [0,(\hat g_H-\hat g_L)/L_B],
\end{align*}
recall $Q$ as defined in \eqref{eq:QFun}, and denote
\begin{align}
	K^0_0(x)&:=\frac{y_A}{L_B}\frac{\partial V^0}{\partial d_A}(\hat g_H-L_Bz,y_A x) = Q(x) + \frac{y_A}{L_B}\Gamma'(y_Ax),\quad x\ge 0,\label{eq:K0FunSlippage0}\\
	K^0_1(z,x)&:=\frac{V^0(\hat g_H-L_Bz,y_Ax)}{L_B},\quad z\ge 0,\;x\ge 0.\label{eq:K1FunSlippage0}
\end{align}
Then, \eqref{eq:DensityEq1Slippage0} and \eqref{eq:FOCEq1Slippage0} become
\begin{align}
	Q(\hat x(z)) = K_0^0(\hat x(z))\int_0^z \hat \theta(\bar z)\dd\bar z,\quad
	\hat \theta(z) = \frac{1-(1-r)\int_0^{z}\hat \theta(\bar z)\dd\bar z}{K^0_1(z,\hat x(z))}.\label{eq:KeyODESlippage0}
\end{align}

The solution to \eqref{eq:KeyODESlippage0} is presented in Appendix \ref{subse:KeyODEEll0}. We now state the main result.

\begin{theorem}\label{thm:MainSlippage0}
	Fix $\ell=0$.
	\begin{enumerate}[leftmargin=2em]
		\item[(i)] Suppose $\Gamma\equiv 0$. Then there exists a unique symmetric mixed Nash equilibrium strategy $(\alpha^*, \Phi^*,D_A^*)\in {\cal A}$, given by
		\begin{align}
			\alpha^* &= \frac{1}{1-r}\left(1-r(1-O^{-1/2})^2\big((1-O^{-1/2})^2 - (1-r)(\hat g_L/L_B)\big)^{-1}\right),\label{eq:Slippage0TradingProb}\\
			\phi^*(g) &= \frac{1}{\alpha^* L_B}r(1-O^{-1/2})^2\big((1-O^{-1/2})^2 - (1-r)g/L_B\big)^{-2},\quad g\in [\hat g_L,\hat g_H],\label{eq:Slippage0GasFee}\\
			D_A^*(g) &= (O^{1/2}-1)y_A,\quad g\in [\hat g_L,\hat g_H],\label{eq:Slippage0TradingAmount}
		\end{align}
		where in the case $r=1$, the value of $\alpha^*$ is obtained by taking the limit $r\to 1$ in \eqref{eq:Slippage0TradingProb}, i.e., $\alpha^* = 1-\hat g_L/\big(L_B(1-O^{-1/2})^2\big)$.
		\item[(ii)] Suppose $\Gamma'(d_A)>0$ and $\Gamma''(d_A)\ge0$ for all $d_A>0$, and assume $z_\infty\ge (1-O^{-1/2})^2$, where $z_\infty$ is given in Lemma \ref{le:ODESlippage0}. Then there exists a unique symmetric mixed equilibrium strategy $(\alpha^*, \Phi^*,D_A^*)\in {\cal A}$, given by
		\begin{align}
			\alpha^* &= \int_0^{(\hat g_H-\hat g_L)/L_B}\hat \theta(z)\dd z,\label{eq:Slippage0rTradingProb}\\
			\phi^*(g) &= \frac{1}{\alpha^*L_B}\hat \theta\big((\hat g_H-g)/L_B\big),\quad g\in [\hat g_L,\hat g_H],\label{eq:Slippage0rGasFee}\\
			D_A^*(g) &= y_A \hat x\big((\hat g_H-g)/L_B\big),\quad g\in [\hat g_L,\hat g_H],\label{eq:Slippage0rTradingAmount}
		\end{align}
		where $(\hat x,\hat \theta)$ is given in Lemma \ref{le:ODESlippage0}-(ii).
	\end{enumerate}
\end{theorem}

Theorem \ref{thm:MainSlippage0}-(i) and (ii) address the cases of no inventory risk and positive inventory risk, respectively. Note that the condition $\Gamma''(d_A)\ge0$ in case (ii) is mild, since inventory risk is typically convex in the size of residual positions. The requirement $z_\infty\ge (1-O^{-1/2})^2$ is technical; Appendix \ref{appx:gammaPositive} shows that it holds for empirically relevant parameter values.

Table \ref{tbl:property_revert} summarizes key properties of the mixed Nash equilibrium strategy in the auto-revert setting. Formal statements and proofs are provided in Appendix \ref{appx:CSSlippage0}. We highlight several main results here.

First, the density function of the gas fee is constant or strictly increasing when inventory risk is zero, in sharp contrast to the no-revert setting. With substantial inventory risk, however, the gas fee density becomes strictly decreasing, as in the no-revert case.

Second, as shown in Proposition \ref{prop:Slippage0}, the arbitrageur always has zero expected profit and a positive probability of abstaining from trade. In the no-revert setting, by contrast, the arbitrageur may earn positive expected profit and always trade.

Third, with zero inventory risk, the arbitrageur selects the maximum trading amount $\hat D_A$. In this case, the DEX transaction that is not executed is reverted, and the completed CEX transaction does not generate inventory risk. Hence, the arbitrageur fully exploits the price discrepancy, assuming their trade is executed first. This contrasts with both the no-revert setting and the auto-revert setting with positive inventory risk, where the arbitrageur trades less than $\hat D_A$ to avoid losses from being the second mover or from exposure to inventory risk.

Fourth, the effects of liquidity $L_B$, price discrepancy $O$, and base gas fee $\hat g_L$ on the gas fee and trading amount are qualitatively similar to those in the no-revert setting.

\begin{table}
	\caption{Properties of the equilibrium strategy $(\alpha^*,\Phi^*,D_A^*)$ in the auto-revert setting.}\label{tbl:property_revert}
	\centering
	\footnotesize
	\begin{tabular}{clc}
		\toprule
		\textbf{Gas Fee Density $\phi^*(g)$} & Zero inventory risk: Constant or strictly increasing in $g$\\
		& Large inventory risk: Strictly decreasing in $g$\\\midrule
		\textbf{Trading amount given gas fee $D_A^*(g)$} & Zero inventory risk: $D_A^*(g)\equiv \hat D_A$\\
		& Positive inventory risk: Strictly increasing in $g$ and less than $\hat D_A$\\\midrule
		\textbf{Arbitrageur's expected profit $\Pi^*$} & 0\\\midrule
		\multicolumn{2}{c}{\textbf{Effect of base gas fee $\hat g_L$}} & \\\midrule
		Trading probability $\alpha^*$ & $<1$ and strictly decreasing in $\hat g_L$\\
		Gas fee paid $\tilde{g}^*$ & Decreasing in $\hat g_L$\\
		Relative trading amount $\tilde{D}^*/y_A$ & Decreasing in $\hat g_L$\\ \midrule
		\multicolumn{2}{c}{\textbf{Effect of pool liquidity $L_B$}}\\\midrule
		Trading probability $\alpha^*$ & $<1$ and strictly increasing in $L_B$\\
		Gas fee paid $\tilde g^*$ & Increasing in $L_B$\\
		Relative trading amount $\tilde D^*/y_A$ & Increasing in $L_B$\\ \midrule
		\multicolumn{2}{c}{\textbf{Effect of price discrepancy $O$}}\\\midrule
		Trading probability $\alpha^*$ & $<1$ and strictly increasing in $O$\\
		Gas fee paid $\tilde g^*$ & Increasing in $O$\\
		Relative trading amount $\tilde D^*/y_A$ & Increasing in $O$\\
		\bottomrule
	\end{tabular}
	\vspace{0.5em}\par\flushleft\small Results for the zero-inventory-risk case, as well as those concerning expected profit and the dependence of trading amount on the gas fee, are proved theoretically. Other results are verified numerically. See Proposition \ref{prop:Slippage0}, Theorem \ref{thm:MainSlippage0}, and Appendix \ref{appx:CSSlippage0}.
\end{table}

\subsection{Mixed Equilibrium in the Selectable-Revert Setting}\label{subsect:TCR}

In the selectable-revert setting, $\ell\in\{0,\infty\}$ becomes part of the arbitrageur's action. Similar to the discussion in Section \ref{subse:MixedEquilibrium}, we consider mixed symmetric equilibrium strategies. Denote by $\alpha$ the probability of trading and by $\lambda$ the randomized strategy of the arbitrageur conditional on trading, which is a probability measure on $\{0,\infty\}\times [\hat g_L,\hat g_H]\times [0,\hat D_A]$. The payoff function can then be written as
\begin{align}\label{eq:GamePayoffChoiceMixed}
	\begin{split}
		\Pi(\alpha,\lambda;\bar \alpha,\bar \lambda):&= \mathbb{E}[R^\ell] = \alpha (1-\bar \alpha)\int R_F(g, d_A)\lambda(\dd \ell \dd g\dd d_A)\\
		&+\alpha \bar \alpha \bigg[ \int R_F(g, d_A)\mathbf 1_{g>\bar g}\lambda (\dd \ell\dd g\dd d_A)\bar \lambda(\dd\bar \ell\dd \bar g\dd\bar d_A)\\
		&+ \int \tfrac{1}{2}\left(R_F(g, d_A)+R_S^\ell(g, d_A;\bar d_A)\right)\mathbf 1_{g=\bar g}\lambda(\dd\ell \dd g\dd d_A)\bar \lambda(\dd\bar \ell \dd\bar g \dd\bar d_A)\\
		&+ \int R_S^\ell(g, d_A;\bar d_A)\mathbf 1_{g<\bar g}\lambda(\dd\ell \dd g\dd d_A)\bar \lambda(\dd \bar \ell \dd\bar g\dd\bar d_A)\bigg].
	\end{split}
\end{align}

The analysis of the mixed symmetric equilibrium strategy is similar to the case when the revert setting $\ell$ is fixed, so we relegate the details to Appendix \ref{appx:slippagechoice}. By Proposition \ref{prop:propertyEqChooseSlippage} therein, we only need to focus on strategies of the form $(\alpha,\Phi,\rho, D_A^0,D_A^\infty)$. Here, $\alpha$ is the probability of trading. Conditional on trading, $\Phi$ is a probability measure on $[\hat g_L,\hat g_H]$ representing the distribution of the randomized gas fee, $\rho$ is a function on $[\hat g_L,\hat g_H]$ taking values in $[0,1]$, where $\rho(g)$ represents the probability of choosing the auto-revert setting $\ell=0$ conditional on gas fee $g$, and $D_A^{\ell}$ is a function on $[\hat g_L,\hat g_H]$ with $D_A^\ell(g)$ denoting the trading amount conditional on gas fee $g$ and transaction reversion setting $\ell$. Note that when $\rho(g)=1$, the value of $D_A^\infty(g)$ is irrelevant, and when $\rho(g)=0$, the value of $D_A^0(g)$ is irrelevant.

The following propositions show that the selectable-revert setting can degenerate to either the auto-revert or the no-revert setting under certain conditions. Proposition \ref{prop:MixedIC0} states that, without inventory risk, choosing to revert with certainty is an equilibrium, so the selectable-revert setting reduces to the auto-revert setting. Proposition \ref{prop:MixedICLarge} shows that, with sufficiently large inventory risk and non-refundable gas fees for reverted transactions, the arbitrageur will always choose not to revert, so the selectable-revert setting reduces to the no-revert setting.

\begin{proposition}\label{prop:MixedIC0}
	Suppose traders can choose the transaction reversion setting $\ell\in \{0,\infty\}$. Assume zero inventory risk, i.e., $\Gamma(d_A)\equiv 0$. Then $(\alpha^*,\Phi^*,\rho^*,D_A^{0,*},D_A^{\infty,*})$ with $\rho^*(g)\equiv 1$ and $(\alpha^*,\Phi^*,D_A^{0,*})$ as given in Theorem \ref{thm:MainSlippage0}(i) is a mixed symmetric equilibrium strategy.
\end{proposition}

\begin{proposition}\label{prop:MixedICLarge}
	Suppose traders can choose the transaction reversion setting $\ell\in \{0,\infty\}$. Assume $r=1$ and that the inventory risk satisfies
	\begin{align}
		\Gamma(d_A)\ge 2(1+f)p_Ay_A^{-1} O^{-1/2}d_A^2,\quad d_A\in [0,\hat D_A].\label{eq:InventoryCostLowerBound}
	\end{align}
	Then, in the mixed symmetric equilibrium strategy, $\rho^*(g)\equiv 0$, and the trading probability, gas fee distribution, and trading amount are uniquely given as in Theorem \ref{thm:Main}.
\end{proposition}

In Appendix \ref{appx:slippagechoice}, we restrict attention to gas fee distributions $\Phi^*$ that admit a density function $\phi^*$, and we derive the equations satisfied by the equilibrium strategy. In the following numerical illustrations, we use the parameter values described below.

First, the base gas fee $\hat g_L$ is equal to the base fee per gas multiplied by the gas used. We use the median of the base fee per gas in blocks with arbitrage and the mode of the gas used in our empirical study (Section \ref{sec:empirical}), yielding a base gas fee of 5.6 USD. Second, we set $L_B$ to the median pool liquidity in blocks with arbitrage in the empirical data, which is 47,021,346 USD. Third, as discussed in Section \ref{subse:Competition}, an arbitrage opportunity exists in the presence of gas fees if and only if $M:=O/(1-(\hat g_L/L_B)^{1/2})^{-2}>1$. We compute $M$ for all blocks with arbitrage in the empirical data and take quantiles from the 5\% to the 95\% level in 10\% increments. We then compute $O=(1-(\hat g_L/L_B)^{1/2})^{-2}M$ using the chosen values of $\hat g_L$ and $L_B$ and the ten quantiles of $M$, resulting in ten values of $O$. Fourth, we set $r=0.2$, consistent with the proportion of gas consumed by reverted transactions in public mempools. Finally, we specify the inventory risk as $\Gamma(d_A) = \tfrac{1}{2}\gamma d_A^2$, where $\gamma\ge 0$ represents the arbitrageur's inventory risk aversion degree. These parameter values are summarized in Table \ref{tab:ParameterValues}.

\begin{table}
	\begin{center}
		\caption{Default Parameter Values Based on Empirical Data}\label{tab:ParameterValues}
		\footnotesize
		\begin{tabular}{c|c|c|c|c|c|c|c|c|c}
			\toprule
			\multicolumn{3}{c}{$\hat g_L$ (USD)} & \multicolumn{3}{c}{$L_B$ (USD)} & \multicolumn{2}{c}{$r$} &  \multicolumn{2}{c}{$\Gamma(d_A)$}\\
			\midrule
			\multicolumn{3}{c}{5.6} & \multicolumn{3}{c}{47,021,346} & \multicolumn{2}{c}{0.2} & \multicolumn{2}{c}{$\tfrac{1}{2}\gamma d_A^2$}\\
			\bottomrule
			\multicolumn{10}{c}{Ten values of $O$}\\
			\midrule
			1.00074 & 1.00081 & 1.00088 & 1.00094 & 1.00101 & 1.00109 & 1.00118 & 1.00132 & 1.00151 & 1.00213\\
			\bottomrule
		\end{tabular}
	\end{center}
\end{table}

\begin{figure}
	\centering
	\includegraphics[width=0.46\linewidth]{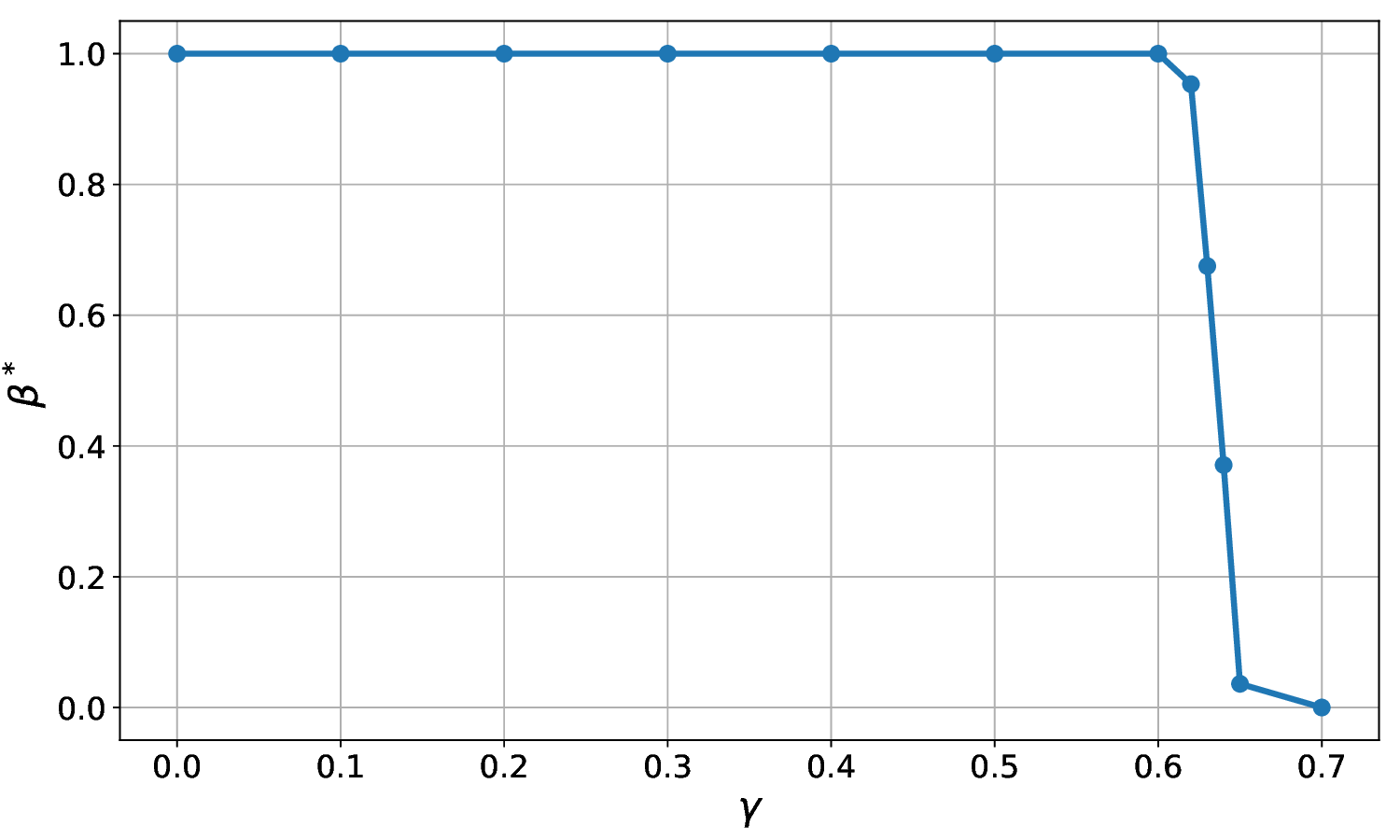}
	\includegraphics[width=0.46\linewidth]{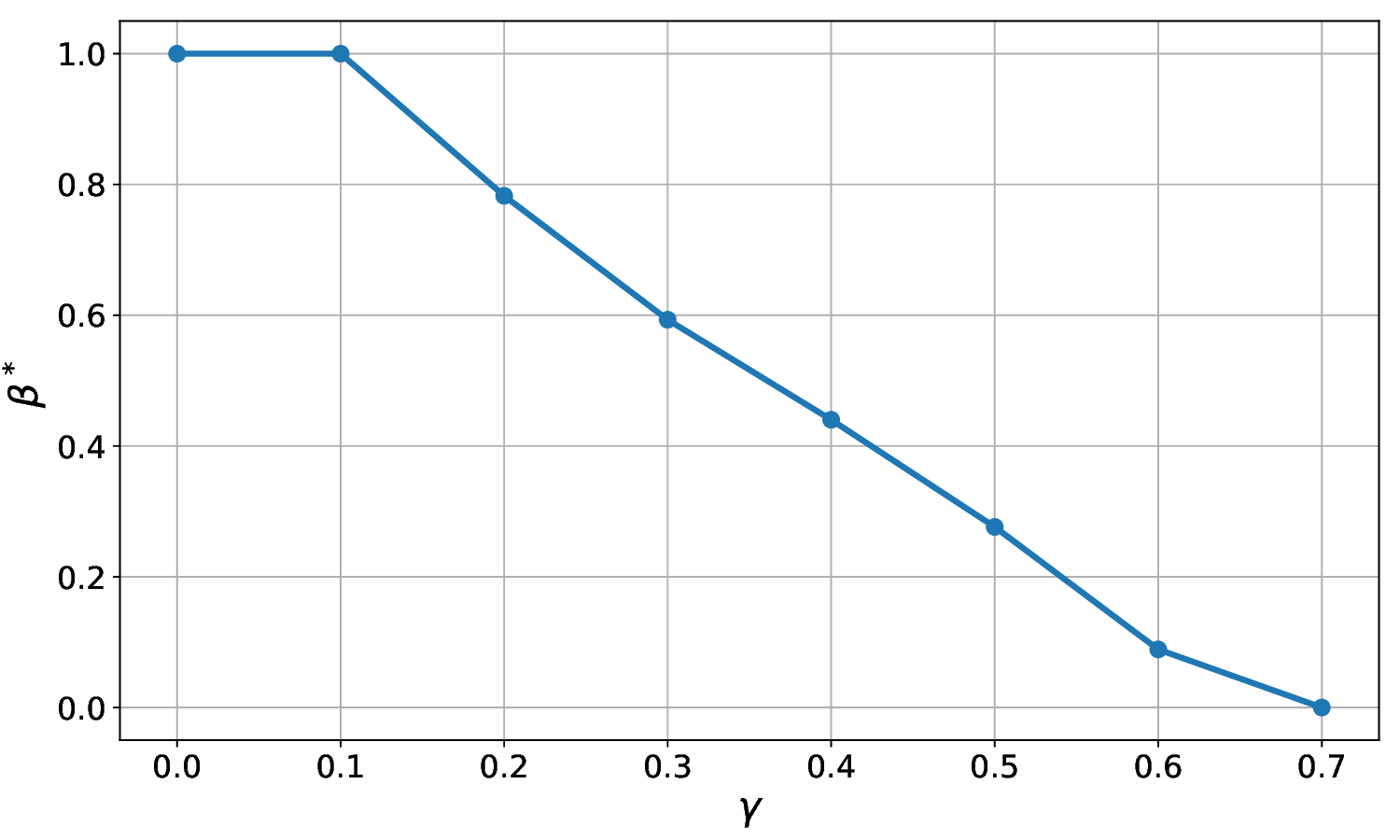}
	\caption{Probability of choosing to revert transactions as a function of inventory risk aversion $\gamma$ in the selectable-revert setting. Parameter values are given in Table \ref{tab:ParameterValues}, with $O=1.00074$ in the left panel and $O=1.00213$ in the right panel.}
	\label{fig:beta_gamma}
\end{figure}

We first compute the arbitrageur's probability of choosing to revert, conditional on trading, as
\begin{align*}
	\beta^* = \int \rho^*(g)\Phi^*(\dd g),
\end{align*}
and plot it against $\gamma$ in Figure~\ref{fig:beta_gamma}. We observe that $\beta^*$ decreases continuously with inventory risk aversion, consistent with Propositions \ref{prop:MixedIC0} and \ref{prop:MixedICLarge}.

\begin{figure}
	\centering
	\includegraphics[width=0.46\linewidth]{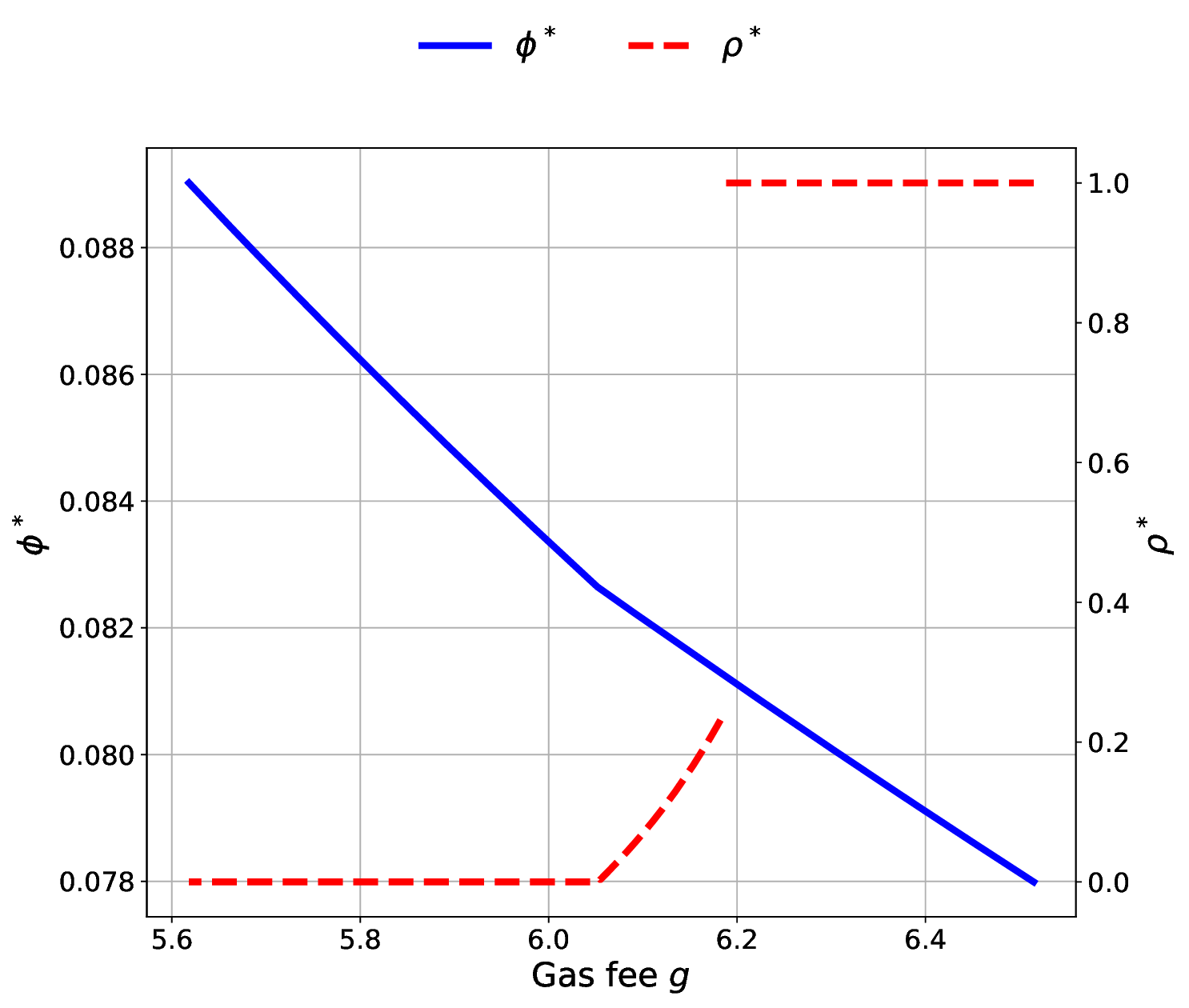}
	\includegraphics[width=0.46\linewidth]{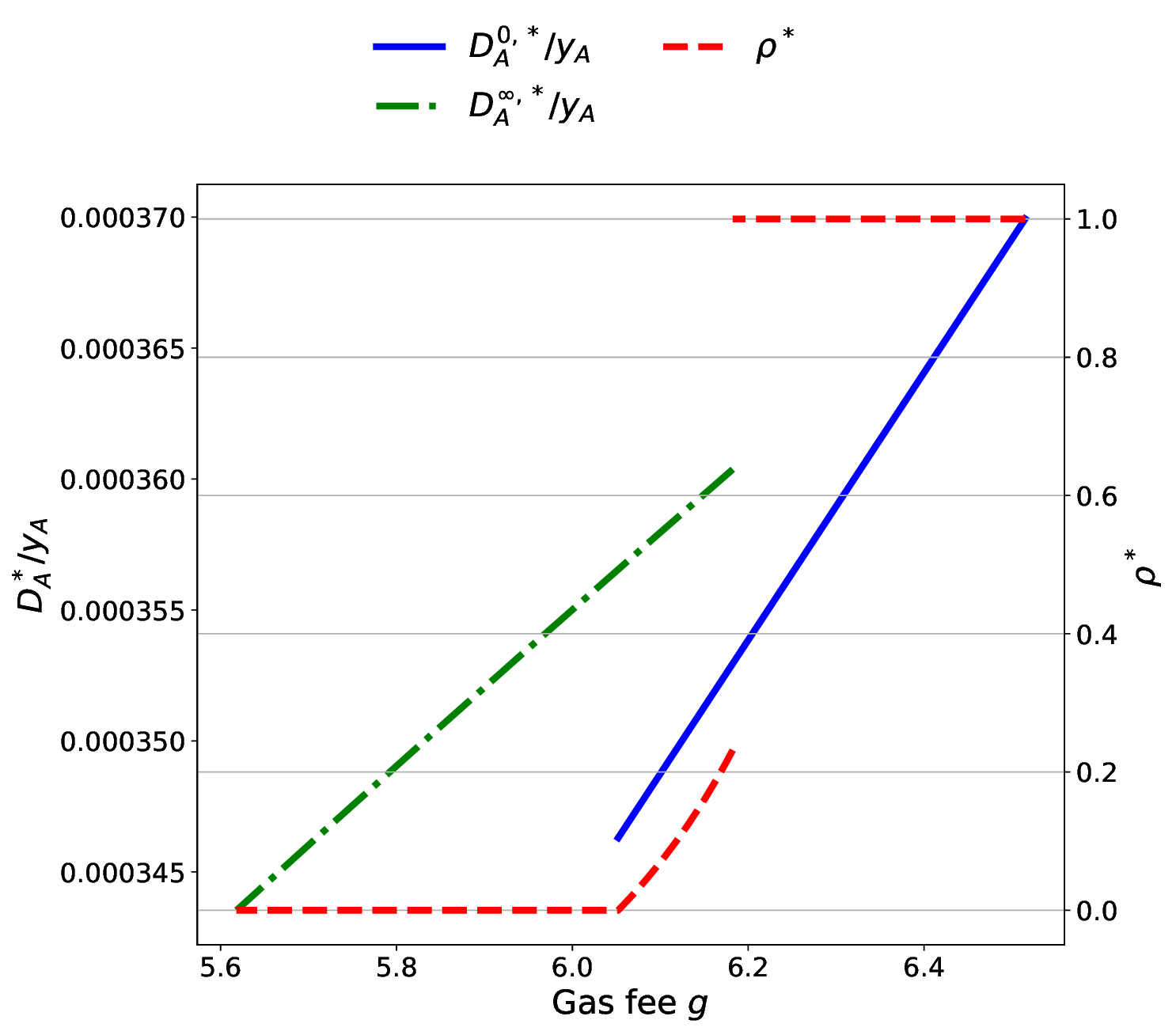}
	\caption{$\rho^*$ (right axis), $\phi^*$ (left panel, left axis), and $D^{\ell,*}/y$, $\ell\in\{0,\infty\}$ (right panel, left axis) as functions of gas fee $g$ in the selectable-revert setting. Parameter values are given in Table \ref{tab:ParameterValues}, with $O=1.00074$ and $\gamma=0.64$.}
	\label{fig:phi_rho}
\end{figure}

Figure \ref{fig:phi_rho} illustrates the equilibrium strategy $(\phi^*,\rho^*,D^{0,*},D^{\infty,*})$ with positive inventory risk. We observe that the choice to revert transactions is coupled with more aggressive gas bidding. This can be explained as follows: In the no-revert setting, the second mover's payoff can still be positive, provided both arbitrageurs avoid aggressive strategies. In the auto-revert setting, however, the second mover's payoff is always negative, since they pay the gas fee and bear inventory risk without extracting value from the arbitrage opportunity. Therefore, arbitrageurs are more incentivized to be the first mover and thus bid more aggressively in the auto-revert setting than in the no-revert setting. Consequently, in the selectable-revert setting, arbitrageurs bid high gas fees when choosing to revert and low gas fees when choosing not to revert.

Because arbitrageurs are more likely to choose to revert when inventory risk is zero or small, and more likely to choose not to revert when inventory risk is high, the properties of the equilibrium in these two cases resemble those in the auto-revert setting (Section \ref{subse:Slippagezero}) and the no-revert setting (Section \ref{subse:Slippageinfinity}), respectively. This conjecture is confirmed in Appendix \ref{appx:propertyTCR}, and we highlight several key properties here:

First, with zero or low inventory risk, the gas fee density function is increasing, while with high inventory risk it is decreasing.

Second, with zero inventory risk, the arbitrageur always trades the maximum amount. With positive inventory risk, the arbitrageur trades less than the maximum, and conditional on the revert choice, the trading amount increases with the gas fee.

Third, with zero or low inventory risk, the arbitrageur's expected profit is zero. With high inventory risk, expected profit can be positive when the price discrepancy is large.

Fourth, the trading amount per unit of reserve in the pool increases with the price discrepancy, consistent with the no-revert and auto-revert cases. Its dependence on liquidity and the base gas fee, however, is not monotone in general, in contrast to the results in the no-revert and auto-revert settings. The gas fee increases with the price discrepancy and liquidity level, consistent with the other settings, but its dependence on the base gas fee  is not monotone.

\section{Empirical Studies}\label{sec:empirical}
In this section, we conduct empirical studies using real market data.

\subsection{Data Description}\label{sec:1}
We focus on the cryptocurrency pair \textit{Wrapped Ether} (WETH) and \textit{Tether} (USDT), where the former represents the ERC-20 tokenized version of \textit{Ether} (ETH) and the latter is a stablecoin pegged to the U.S. dollar (USD). This pair is among those with the highest total value locked in Uniswap V2. In our model, we treat WETH and USDT as assets $A$ and $B$, respectively.

Our dataset consists of three parts. First, we collect 100,000 swap transactions for the WETH/USDT pair on Uniswap V2 from October 27, 2024 to December 1, 2024.\footnote{The specific time range is from 10:26:35 UTC, October 27, 2024 to 05:08:47 UTC, December 1, 2024. The WETH/USDT pair address on Uniswap V2 is obtained via the subgraph \url{https://thegraph.com/explorer/subgraph/ianlapham/uniswapv2}.} Each swap record contains the timestamp, block number, swap amount, gas used, priority fee per gas (in ETH), and the total gas fee (in ETH) paid for the swap. Note that all swap transactions in this period, whether they were submitted in public mempools or private pools, are included in these 100,000 transactions. We, however, are not able to identify the submitting venue for each transaction.

Second, we collect blockchain data for the blocks containing these 100,000 swaps.\footnote{The block numbers range from 21,056,358 to 21,305,999, collected via Infura (\url{https://www.infura.io/zh}).} Each record includes the block number, block generation time, WETH/USDT reserve amounts on Uniswap V2 at block generation, and the base fee per gas.

Third, we collect second-by-second ETH price data from Binance, the largest centralized exchange (CEX) for cryptocurrency trading. Since WETH is pegged to ETH, they share the same price. As USDT is pegged to USD, its price is fixed at \$1.

Finally, the unit transaction fee $f$ on Uniswap V2 is 0.3\%.

\subsection{Identification of Arbitrage Opportunities}
As discussed in Section \ref{subse:Competition}, when asset $B$ is overpriced relative to asset $A$ on the CEX compared to the DEX, i.e., when $(p_B/p_A)/(y_A/y_B)>1$, an arbitrage opportunity exists (accounting for gas fees) if and only if
\begin{align*}
	M:=\frac{O_B}{(1-(\hat g_L/L_B)^{1/2})^{-2}}>1,
\end{align*}
where $O_B:=(y_Bp_B)/\big(y_Ap_A(1+f)\big)$. Symmetrically, when asset $A$ is overpriced relative to asset $B$ on the CEX compared to the DEX, i.e., when $(p_B/p_A)/(y_A/y_B)<1$, an arbitrage opportunity exists if and only if
\begin{align*}
	M:=\frac{O_A}{(1-(\hat g_L/L_A)^{1/2})^{-2}}>1,
\end{align*}
where $O_A:=(y_Ap_A)/\big(y_Bp_B(1+f)\big)$ and $L_A:=y_Ap_A$.

To identify arbitrage opportunities in our data, we compute $M$ for each block. This requires $p_i$, $y_i$ for $i\in\{A,B\}$, and $\hat g_L$. The reserves $y_A$ and $y_B$ are obtained from blockchain data and remain constant between block generations (approximately 12 seconds). The CEX price, however, fluctuates within each block. Using second-by-second price data, we select the second in each block where the price discrepancy, $\max(O_A,O_B)$, is maximized. We then set $p_i$ to the CEX price of asset $i$ at that second. This price is also used to convert gas fees denominated in ETH into USD. Notably, in every block in our dataset, either $O_A\le 1$ for all 12 seconds or $O_B\le 1$ for all 12 seconds. Thus, no block contains arbitrage opportunities in opposite directions.

The base gas fee $\hat g_L$ equals the base fee per gas multiplied by the gas used. The base fee per gas depends on network traffic and is known once the block is generated. We assume traders' estimates are precise and use the actual base fee per gas in each block. Gas usage varies across transactions, but the most frequent value in our data is 107,176, suggesting a standard transaction consumes this amount. We therefore assume this value in calculating $\hat g_L$.

We classify a block as an {\em arbitrage block} if $M>1$, and a {\em non-arbitrage block} otherwise. In an arbitrage block, asset $B$ may be overpriced relative to $A$ on the CEX compared to the DEX or vice versa. To exploit arbitrage, one swaps $A$ for $B$ on the DEX in the former case and the reverse in the latter. Within an arbitrage block, we classify a swap as an {\em arbitrage swap} if its direction matches the arbitrage strategy, its priority fee is positive, and its first-mover profit, defined as the profit assuming the swap is executed first, is non-negative.

Table \ref{tab:arbitrage_swap_block} reports the number of arbitrage blocks. Overall, only 3.356\% of blocks contain arbitrage opportunities. Breaking down by the number of swaps per block, we find that the fraction of arbitrage blocks increases with the number of swaps, suggesting that arbitrage blocks tend to attract more trading activity, consistent with the intuition that arbitrage opportunities draw arbitrageurs.

We also break down arbitrage blocks by the number of arbitrage swaps per block (see the last four columns of Table \ref{tab:arbitrage_swap_block}). None of the arbitrage blocks contains more than two arbitrage swaps. This justifies our assumption of modeling two arbitrageurs.

\begin{table}[htbp]
	\centering
	\caption{Block counts by number of swaps and arbitrage swaps.}
	\footnotesize
	\begin{tabular}{ccccccccccc}
		\toprule
		& \multicolumn{5}{c}{\textbf{Broken down by swap count}} && \multicolumn{4}{c}{\textbf{By arbitrage swap count}} \\\cmidrule{2-6}\cmidrule{8-11}
		& \textbf{0} & \textbf{1} & \textbf{2} & \textbf{$\ge$3} & \textbf{Total} && \textbf{0} & \textbf{1} & \textbf{2} & \textbf{$\ge$3}\\
		\midrule
		\textbf{Block count} &170,662 &61,784&14,183&3,013 &249,642 & & --& -- & -- &--\\
		\textbf{Arbitrage block count} &2,376 &4,165&1,403&433 &8,377 & & 6,640 & 1,725 & 12 & 0 \\
		\midrule
		\textbf{\% of arbitrage blocks} &1.392$\%$ & 6.741$\%$ & 9.892$\%$ & 14.371$\%$ &3.356$\%$ & & --& -- & -- &--\\
		\bottomrule
	\end{tabular}%
	\label{tab:arbitrage_swap_block}
\end{table}

Finally, we identify sequences of consecutive blocks with the same arbitrage direction and define the {\em arbitrage block duration} as the length of such sequences. We find 3,598 sequences of duration 1, 1,214 of duration 2, and 635 of duration longer than 2. These results indicate that some arbitrage opportunities persist across multiple blocks, consistent with our model's implication that arbitrageurs may choose not to trade with positive probability even when arbitrage opportunities exist.

\subsection{Inventory Risk Aversion and Revertible Transactions}

On Uniswap V2, investors can set a slippage tolerance when submitting transactions in public mempools. A very low tolerance causes the transaction to be reverted if it is not executed first, while a very high tolerance ensures that it is not reverted. Investors can also submit transactions privately with execution conditions that automatically trigger a revert if prespecified price or profitability requirements are not met. As shown in Section \ref{subsect:TCR}, the likelihood that arbitrageurs choose to revert transactions depends on their inventory risk aversion. Specifically, with zero inventory risk aversion, arbitrageurs revert transactions that are not executed first, whereas with high inventory risk aversion, they do not revert. In the data, we cannot directly observe investors' inventory risk aversion or their slippage tolerances. Instead, we provide indirect evidence suggesting that arbitrageurs exhibit positive inventory risk aversion and do not always choose to revert transactions.

First, if all arbitrageurs reverted transactions not executed first, we would observe at most one arbitrage swap in each arbitrage block. However, as shown in Table \ref{tab:arbitrage_swap_block}, 12 arbitrage blocks contain at least two arbitrage swaps.
This indicates that arbitrageurs have a positive probability of choosing not to revert transactions. In particular, they must have positive inventory risk aversion; otherwise, they would revert transactions with certainty (see Proposition \ref{prop:MixedIC0}).

Second, if arbitrageurs had zero inventory risk aversion, they would always take the maximum trading amount, which is optimal when transactions are guaranteed to be executed first (see Proposition \ref{prop:MixedIC0} and Theorem \ref{thm:MainSlippage0}). We computed the maximum trading amount in each arbitrage block and compared it to the actual amounts of arbitrage swaps in the block. We found that 88.85\% of arbitrage swaps involve transaction amounts smaller than the maximum amount. This again implies that arbitrageurs exhibit positive inventory risk aversion.

\begin{figure}
	\centering
	\includegraphics[width=0.68\linewidth]{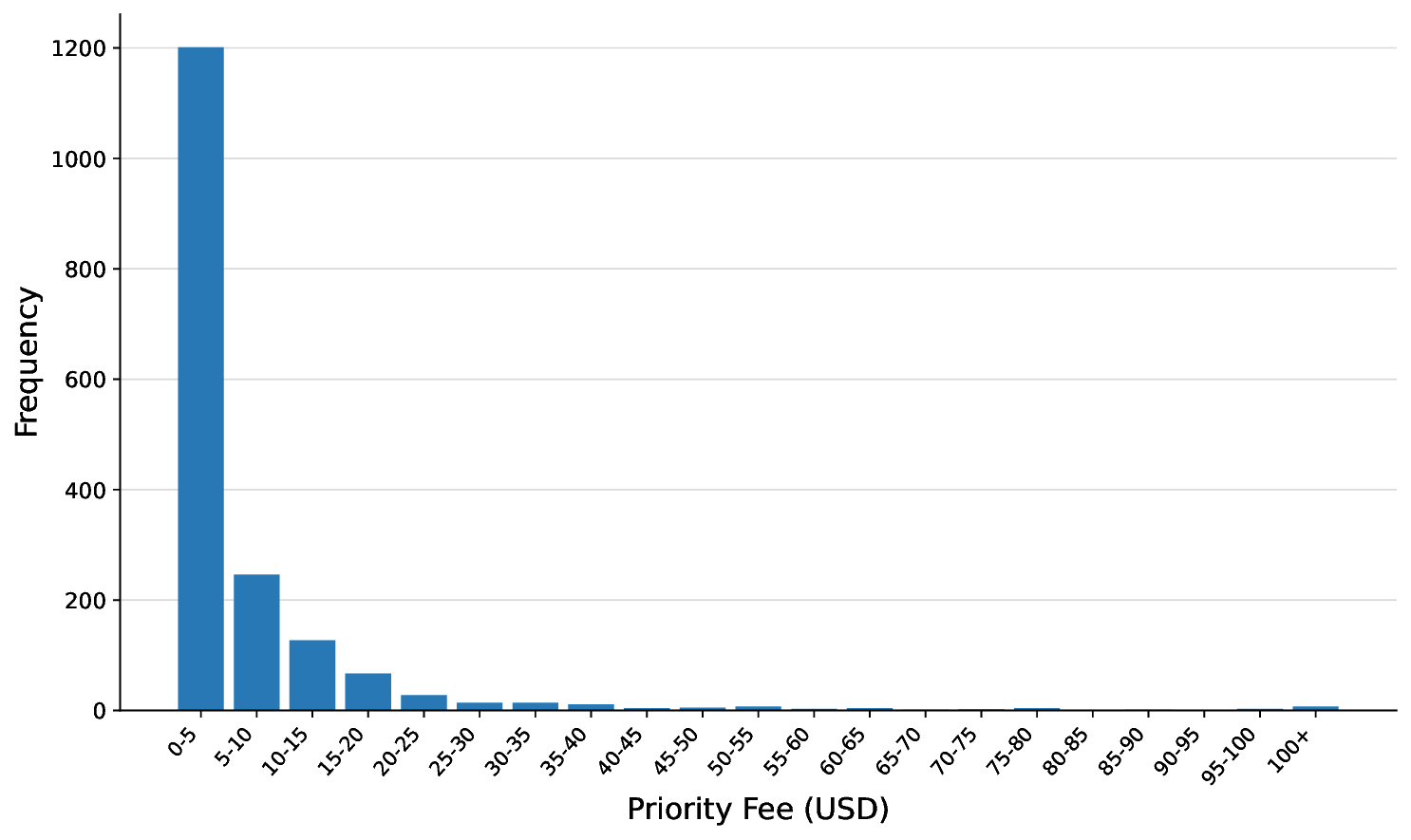}
	\caption{Histogram of priority fees in arbitrage swaps.}
	\label{fig:priority}
\end{figure}

Third, if arbitrageurs had zero inventory risk aversion, they would revert transactions not executed first (see Proposition \ref{prop:MixedIC0}). As discussed in Section \ref{subse:Slippagezero}, this would imply an increasing gas fee density function, meaning arbitrageurs would be more likely to choose higher gas fees (i.e., specify higher priority fees). By contrast, if arbitrageurs do not revert transactions, they are more likely to choose lower priority fees (see Section \ref{subse:Slippageinfinity}). Figure \ref{fig:priority} plots the histogram of priority fees in arbitrage swaps, showing a decreasing probability density. This provides further evidence that arbitrageurs have positive inventory risk aversion.

\subsection{Test of Model Implications}

We compute the average size of arbitrage opportunities, $M$, in arbitrage blocks with 0, 1, and 2 arbitrage swaps, as reported in the first row of Table \ref{tab:TradingProbProfit}. From these values and the Welch one-sided $t$-test results in the second row, we observe that larger arbitrage opportunities are associated with a higher likelihood of trading. This finding is consistent with our model implication that the trading probability increases with the degree of mispricing and pool liquidity, and decreases with the base gas fee.

\begin{table}
	\centering
	\caption{Average arbitrage size $M$ across blocks and ex ante profitability of arbitrage swaps by arbitrage swap count.}\label{tab:TradingProbProfit}
	\footnotesize 
	\begin{tabular}{clll}
		\toprule
		& \multicolumn{3}{c}{\textbf{Broken down by arbitrage swap count}} \\
		\cmidrule(lr){2-4}
		& \textbf{0} & \textbf{1} & \textbf{2}  \\
		\midrule
		\textbf{Average arbitrage size $M$} & 1.000383 & 1.000489 & 1.001160  \\
		\multicolumn{1}{r}{\em Difference} & -- & 0.000106*** & 0.000671***  \\
		\midrule
		\textbf{Profitability of arbitrage swaps} & -- & 94.145\% & 54.167\% \\
		\multicolumn{1}{r}{\em Difference} & -- & -- & $-$39.978\%***  \\
		\bottomrule
	\end{tabular}
	\vspace{0.5em}\par\flushleft\footnotesize {\em Note:} Rows 4 and 6 report differences across groups and statistical test results. Symbols `*', `**', and `***' denote significance at the 5\%, 1\%, and 0.1\% levels, respectively.
\end{table}

For each arbitrage swap, assuming the arbitrageur takes the reverse trade on the CEX, we compute the {\em ex ante profit and loss} (P\&L) using equation \eqref{eq:FMNetProfit}, where the gas fee $g$ and trading amount $d_i$ are extracted from the swap data, and the prices $p_i$ are taken from Binance, $i\in\{A,B\}$. The ex ante P\&L corresponds to the arbitrageur's payoff in our model, conditional on trading. The third row of Table \ref{tab:TradingProbProfit} reports the {\em ex ante profitability of arbitrage swaps}, i.e., the percentage of arbitrage swaps with positive ex ante P\&L, broken down by arbitrage blocks with 1 and 2 arbitrage swaps. The corresponding two-proportion $z$-test results are shown in the fourth row. Profitability decreases with the number of arbitrage swaps in a block, consistent with our model implication that greater competition reduces profit.

Finally, recall two robust implications of our model across different transaction reversion settings: (i) the relative trading amount (trading amount per unit of reserve) increases with the gas fee and price discrepancy; and (ii) the gas fee increases with liquidity and price discrepancy (see Sections \ref{subse:Slippageinfinity}, \ref{subse:Slippagezero}, and \ref{subsect:TCR}). To test these implications, we run linear regressions with the relative swap amount $D_i/y_i$ and gas fee $g$ as dependent variables, and base gas fee $\hat g_L$, liquidity $y_ip_i$, price discrepancy $O_i$, and gas fee $g$ as explanatory variables across all arbitrage swaps, where $i$ denotes the asset deposited into the liquidity pool. Unlike the comparative statics in our model, the relative swap amount and gas fee in the data are conditional on trading and thus exclude the case of no trading. Consequently, they do not correspond exactly to the comparative statics. Nonetheless, the regression results provide evidence on whether our model implications align with empirical data. 
We find that the relative trading amount, conditional on trading, increases with price discrepancy and gas fee, and that the gas fee, conditional on trading, increases with price discrepancy and liquidity. Both findings are consistent with our model implications.

\begin{table}[htbp!]
	\centering
	\caption{Regression results across arbitrage swaps.}\label{tab:regression}
	\footnotesize
	\begin{tabular}{lll}
		\toprule
		& \textbf{Relative Swap Amount} & \textbf{Gas Fee} \\
		\midrule
			\textbf{Base Gas Fee} & 0.143*** & 0.053** \\
			\textbf{Liquidity} & $-$0.068*** & 0.056*** \\
			\textbf{Price Discrepancy} & 0.318*** & 0.748*** \\
			\textbf{Gas Fee} & 0.536*** & \\
		\midrule
		$R^2$ & 0.800 & 0.594 \\
			\textbf{Observations} & 1,749 & 1,749 \\
		\bottomrule
	\end{tabular}
	\vspace{0.5em}\par\flushleft\footnotesize {\em Note:} All regression variables are mean-centered and scaled to unit variance. Dependent variables are relative swap amount $D_i/y_i$, conditional on trading, and gas fee $g$. Explanatory variables are base gas fee $\hat g_L$, liquidity $L_i$, price discrepancy $O_i$, and gas fee $g$, where the arbitrage swap deposits asset $i$ to the liquidity pool. Symbols ``*", ``**", and ``***" indicate significance at the 5\%, 1\%, and 0.1\% levels, respectively.
\end{table}%

\section{Comparison of Different Settings of Transaction Reversion}\label{sec:compareSettings}
In practice, investors can choose to revert transactions that are not executed. However, it is unclear whether the selectable-revert setting is better than the no-revert and auto-revert settings. In this section, we compare these three settings in terms of trading behavior, arbitrageur incentives, and market efficiency.

\begin{figure}[htbp!]
	\centering
	\includegraphics[width=0.45\linewidth]{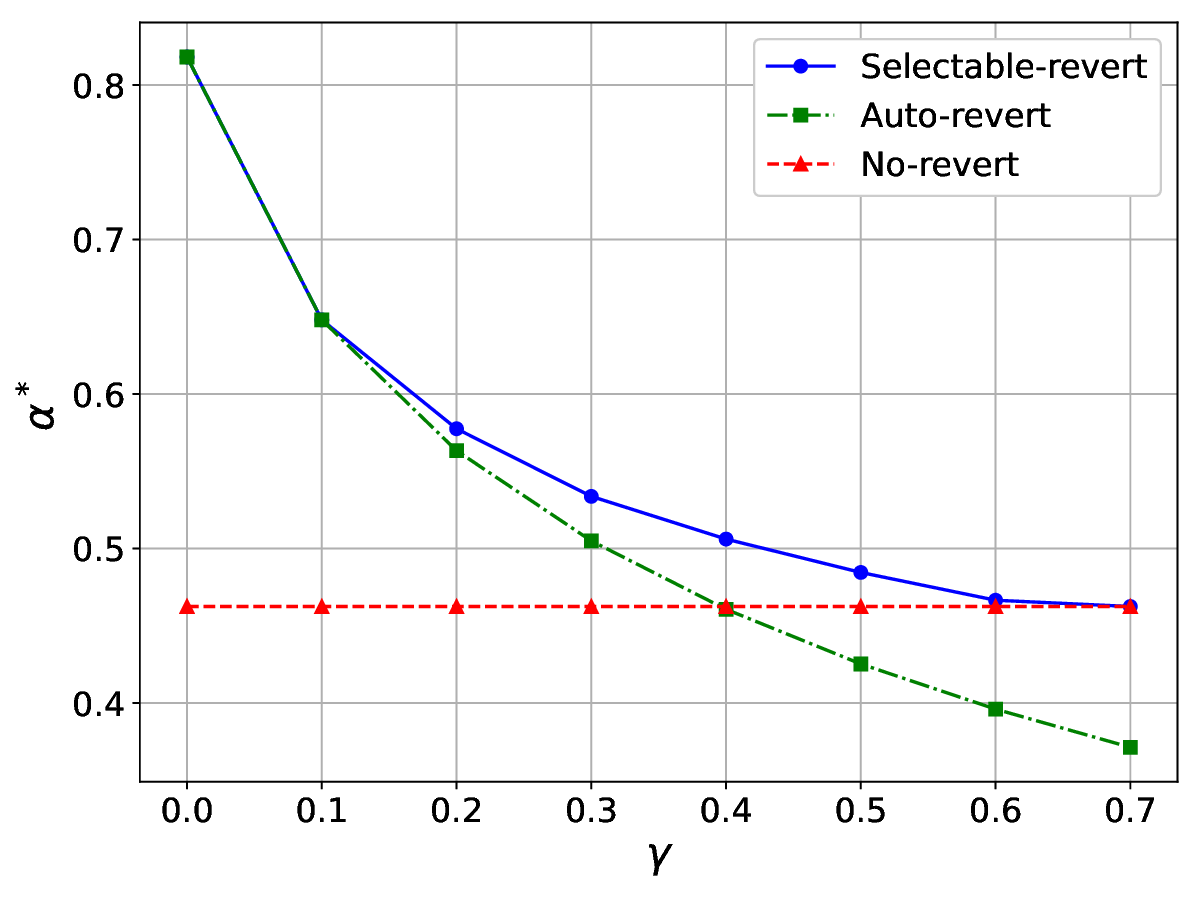}
	\includegraphics[width=0.45\linewidth]{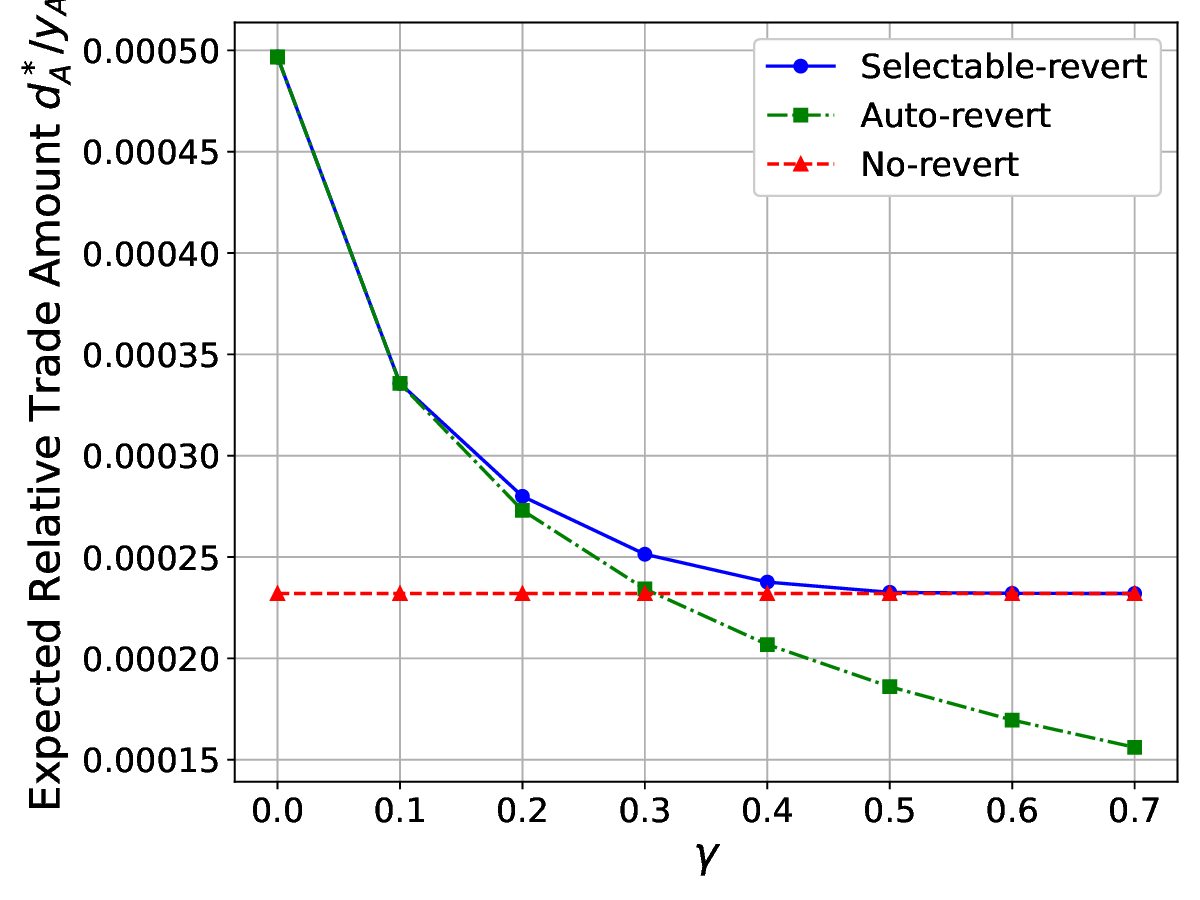}\\
	\includegraphics[width=0.45\linewidth]{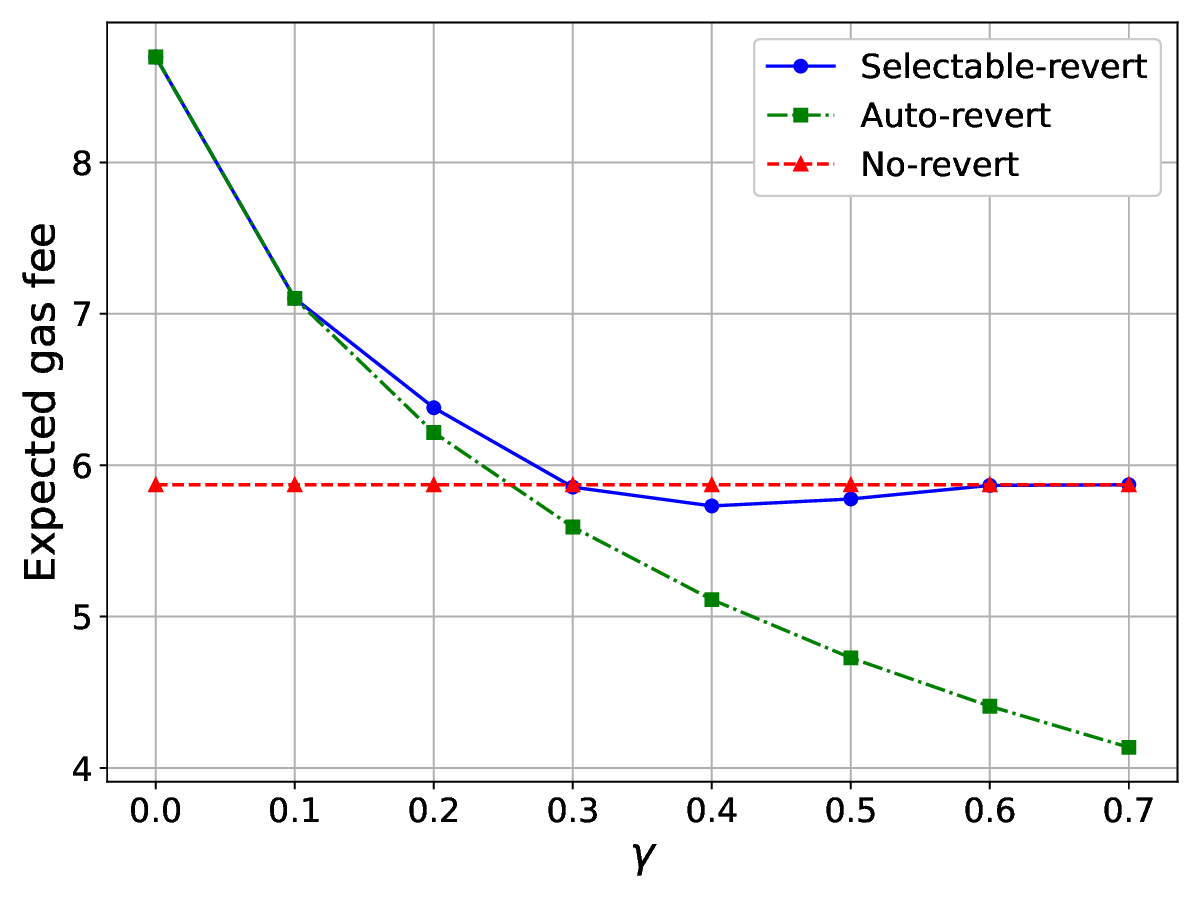}
	\includegraphics[width=0.45\linewidth]{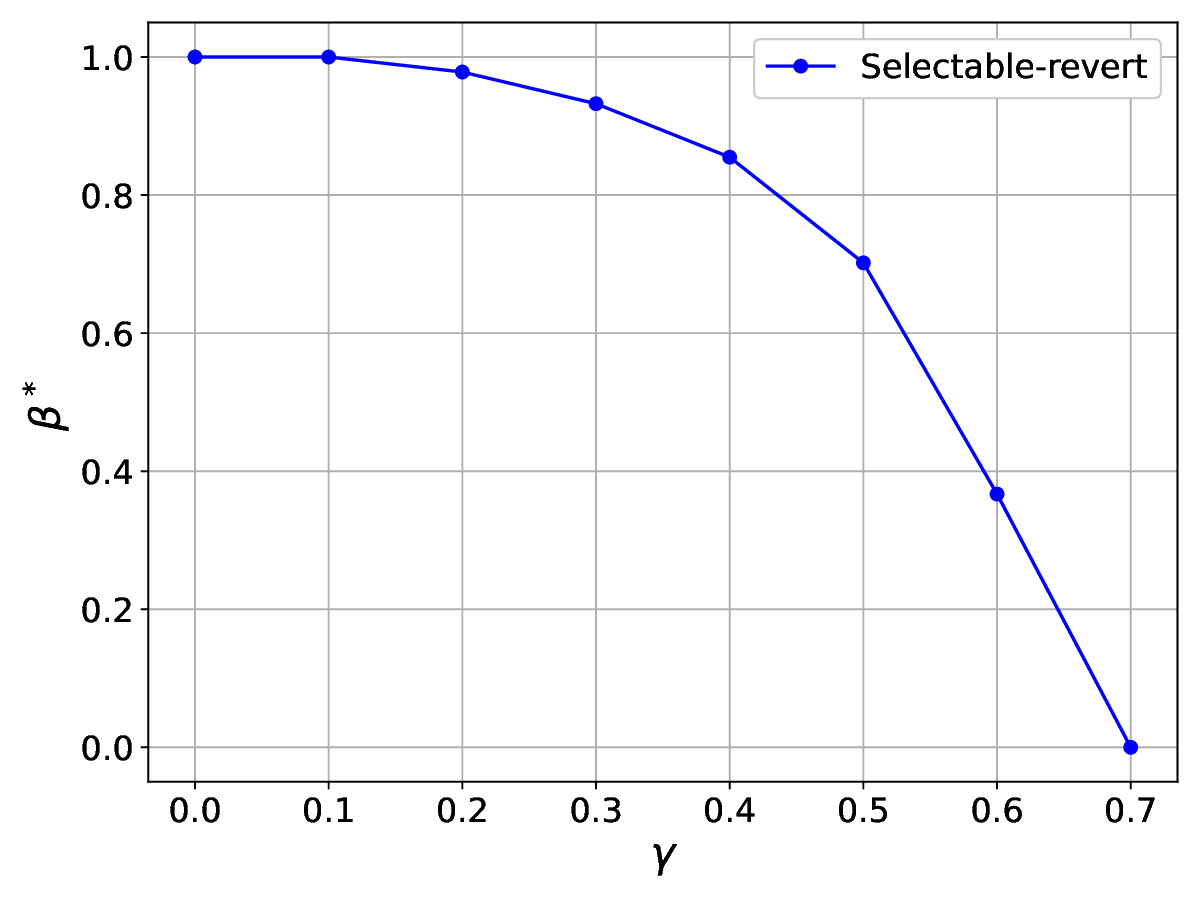}\\
	\includegraphics[width=0.45\linewidth]{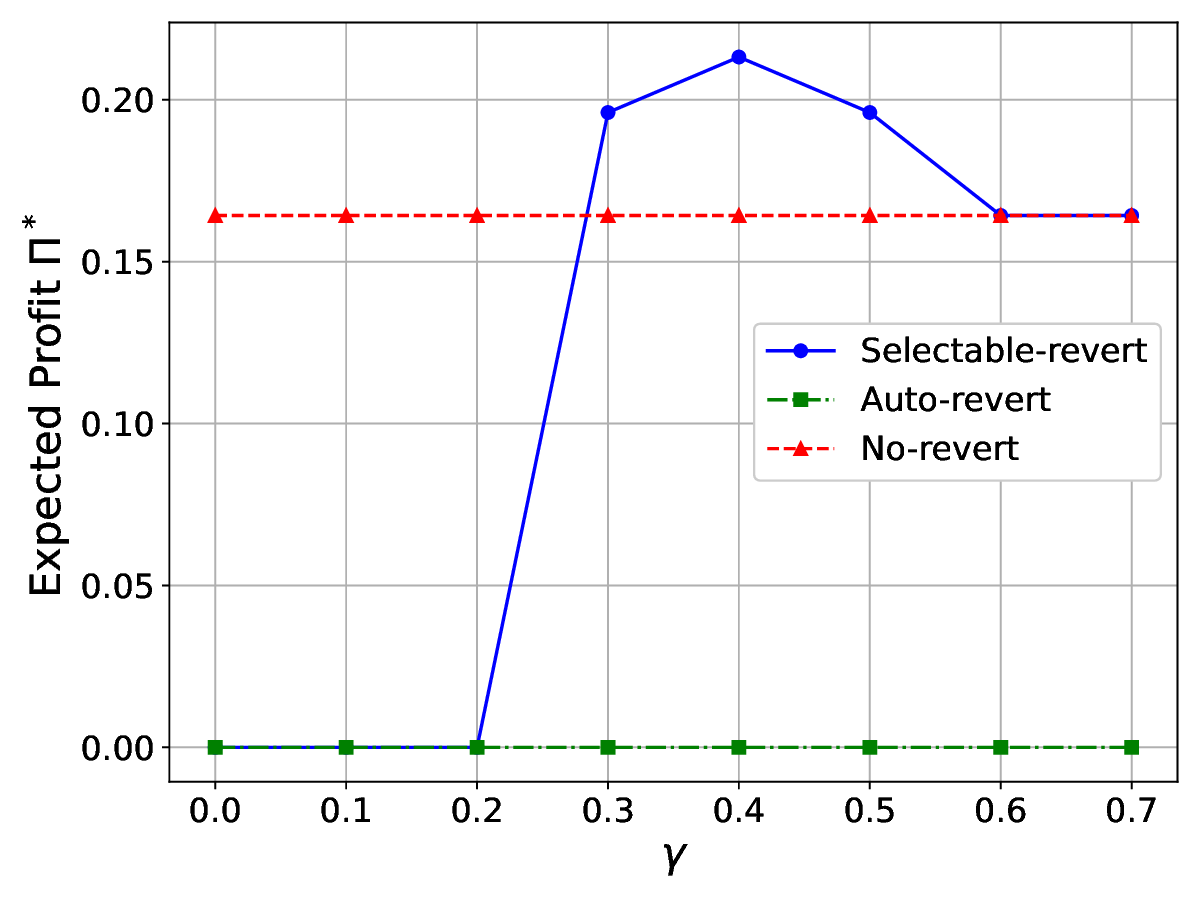}
	\includegraphics[width=0.45\linewidth]{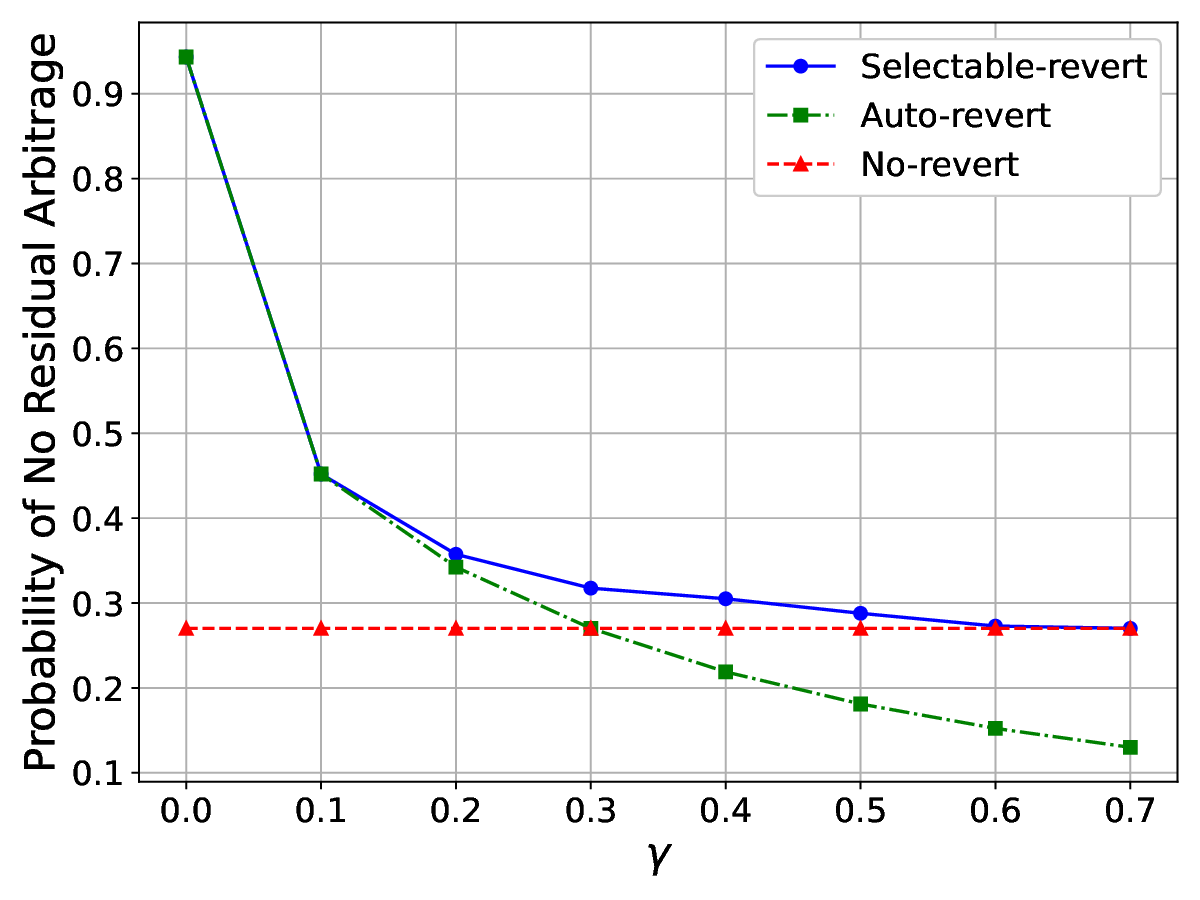}
	\caption{Trading probability $\alpha^*$ (top-left), expected relative trading amount $D^*_A/y_A$ (top-right), expected gas fee paid (middle-left), probability $\beta^*$ of choosing to revert transactions in the selectable-revert setting (middle-right), expected profit (bottom-left), and probability of no residual arbitrage (bottom-right) as functions of inventory risk aversion $\gamma$. Parameter values are given in Table \ref{tab:ParameterValues}. For each value of $O$, we compute each of the six quantities and then average across the ten values of $O$.}
	\label{fig:expected_profit_0_all}
\end{figure}

We first compute the trading probability $\alpha^*$, the expected relative trading amount $D^*/y_A$ (accounting for the possibility of not trading, in which case the trading amount is zero), the expected gas fee paid (accounting for the possibility of not trading, in which case the gas fee is zero, and for reverted transactions in which a proportion $r$ of the selected gas fee is paid), and the probability of choosing to revert transactions in the selectable-revert setting. We use the parameter values in Table \ref{tab:ParameterValues}. For each value of $O$ in the table, we compute the trading probability and then average across the ten values of $O$. Other quantities are computed similarly. The trading probability, expected relative trading amount, and probability of choosing to revert transactions are plotted in the top-left, top-right, and middle-right panels of Figure \ref{fig:expected_profit_0_all}, respectively.

We observe that both the trading probability and the trading amount in the selectable-revert setting are consistently higher than in the no-revert and auto-revert settings, indicating that the flexibility of allowing arbitrageurs to choose whether to revert transactions encourages trading. When inventory risk aversion $\gamma$ is low, arbitrageurs are more likely to trade and tend to trade more in the auto-revert setting than in the no-revert setting, because the potential loss from non-reverted transactions in the no-revert setting outweighs the inventory risk of reverted transactions in the auto-revert setting. When inventory risk aversion is high, the opposite holds.

From the middle panels, we see that the expected gas fee in the selectable-revert setting is not universally decreasing in $\gamma$, although the probability of choosing to revert transactions decreases with $\gamma$. When inventory risk aversion is low, arbitrageurs in the auto- and selectable-revert settings choose to revert transactions, either mandatorily or voluntarily. As explained in Section \ref{subsect:TCR}, in this case arbitrageurs bid more aggressively for gas than in the no-revert setting. Consequently, the expected gas fee paid is higher in the auto- and selectable-revert settings than in the no-revert setting. When inventory risk aversion is higher, arbitrageurs in the auto-revert setting become less likely to trade. Moreover, reverted transactions pay only a proportion of the selected gas fee. Both effects imply a lower expected gas fee paid than in the no-revert and selectable-revert settings.

Next, we compute and plot the arbitrageur's expected profit as a function of inventory risk aversion in the bottom-left panel of Figure \ref{fig:expected_profit_0_all}. With low inventory risk aversion, expected profit is higher in the no-revert setting than in the auto- and selectable-revert settings. With high inventory risk aversion, expected profit is higher in the no-revert and selectable-revert settings than in the auto-revert setting.

Finally, we compute and plot market efficiency, measured by the probability that the arbitrage opportunity disappears after arbitrageurs act, in the bottom-right panel of Figure \ref{fig:expected_profit_0_all}. With low inventory risk aversion, market efficiency is higher in the auto- and selectable-revert settings than in the no-revert setting. With high inventory risk aversion, market efficiency is higher in the no-revert and selectable-revert settings than in the auto-revert setting.

In conclusion, the preferred setting, no-revert, auto-revert, or selectable-revert, depends on whether the focus is on arbitrageur profit or market efficiency, and on whether inventory risk aversion is low or high, as summarized in Table \ref{tab:PreferredSetting}.

\begin{table}
	\begin{center}
		\caption{Preferred settings for transaction reversion, from the perspectives of arbitrageurs and market efficiency.}\label{tab:PreferredSetting}
		\footnotesize
		\begin{tabular}{c@{\hskip 2em}c@{\hskip 2em}c}
			\toprule
			&\multicolumn{2}{c}{\textbf{Inventory risk aversion}}\\
			\cline{2-3}
			& {\em Low} & {\em High}\\
			\midrule
			\textbf{Arbitrageurs' profit} & no-revert   & selectable-revert or no-revert \\
			\midrule
			\textbf{Market efficiency} & selectable-revert or auto-revert  & selectable-revert or no-revert\\
			\bottomrule
		\end{tabular}
	\end{center}
\end{table}

\section{Conclusion}\label{sec:conclude}

This paper develops the first equilibrium model of gas fee competition between arbitrageurs on DEXs, highlighting the strategic interaction that arises when multiple traders pursue the same arbitrage opportunity. By analyzing three distinct transaction reversion settings, no-revert, auto-revert, and selectable-revert, we provide a unified framework for understanding how inventory risk, gas fee bidding, and execution uncertainty shape arbitrage behavior and market outcomes.

Our theoretical results demonstrate that pure symmetric equilibria do not exist in any of the three settings, but unique mixed symmetric equilibria can be characterized. In the no-revert and auto-revert settings, equilibrium strategies are derived from nonlinear ODEs, while in the selectable-revert setting, equilibrium strategies are characterized through a system of equations and verified numerically. These equilibria reveal important trade-offs: arbitrageurs balance the probability of execution against expected profit, and their willingness to revert transactions depends critically on inventory risk aversion.

	Our empirical study, using data from Binance and Uniswap V2, confirms several key implications of the model. We find that arbitrageurs frequently trade less than the maximum amount and that priority gas fees are more often chosen at lower levels, both suggesting positive inventory risk aversion. Moreover, the data validate our comparative statics: gas fees increase with price discrepancies and liquidity, while relative trading amounts rise with both price discrepancies and gas fees. These findings provide robust evidence that the theoretical framework captures essential features of real-world arbitrage behavior.

Comparative analysis across settings shows that the arbitrageur's expected profit and market efficiency depend jointly on inventory risk and the reversion mechanism. With low inventory risk, the no-revert setting yields higher expected profits, while the auto-revert and selectable-revert settings enhance market efficiency. With moderate to high inventory risk, however, the no-revert and selectable-revert settings dominate both in profitability and efficiency.

\bibliography{LongTitles,BibFile}

\clearpage

\pagenumbering{arabic}
\renewcommand*{\thepage}{A-\arabic{page}}

\setlength{\abovedisplayskip}{5pt}
\setlength{\belowdisplayskip}{5pt}

\begin{center}
	\Large \textbf{Online Appendix}
\end{center}

\begin{appendices}

\setcounter{equation}{0}
\renewcommand{\theequation}{A.\arabic{equation}}

\section{Theoretical Results}\label{appendix:theoreticalResults}

In this section, we present theoretical results that support the analyses in Section \ref{sec:model}.

\subsection{Properties of First-Mover Advantage}
\begin{lemma}\label{le:Discontinuity}
	Fix the transaction reversion setting $\ell \in \{0,\infty\}$. Let $y_i>0$, $p_i>0$, $i\in \{A,B\}$ with $(p_B/p_A)/(y_A/y_B)\in [1/(1+f),1+f]$. Then the following hold:
	\begin{enumerate}[leftmargin=2em]
		\item[(i)] For each fixed $g\ge 0$ and $\bar d_A\ge 0$, $R_F(g, d_A)$ is continuous and strictly concave in $d_A\in [0,\infty)$, while $R_S^\ell(g,d_A;\bar d_A)$ is continuous and concave in $d_A \in [0,\infty)$.
		\item[(ii)] For each $g\ge 0$, $\bar d_A>0$, and $d_A>0$, we have $V^\infty(g,d_A;\bar d_A)>0$.
	\end{enumerate}
\end{lemma}

\subsection{Key ODE in the No-Revert Setting}

\begin{lemma}\label{le:ODE}
	Suppose $O\le 3$. Then there exists a unique positive continuous function $\hat x(\cdot)$ satisfying
	\eqref{eq:KeyODENormalized} up to a finite explosion point $z_\infty\in (0,\infty)$ with $\lim_{z\uparrow z_\infty}\hat x(z)=0$. Moreover, $\hat x$ is continuously differentiable and strictly decreasing on $[0,z_\infty)$. Furthermore,
	\begin{align}
		\int_0^{z_\infty} \hat x^{-2}(\bar z)\dd \bar z=+\infty.\label{eq:ODESolutionSqIntegralExpl}
	\end{align}
	Consequently, for any $\ell \in (0,\infty)$, there exists a unique $\hat z\in (0,z_\infty)$ such that
	\begin{align}
		\int_0^{\hat z}v(\hat x(\bar z))\dd \bar z=1,\label{eq:Z0ell}
	\end{align}
	where
	\begin{align}
		v(x):=\frac{L_B}{V(y_A x, y_Ax)} =  \frac{(1+x)(1+2x)}{2x^2},\quad x>0.\label{eq:FMAInvNormalized}
	\end{align}
\end{lemma}

\subsection{Comparative Statics in the No-Revert Setting}\label{subse:ComparativeStaticsNoRevert}

This section presents the comparative statics of the equilibrium strategy in the no-revert setting.

\begin{proposition}\label{prop:EquilibriumMonotonocity}
	Let $(\alpha^*,\Phi^*,D_A^*)$ be the unique symmetric mixed Nash equilibrium in the setting $\ell=\infty$.
	\begin{enumerate}[leftmargin=2em]
		\item[(i)] The optimal trading amount $D_A^*(g)$ is strictly increasing in the gas fee $g\in [\hat g_L,g_H]$. Moreover, $D_A^*(g_H)=\hat D_A$.
		\item[(ii)] The density function $\phi^*(g)$ of the randomized gas fee is strictly decreasing in $g\in [\hat g_L,g_H]$.
	\end{enumerate}
\end{proposition}

We denote by $\tilde g$ and $\tilde D$ the random gas fee paid and the trading amount chosen by the arbitrageur, respectively. Conditional on not trading, $\tilde g=\tilde D=0$. Conditional on trading, $\tilde D = D_A^*(\tilde g)$. We call $\tilde D/y_A$ the {\em relative trading amount}. Denote by $F_{\tilde g}$ and $F_{\tilde D/y_A}$ the {\em decumulative distribution functions} of $\tilde g$ and $\tilde D/y_A$, respectively. For notational convenience, we extend $\phi^*$ from $[\hat g_L,g_H]$ to $(0,\infty)$ by setting $\phi^*(g)=0$ for $g\notin [\hat g_L,g_H]$. Then,
\begin{align}
	F_{\tilde g}(g)= \alpha^*\left(\int_g^{\infty}\phi^*(\bar g)\dd \bar g\right),\quad g\ge 0.\label{eq:GasDDF}
\end{align}
Note that $D^*_A(g)/y_A=\hat x((g_H-g)/L_B)$, which is strictly increasing in $g\in[\hat g_L,g_H]$, and that $\tilde D/y_A\le D^*_A(g_H)/y_A=\hat x(0)=O^{1/2}-1$. Consequently,
\begin{align}
	F_{\tilde D/y_A}(d) = F_{\tilde g}\left(g_H-L_B\hat x^{-1}(d)\right)\mathbf 1_{d<\hat x(0)},\quad d> 0.\label{eq:TradingAmountDDF}
\end{align}

\begin{proposition}\label{prop:BaseGasFee}
	Let $(\alpha^*,\Phi^*,D_A^*)$ be the unique symmetric mixed Nash equilibrium in Theorem \ref{thm:Main}.
	\begin{enumerate}[leftmargin=2em]
		\item[(i)] The trading probability $\alpha^*$ decreases in $\hat g_L$, with strict monotonicity when $\alpha^*<1$.
		\item[(ii)] The arbitrageur's expected profit decreases in $\hat g_L$, with strict monotonicity when the expected profit is positive.
		\item[(iii)] For $\hat g_L$ with $\alpha^*=1$, i.e., $\hat g_L\le \hat g_H- L_B\hat z$, $F_{\tilde g}(g)$ increases in $\hat g_L$ for every $g$. For $\hat g_L$ with $\alpha^*<1$, i.e., $\hat g_L> \hat g_H- L_B\hat z$, $F_{\tilde g}(g)$ decreases in $\hat g_L$ for every $g$.
		\item[(iv)] For $\hat g_L$ with $\alpha^*=1$, i.e., $\hat g_L\le \hat g_H- L_B\hat z$, $F_{\tilde D/y_A}(d)$ is independent of $\hat g_L$ for every $d$. For $\hat g_L$ with $\alpha^*<1$, i.e., $\hat g_L> \hat g_H- L_B\hat z$, $F_{\tilde D/y_A}(d)$ decreases in $\hat g_L$ for every $d$.
	\end{enumerate}
\end{proposition}

\begin{proposition}\label{prop:Liquidity}
	Let $(\alpha^*,\Phi^*,D_A^*)$ be the unique symmetric mixed Nash equilibrium in Theorem \ref{thm:Main}.
	\begin{enumerate}[leftmargin=2em]
		\item[(i)] The trading probability $\alpha^*$ increases in $L_B$, with strict monotonicity when $\alpha^*<1$.
		\item[(ii)] The arbitrageur's expected profit increases in $L_B$, with strict monotonicity when the expected profit is positive.
		\item[(iii)] $F_{\tilde g}(g)$ increases in $L_B$ for every $g$.
		\item[(iv)] For $L_B$ with $\alpha^*=1$, i.e., $L_B\ge \hat g_L/\big((1-O^{-1/2})^2-\hat z\big)$, $F_{\tilde D/y_A}(d)$ is independent of $L_B$ for every $d$. For $L_B$ with $\alpha^*<1$, i.e., $L_B< \hat g_L/\big((1-O^{-1/2})^2-\hat z\big)$, $F_{\tilde D/y_A}(d)$ increases in $L_B$ for every $d$.
	\end{enumerate}
\end{proposition}

\begin{figure}
	\centering
	\includegraphics[width=0.46\textwidth]{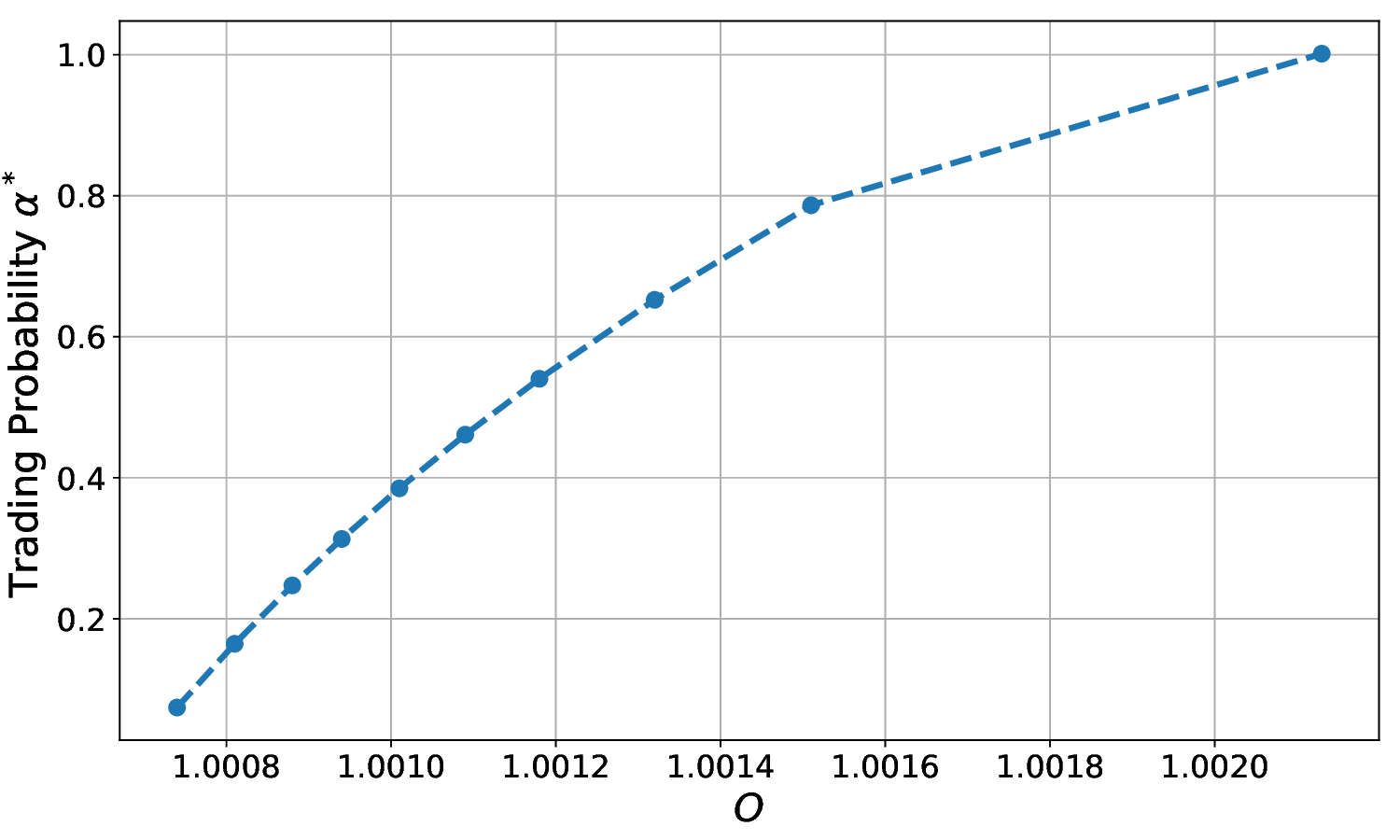}
	\includegraphics[width=0.46\textwidth]{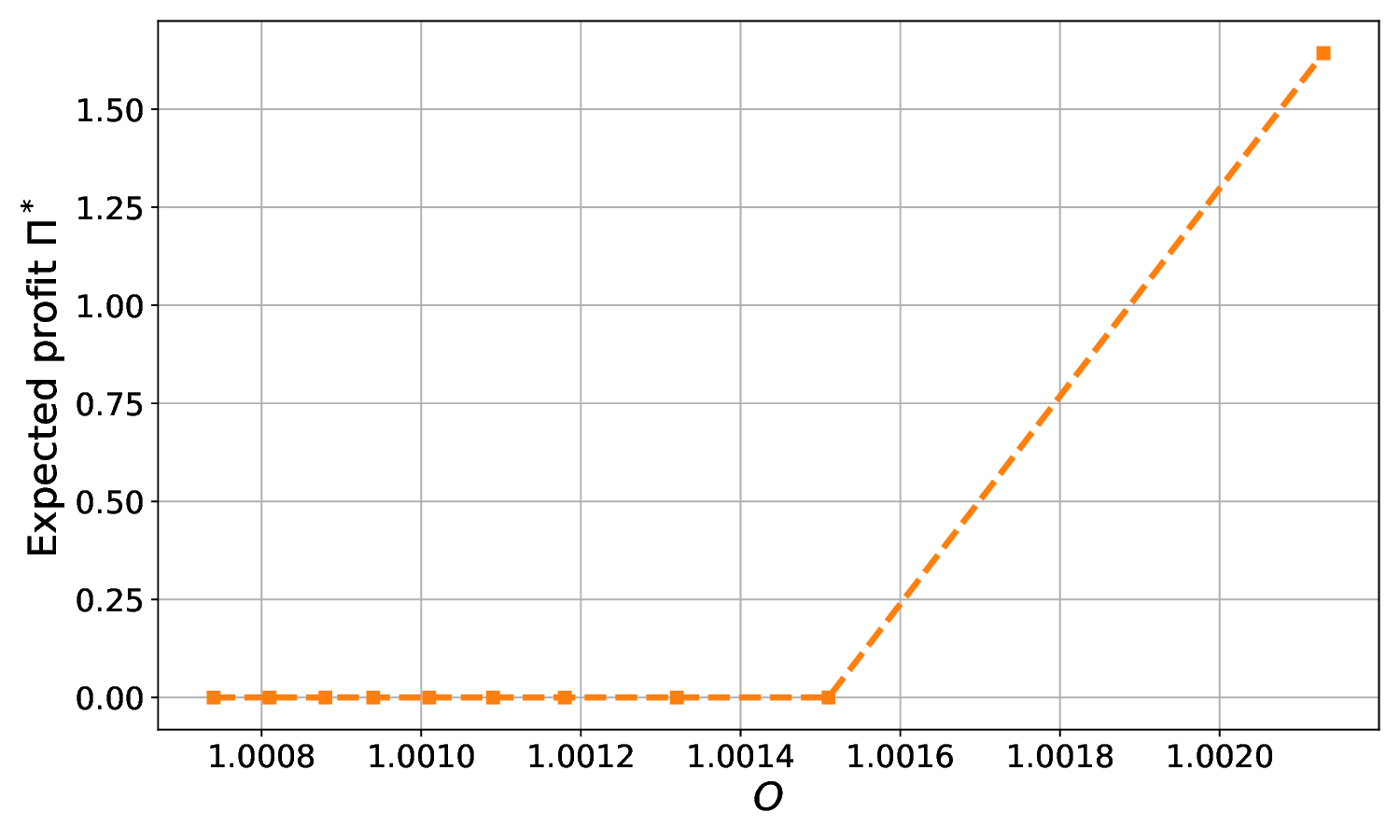}\\
	\includegraphics[width=0.46\textwidth]{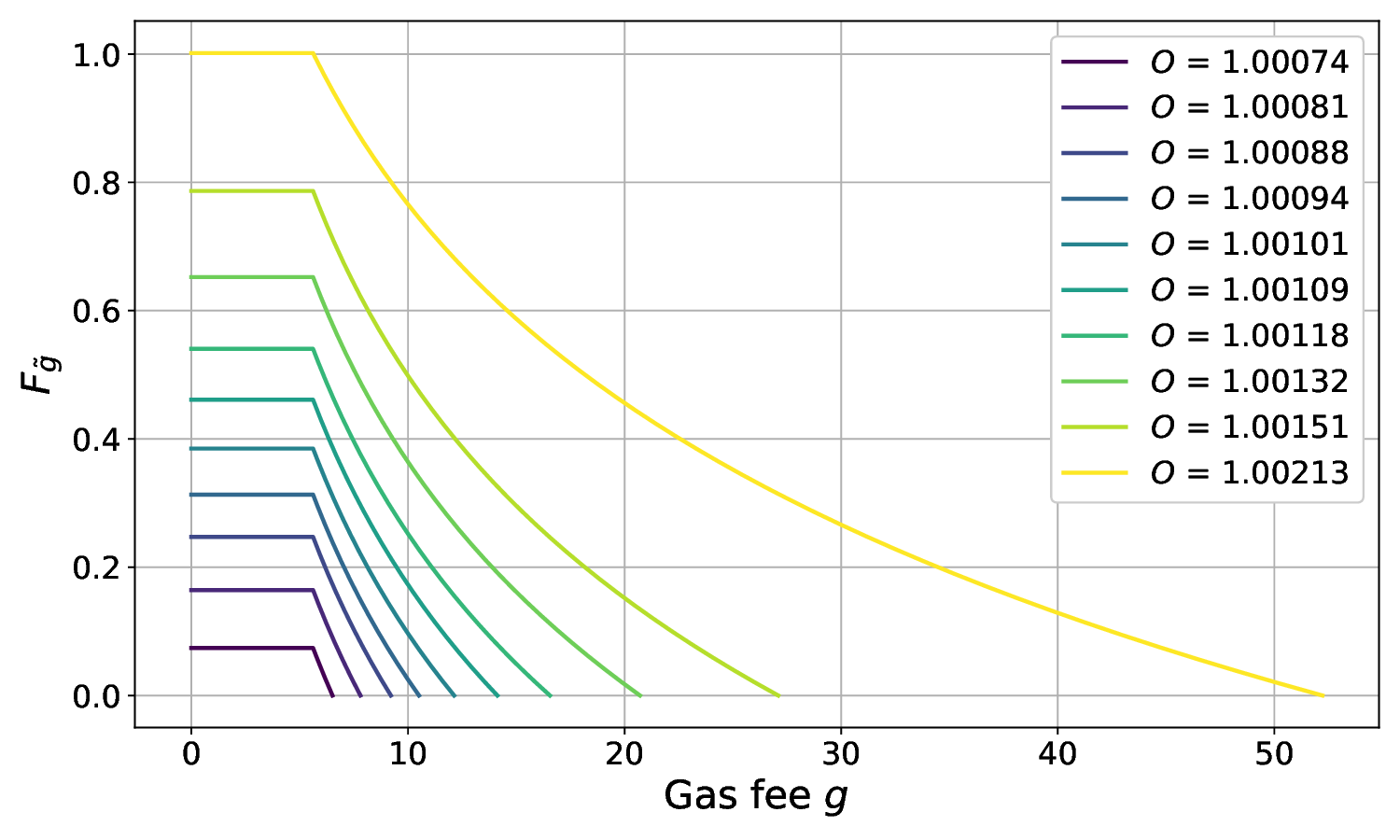}
	\includegraphics[width=0.46\textwidth]{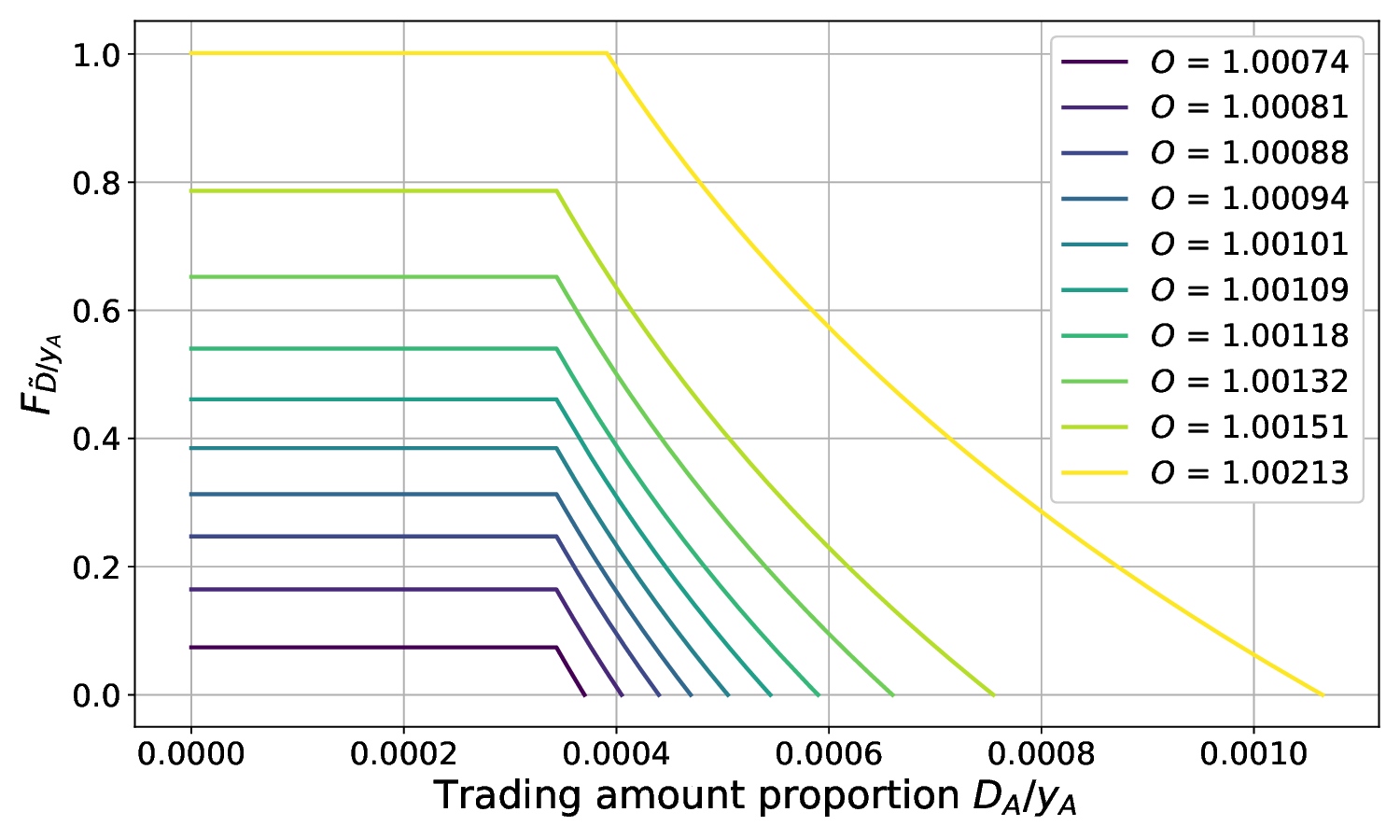}
	\caption{Probability of trading (top-left), expected profit (top-right), decumulative distribution of gas fee (bottom-left), and decumulative distribution of trading amount (bottom-right) in the no-revert setting. The values of $\hat g_L$ and $L_B$ are given in Table \ref{tab:ParameterValues}.}\label{alphaprofit_o}
\end{figure}

Finally, using the parameter values in Table \ref{tab:ParameterValues}, we numerically verify the impact of the degree of mispricing $O$ in Figure \ref{alphaprofit_o}. We observe that the trading probability and expected profit increase with $O$. The decumulative distribution functions $F_{\tilde D/y_A}$ and $F_{\tilde g}$ decrease with $O$, showing that the relative trading amount and gas fee both increase with $O$.

\subsection{Key ODE in the Auto-Revert Setting}\label{subse:KeyODEEll0}

We now consider the auto-revert setting. By Proposition \ref{le:RandomizedGasDensity}, the trading amount $D_A^*(g)$ must lie in $(0,\hat D_A]$, and the first-mover advantage $V^0(g,D_A^*(g);D_A^*(g))$ must be positive. In other words, $\hat x(z)$ must lie in $(0,O^{1/2}-1]$, and $K^0_1(z,\hat x(z))$ must be positive. Moreover, $\hat \theta$ must be positive.

Note that
% \begin{align*}
$
	K^0_1(z,x)=x\left((1+x)^{-1}-O^{-1}\right)+\Gamma(y_Ax)/L_B-(1-r)(\hat g_H/L_B-z).
$
% \end{align*}
It is straightforward to see that $x\left((1+x)^{-1}-O^{-1}\right)+\Gamma(y_Ax)/L_B$ is strictly increasing in $x\in [0,O^{1/2}-1]$. Therefore, there exists a continuous, strictly increasing function $\varphi$ on $[0,\infty)$ such that for any $c\ge 0$, $x\left((1+x)^{-1}-O^{-1}\right)+\Gamma(y_Ax)/L_B\le c$ if and only if $x\le \varphi(c)$. Moreover, since $\Gamma(0)=0$, we must have $\varphi(0)=0$. Define $\varphi(c)=0$ for $c<0$. Thus, for any $z\ge 0$, $K^0_1(z,x)>0$ if and only if
$x\in \big(\varphi((1-r)(\hat g_H/L_B-z)),O^{1/2}-1\big]$. Therefore, for any $T>0$, we seek a solution $(\hat x,\hat \theta)$ to \eqref{eq:KeyODESlippage0} such that (i) $\hat x$ lies in
\begin{align*}
	\mathcal{C}([0,T]):=\big\{y:y \text{ is continuous on } [0,T], \; y(z)\in\big(\varphi((1-r)(\hat g_H/L_B-z)),O^{1/2}-1\big], \; \forall z\in[0,T]\big\},
\end{align*}
and (ii) $\hat \theta$ is a positive, continuous function on $[0,T]$. Note that
\begin{align*}
	&(O^{1/2}-1)\left((1+(O^{1/2}-1))^{-1}-O^{-1}\right)+\Gamma(y_A(O^{1/2}-1))/L_B \\
	&= \hat g_H/L_B +\Gamma(y_A(O^{1/2}-1))/L_B \ge \hat g_H/L_B,
\end{align*}
which implies $\varphi(\hat g_H/L_B)\le O^{1/2}-1$. Because $r>0$, we have $\varphi((1-r)\hat g_H/L_B)<\varphi(\hat g_H/L_B)\le O^{1/2}-1$. Therefore, $\mathcal{C}([0,T])$ is nonempty.

\begin{lemma}\label{le:ODESlippage0}
	\begin{enumerate}[leftmargin=2em]
		\item[(i)] Suppose $\Gamma\equiv 0$. Then, for any $T>0$, \eqref{eq:KeyODESlippage0} admits a unique solution $(\hat x,\hat \theta)$ on $[0,T]$ with $\hat x\in \mathcal{C}([0,T])$ and $\hat \theta$ positive and continuous on $[0,T]$. The solution takes the form:
		\begin{align}
			\hat x(z) = O^{1/2}-1,\quad
			\hat \theta(z) = (r\hat g_H/L_B)\big(r\hat g_H/L_B + (1-r)z\big)^{-2},\quad z\ge 0.\label{eq:ODESolutionSlippage0Gamma0phi}
		\end{align}
		Moreover,
		\begin{align}
			\int_0^z \hat \theta(\bar z)\dd\bar z =
			(r\hat g_H/L_B + (1-r)z)^{-1}z,\quad  r\le1,\; z\ge 0.\label{eq:ODESolutionSlippage0Gamma0phiIntSol}
		\end{align}
		\item[(ii)] Suppose $\Gamma(d_A)$ is twice continuously differentiable, with $\Gamma'(d_A)>0$ and $\Gamma''(d_A)\ge0$ for all $d_A>0$. Then \eqref{eq:KeyODESlippage0} admits a unique solution $(\hat x,\hat \theta)$ up to
		\begin{align}
			z_\infty:=\sup\{z\ge 0:\forall 0<u<z, \hat x(u)>\varphi((1-r)(\hat g_H/L_B-u))\},\label{eq:Slippage0Explosion}
		\end{align}
		such that $\hat x\in \mathcal{C}([0,T])$ and $\hat \theta$ is positive on $[0,T]$ for every $T<z_\infty$. Moreover, $\hat x(z)$ is strictly decreasing in $z$.
	\end{enumerate}
\end{lemma}

\subsection{Comparative Statics in the Auto-Revert Setting}\label{appx:CSSlippage0}

In this section, we conduct comparative statics of the equilibrium strategy in the auto-revert setting.

\subsubsection{Theoretical Results}
We first establish some properties of the equilibrium strategy.

\begin{proposition}\label{prop:phi_da_slippage0}
	Consider the auto-revert setting.
	\begin{enumerate}[leftmargin=2em]
		\item[(i)] Suppose $\Gamma\equiv 0$. Then the gas fee density $\phi^*(g)$ is constant in $g$ when $r=1$, and strictly increasing in $g$ when $r<1$.
		\item[(ii)] Suppose $\Gamma'(d_A)>0$, $\forall d_A>0$. Then the trading amount $D_A^*(g)$ is strictly increasing in $g$.
	\end{enumerate}
\end{proposition}

\begin{proposition}\label{prop:CSNoSlippageO}
	Consider the auto-revert setting and suppose $\Gamma\equiv 0$. Fix $\hat g_L$ and $L_B$. Then:
	\begin{enumerate}[leftmargin=2em]
		\item[(i)] $\alpha^*$ is strictly increasing in $O$, and
		\begin{align}
			\lim_{O\uparrow \infty}\alpha^* = \frac{1-\hat g_L/L_B}{1-(1-r)(\hat g_L/L_B)}<1.\label{eq:TradingProbInfOSetting0}
		\end{align}
		\item[(ii)] $F_{\tilde g}(g)$ is increasing in $O$ for every $g\ge 0$, and strictly increasing in $O$ for $g\in[0,\hat g_H)$.
		\item[(iii)] $F_{\tilde D/y_A}(d)$ is increasing in $O$ for every $d\ge 0$, and strictly increasing in $O$ for $d\in[0,\hat D)$.
	\end{enumerate}
\end{proposition}

\begin{proposition}\label{prop:CSgLSlippageO}
	Consider the auto-revert setting and suppose $\Gamma\equiv 0$. Fix $O$ and $L_B$. Then:
	\begin{enumerate}[leftmargin=2em]
		\item[(i)] $\alpha^*$ is strictly decreasing in $\hat g_L$, and
		% \begin{align}
			$\lim_{\hat g_L\downarrow 0}\alpha^* = 1.$
			% \label{eq:TradingProbzerogLSetting0}
		% \end{align}
		\item[(ii)] $F_{\tilde g}(g)$ is decreasing in $\hat g_L$ for every $g\ge 0$, and strictly decreasing in $\hat g_L$ for $g\in[0,\hat g_L)$.
		\item[(iii)] $F_{\tilde D/y_A}(d)$ is decreasing in $\hat g_L$ for every $d\ge 0$, and strictly decreasing in $\hat g_L$ for $d\in[0,\hat D)$.
	\end{enumerate}
\end{proposition}

\begin{proposition}\label{prop:CSLBSlippageO}
	Consider the auto-revert setting and suppose $\Gamma\equiv 0$. Fix $O$ and $\hat g_L$. Then:
	\begin{enumerate}[leftmargin=2em]
		\item[(i)] $\alpha^*$ is strictly increasing in $L_B$, and
		% \begin{align}
			$\lim_{L_B\uparrow \infty}\alpha^* =1.$
			% \label{eq:TradingProbInfLBSetting0}
		% \end{align}
		\item[(ii)] $F_{\tilde g}(g)$ is increasing in $L_B$ for every $g\ge 0$, and strictly increasing in $L_B$ for $g\in[0,\hat g_H)$.
		\item[(iii)] $F_{\tilde D/y_A}(d)$ is increasing in $L_B$ for every $d\ge 0$, and strictly increasing in $L_B$ for $d\in[0,\hat D)$.
	\end{enumerate}
\end{proposition}

\subsubsection{Numerical Results}\label{appx:gammaPositive}

In this section, we use the parameter values in Table \ref{tab:ParameterValues} to conduct comparative statics of the equilibrium strategy when inventory risk  is positive.

\begin{figure}
	\centering
	\includegraphics[width=0.46\textwidth]{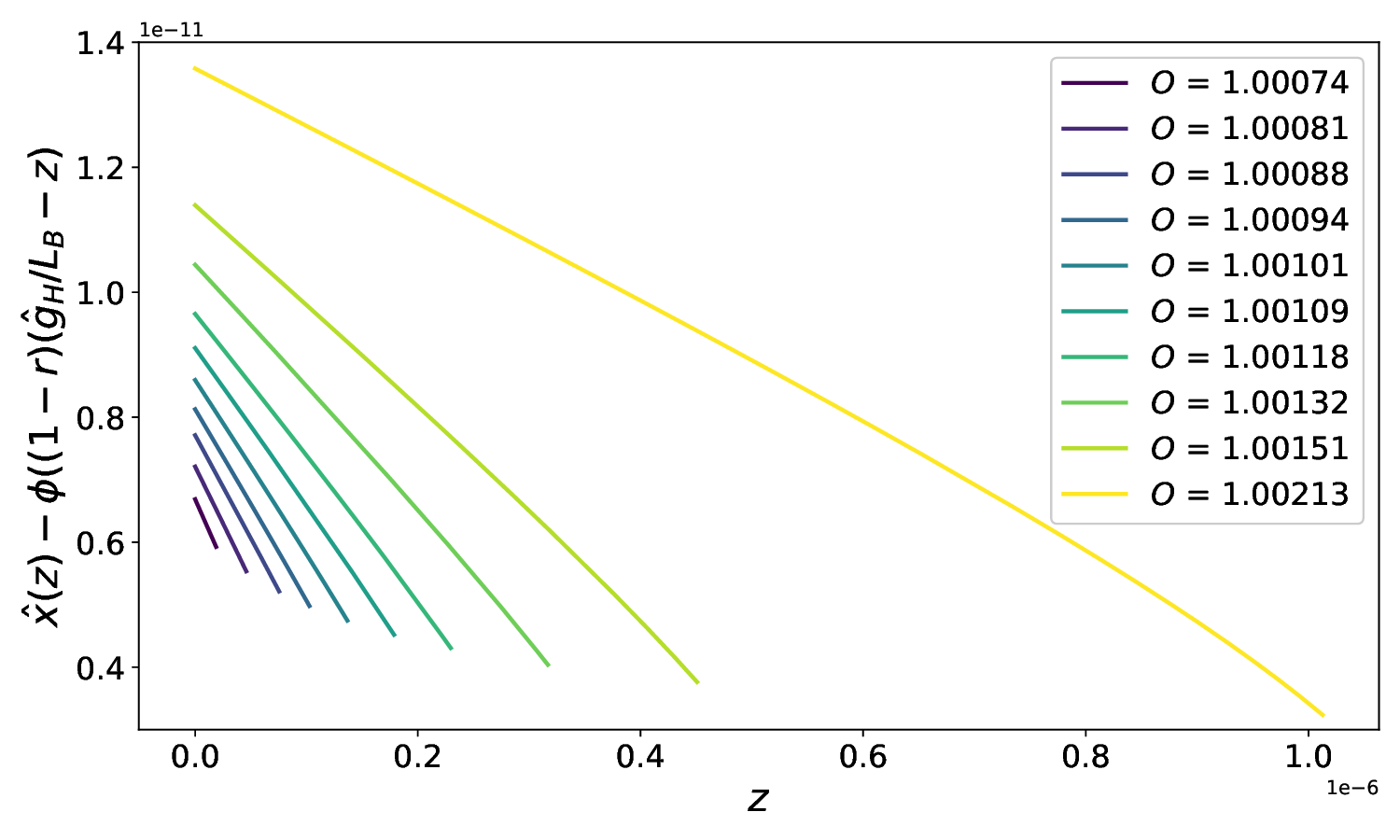}
	\caption{Plot of $\hat x(z)-\varphi((1-r)(\hat g_H/L_B-z))$ over $z\in[0,(1-O^{-1/2})^2]$. The values of $\hat g_L$, $L_B$, and the form of $\Gamma$ are given in Table \ref{tab:ParameterValues} with $r=0.2$.}\label{fig:condition_slippage0}
\end{figure}

First, we verify the condition assumed in Theorem \ref{thm:MainSlippage0}, namely $z_\infty\ge(1-O^{-1/2})^2$. This condition is equivalent to requiring that $\hat x(z)-\varphi((1-r)(\hat g_H/L_B-z))>0$ for all $z\in[0,(1-O^{-1/2})^2]$; see Lemma \ref{le:ODESlippage0}. Figure \ref{fig:condition_slippage0} plots $\hat x(z)-\varphi((1-r)(\hat g_H/L_B-z))$ over $z\in[0,(1-O^{-1/2})^2]$ for various values of $O$, showing that the condition indeed holds.

\begin{figure}
	\centering
	\includegraphics[width=0.46\textwidth]{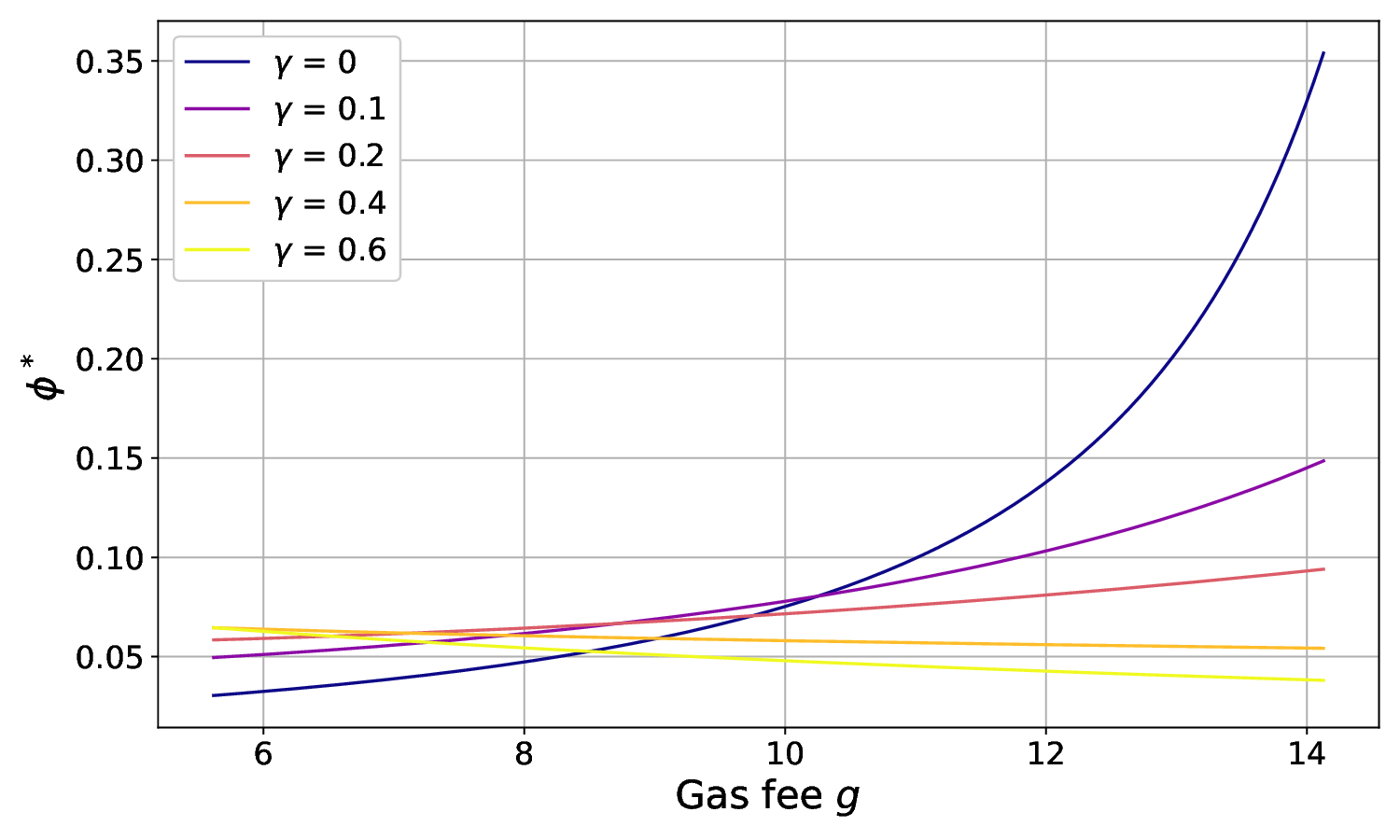}
	\caption{Density function $\phi^*$ of the gas fee in the auto-revert setting. Parameter values are given in Table \ref{tab:ParameterValues} with $O=1.00109$.}\label{fig:phi_gamma_slippage0}
\end{figure}

\begin{figure}
	\centering
	\includegraphics[width=0.32\textwidth]{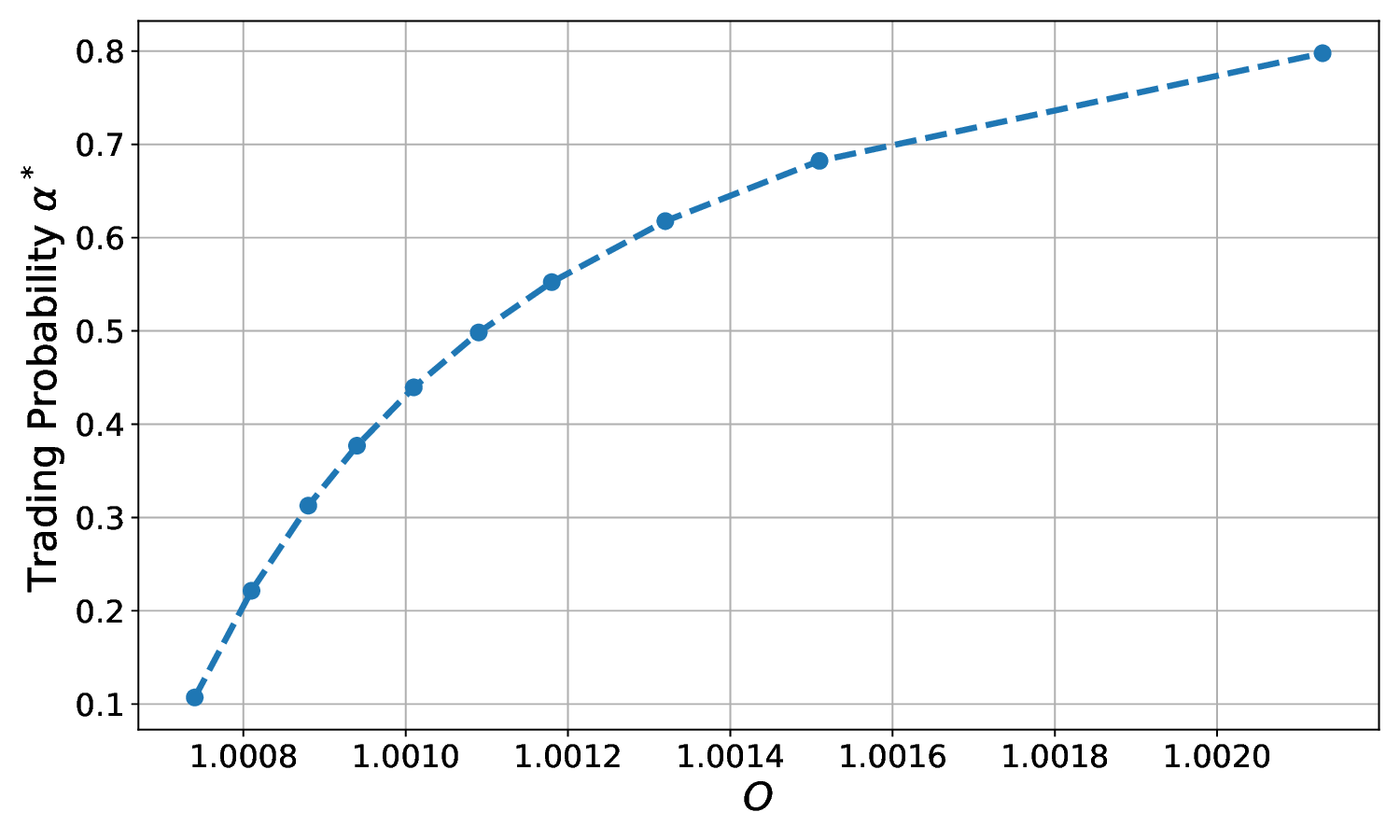}
	\includegraphics[width=0.32\textwidth]{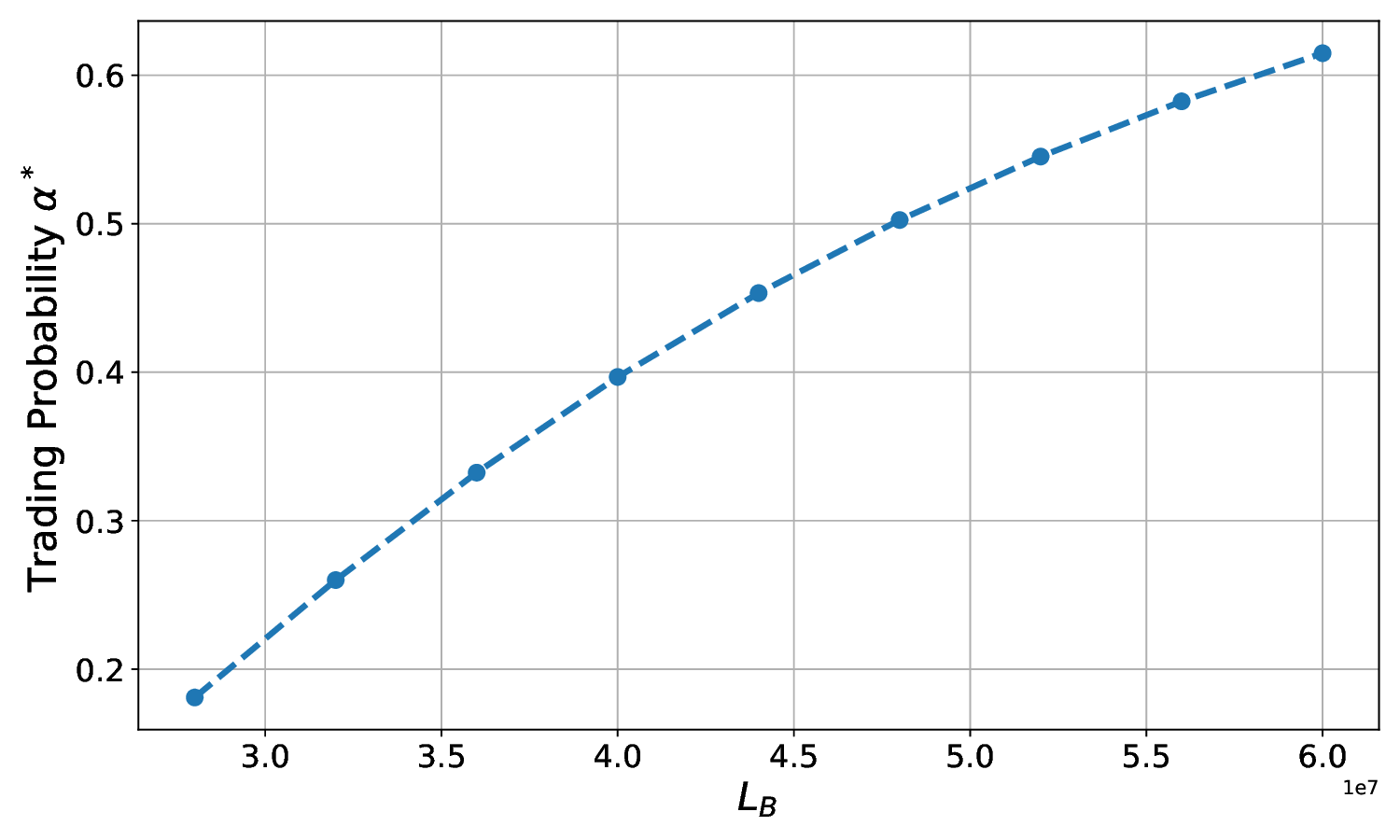}
	\includegraphics[width=0.32\textwidth]{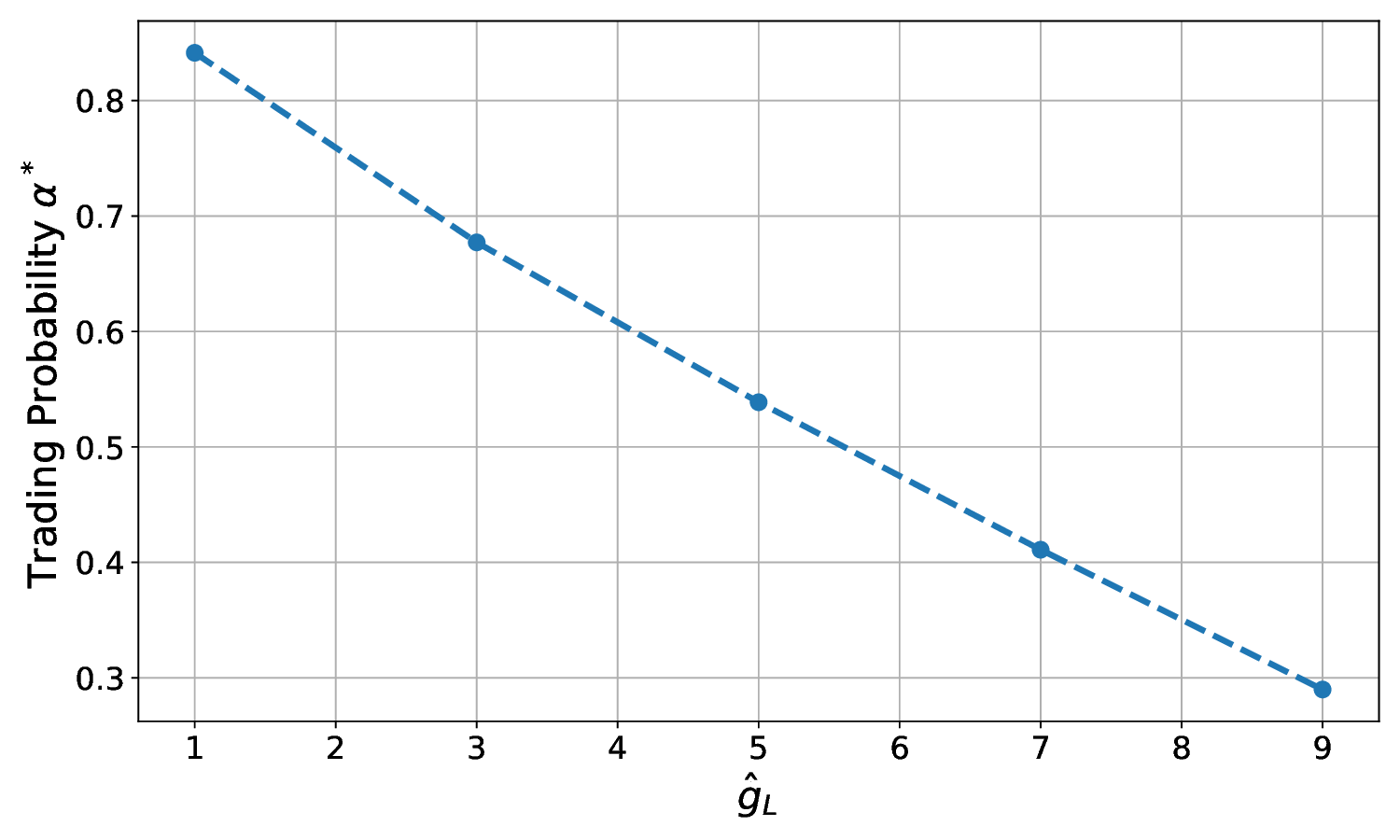}
	\caption{Probability of trading with respect to $O$, $L_B$, and $\hat g_L$ in the auto-revert setting. Default parameter values are given in Table \ref{tab:ParameterValues} with $O=1.00109$ and $\gamma=0.4$.}\label{fig:alpha_slippage0}
\end{figure}

\begin{figure}
	\centering
	\includegraphics[width=0.46\linewidth]{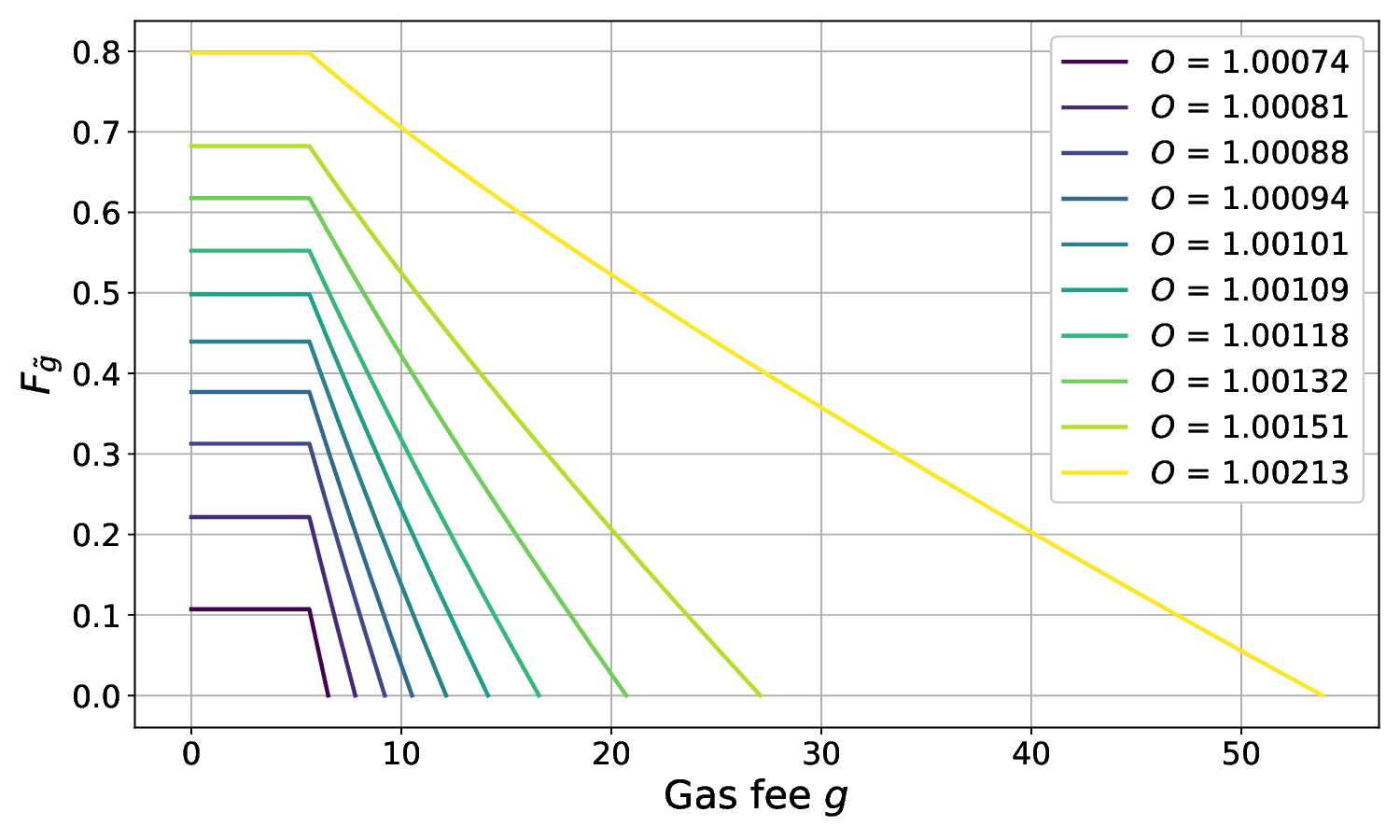}
	\includegraphics[width=0.46\linewidth]{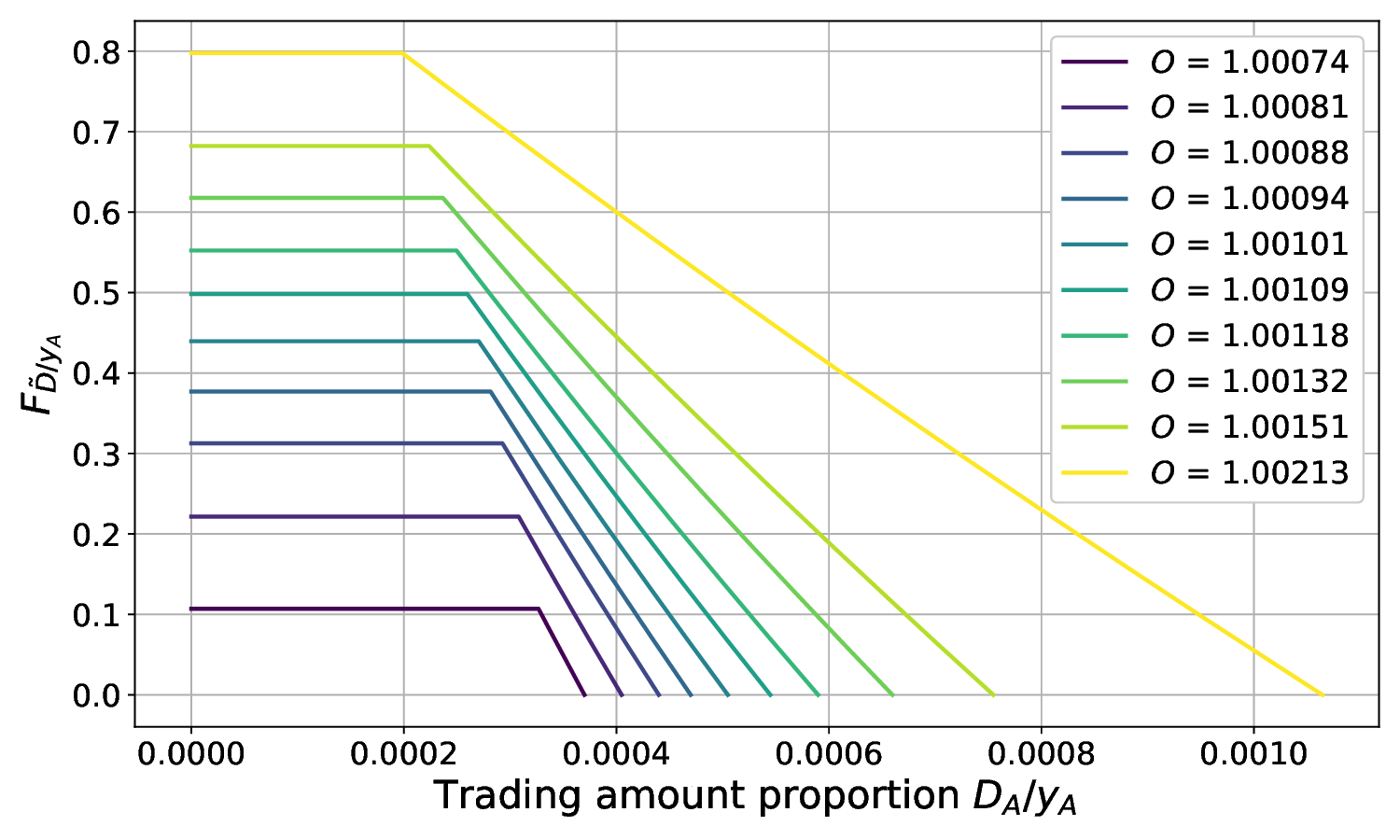}\\
	\includegraphics[width=0.46\linewidth]{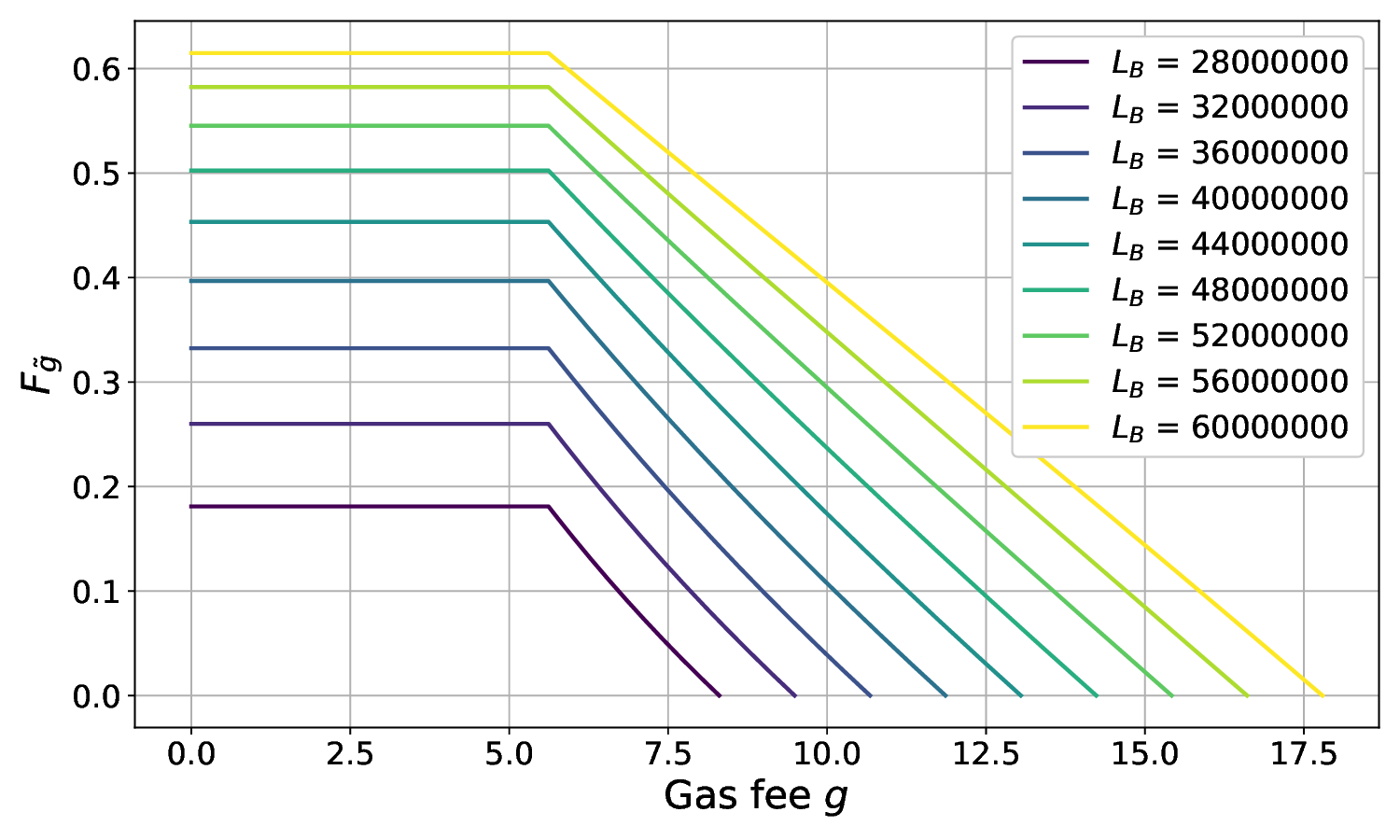}
	\includegraphics[width=0.46\linewidth]{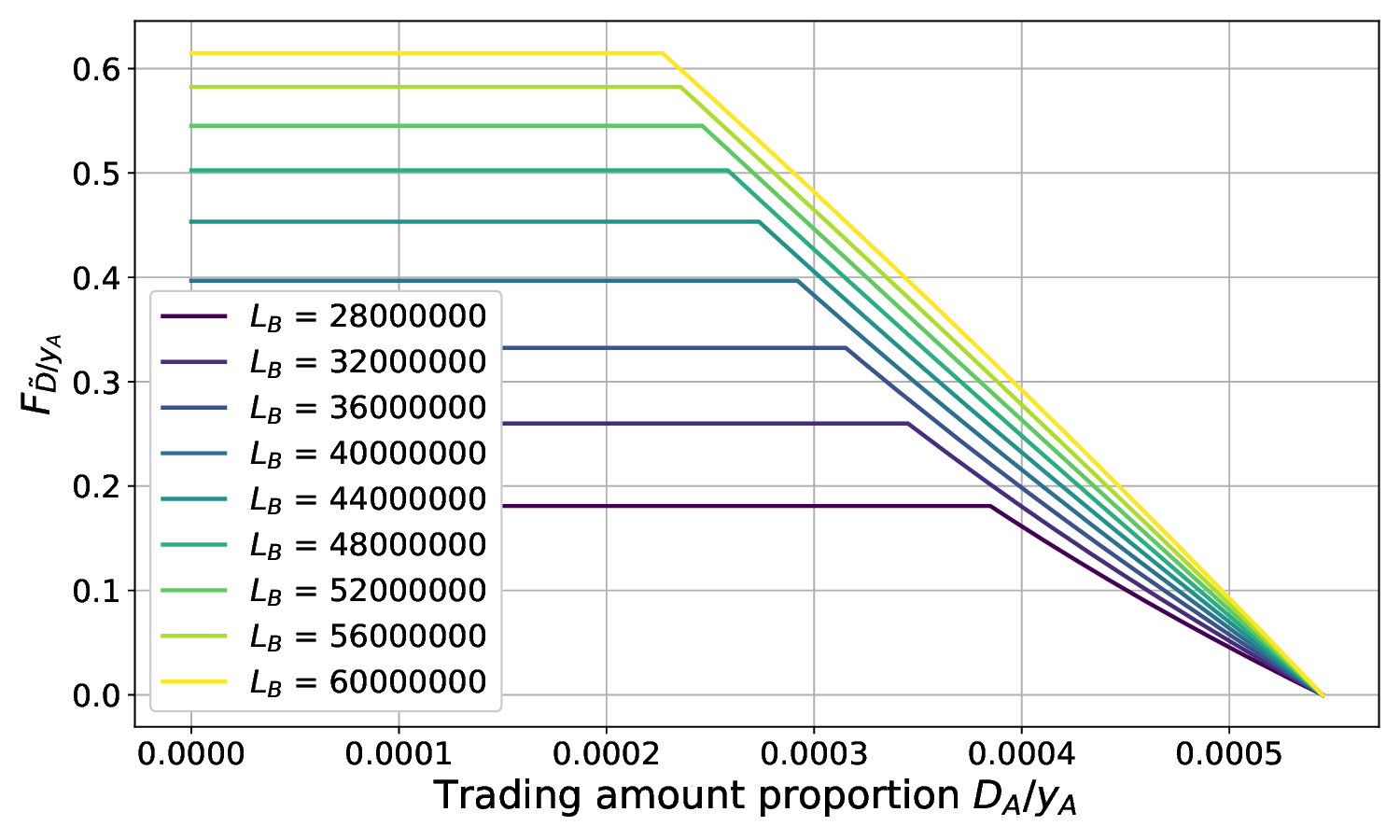}\\
	\includegraphics[width=0.46\linewidth]{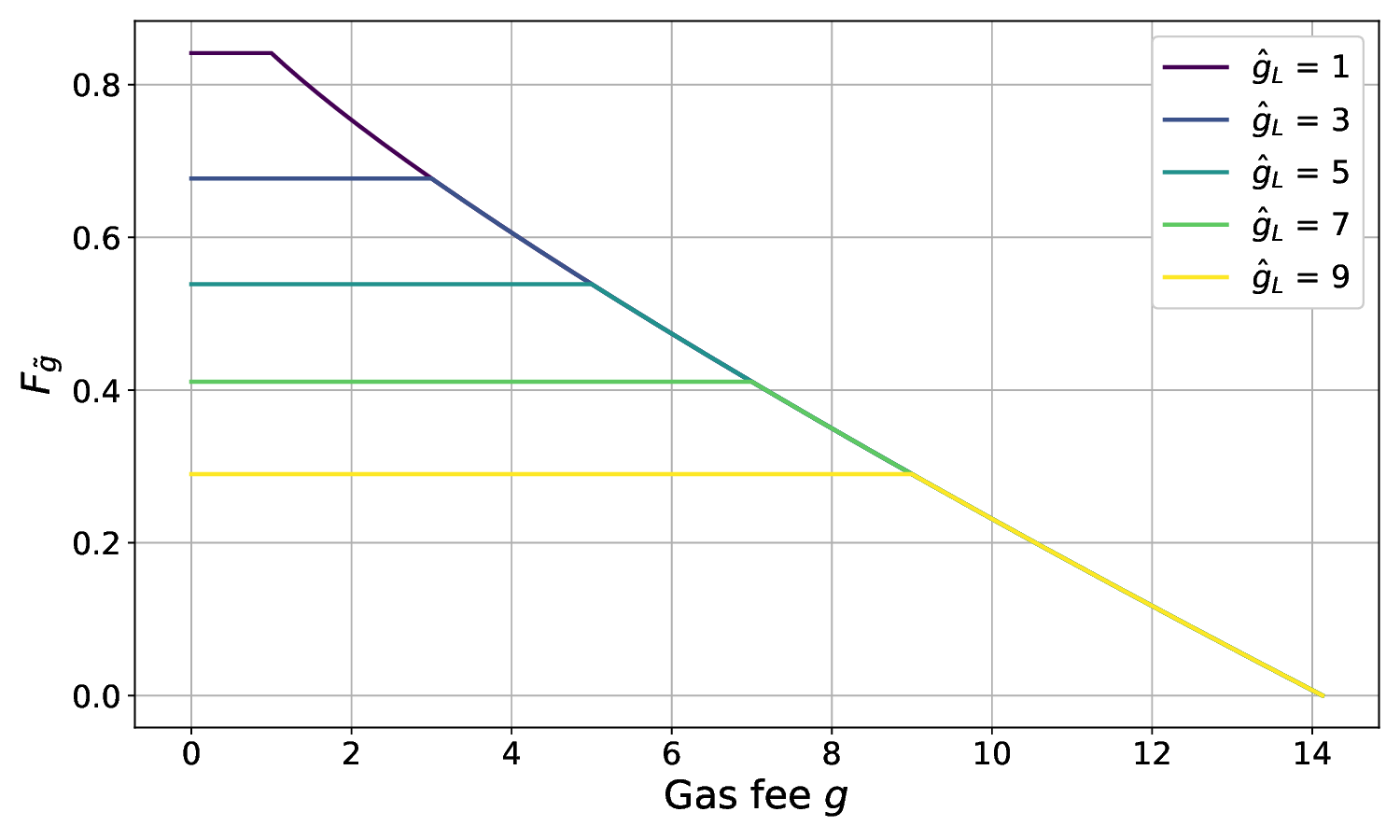}
	\includegraphics[width=0.46\linewidth]{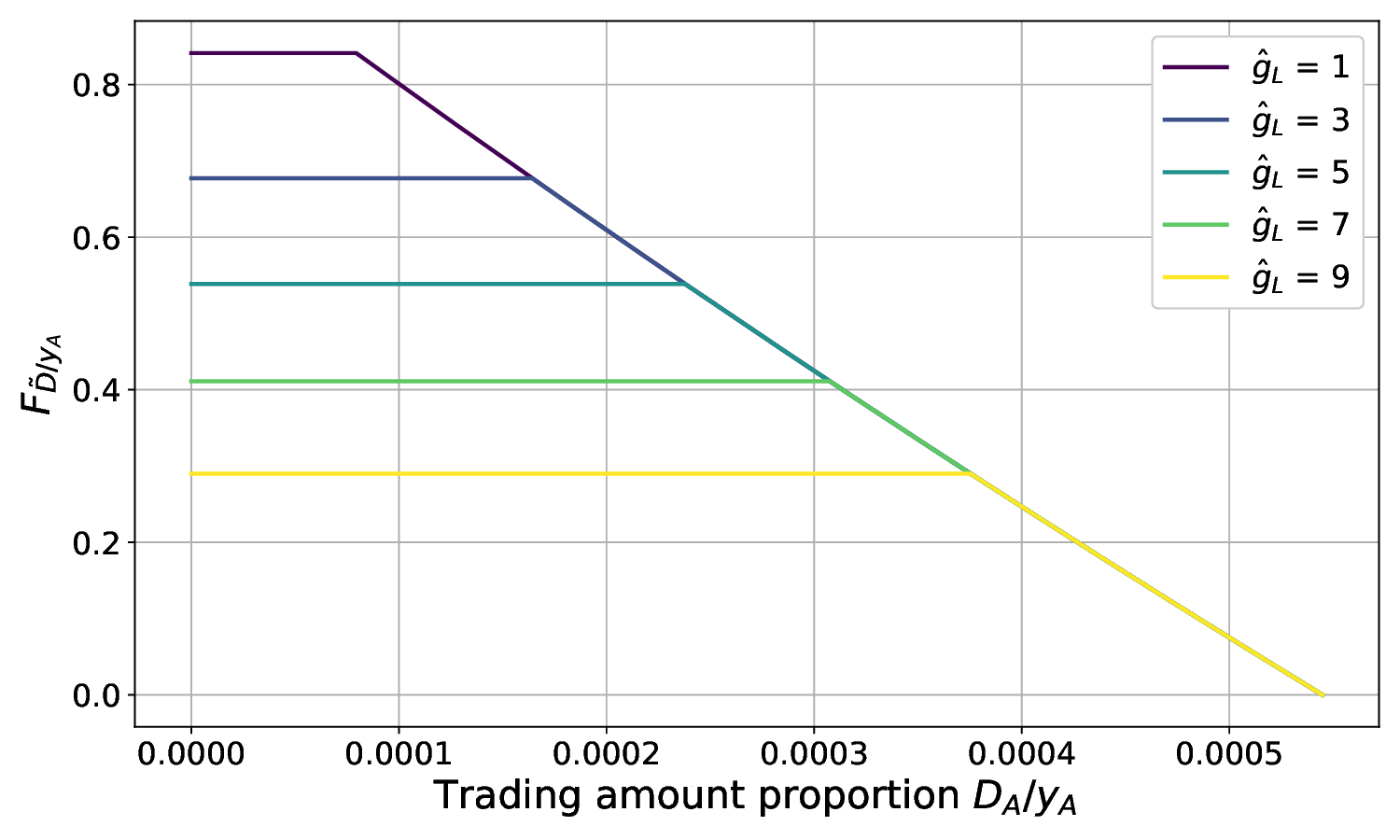}
	\caption{Decumulative distribution function of the gas fee (left panels) and relative trading amount $\tilde D/y_A$ (right panels) with respect to price discrepancy $O$, liquidity $L_B$, and base gas fee $\hat g_L$ in the auto-revert setting. Default parameter values are given in Table \ref{tab:ParameterValues} with $O=1.00109$ (for middle and bottom panels) and $\gamma=0.4$.}\label{fig:Fg_slippage0}
\end{figure}

In the numerical illustrations that follow, we set $O=1.00109$ and $\gamma=0.4$, as other values of $O$ in Table \ref{tab:ParameterValues} and other values of $\gamma$ yield qualitatively similar results. Figure \ref{fig:phi_gamma_slippage0} plots the density function of the gas fee. We observe that the density function is increasing when inventory risk aversion $\gamma$ is small, but becomes decreasing when $\gamma$ is large. Figure \ref{fig:alpha_slippage0} shows the trading probability $\alpha^*$ with respect to $O$, $L_B$, and $\hat g_L$. We observe that $\alpha^*$ increases with $O$ and $L_B$, and decreases with $\hat g_L$. The left panels of Figure \ref{fig:Fg_slippage0} plots the decumulative distribution function of the gas fee with respect to $O$, $L_B$, and $\hat g_L$. We observe that the gas fee increases with $O$ and $L_B$, and decreases with $\hat g_L$. Finally, the right panels in Figure \ref{fig:Fg_slippage0} plots the decumulative distribution function of the relative trading amount $\tilde D/y_A$ with respect to $O$, $L_B$, and $\hat g_L$. We observe that the relative trading amount increases with $O$ and $L_B$, and decreases with $\hat g_L$.

% \begin{figure}
% 	\centering
% 	\includegraphics[width=0.46\linewidth]{FD_O_slippage0_04_02.eps}
% 	\includegraphics[width=0.46\linewidth]{FD_LB_slippage0_04_02_100109.eps}
% 	\includegraphics[width=0.46\linewidth]{FD_gL_slippage0_04_02_100109.eps}
% 	\caption{Decumulative distribution function of the relative trading amount $\tilde D/y_A$ with respect to price discrepancy $O$ (top-left), liquidity $L_B$ (top-right), and base gas fee $\hat g_L$ (bottom) in the auto-revert setting. Default parameter values are given in Table \ref{tab:ParameterValues} with $O=1.00109$ and $\gamma=0.4$.}\label{fig:FD_slippage0}
% \end{figure}

\subsection{The Selectable-Revert Setting}\label{appx:slippagechoice}

In this subsection, we derive the equilibrium strategy in the selectable-revert setting.

The following proposition parallels Proposition \ref{prop:propertyEq}.

\begin{proposition}\label{prop:propertyEqChooseSlippage}
	Consider the selectable-revert setting. Suppose $(\alpha^*,\lambda^*)$ is a symmetric mixed Nash equilibrium. Then:
	\begin{enumerate}[leftmargin=2em]
		\item[(i)] $\alpha^*>0$;
		\item[(ii)] For each $\ell\in \{0,\infty\}$, define
		\begin{align}\label{eq:ResponseFunctionChoice}
			\begin{split}
				h^\ell(g,d_A):&=(1-\alpha^*)R_F(g, d_A)+\alpha^* \bigg[ \int R_F(g, d_A)\mathbf 1_{g>\bar g}\lambda^*(\dd \bar \ell \dd\bar g\dd\bar d_A)\\
				&\quad+ \int \tfrac{1}{2}\left(R_F(g, d_A)+R_S^\ell(g, d_A;\bar d_A)\right)\mathbf 1_{g=\bar g}\lambda^*(\dd \bar \ell \dd\bar g \dd\bar d_A)\\
				&\quad+ \int R_S^\ell(g, d_A;\bar d_A)\mathbf 1_{g<\bar g}\lambda^*(\dd \bar \ell \dd\bar g \dd\bar d_A)\bigg],\quad g\in[\hat g_L,\hat g_H],\; d_A\in[0,\hat D_A].
			\end{split}
		\end{align}
		Then, for almost every $g$ under the measure $\lambda^*$, $h^\ell(g,d_A)$ is continuous and strictly concave in $d_A\in[0,\hat D_A]$, and hence admits a unique maximizer $D_A^{\ell,*}(g)$, $\ell\in\{0,\infty\}$. Moreover,
		\begin{align*}
			\lambda^*\big(\{(\ell,g,d):d=D_A^{\ell,*}(g),\ell\in\{0,\infty\},g\in[\hat g_L,\hat g_H]\}\big)=1.
		\end{align*}		
	\end{enumerate}
\end{proposition}

In view of Proposition \ref{prop:propertyEqChooseSlippage}, we can restrict attention to strategies in which the trading amount is a deterministic function of the transaction reversion setting $\ell$ and gas fee $g$. The joint distribution of $g$ and $\ell$ can be represented by the marginal distribution of $g$ and the conditional distribution of $\ell$ given $g$. Thus, the arbitrageur's strategy can be represented by a 5-tuple $(\alpha,\Phi,\rho,D_A^0,D_A^\infty)$; see the discussion in Section \ref{subsect:TCR}.

\begin{proposition}\label{le:ChooseRandomizedGasDensity}
	Consider the selectable-revert setting. Suppose $(\alpha^*,\Phi^*,\rho^*,D_A^{0,*},D_A^{\infty,*})$ is a symmetric mixed Nash equilibrium. Then:
	\begin{enumerate}[leftmargin=2em]
		\item[(i)] $\Phi^*\big(g\in[\hat g_L,\hat g_H]:\rho^*(g)>0,D_A^{0,*}(g)=0\big)=0$, $\Phi^*\big(g\in[\hat g_L,\hat g_H]:\rho^*(g)<1,D_A^{\infty,*}(g)=0\big)=0$, and $\Phi^*\big(g\in[\hat g_L,\hat g_H]:\rho^*(g)>0,V^{0}(g,D_A^{0,*}(g))\le 0\big)=0$.
		\item[(ii)] $\Phi^*$ has no atoms on $[\hat g_L,\hat g_H]$.
		\item[(iii)] The reward function defined in \eqref{eq:ResponseFunctionChoice} can be written as
		\begin{align}
			h^\ell(g,d_A) = R_F(g, d_A)-\alpha^* \int_{g}^{g_H}\big[\rho^*(\bar g)V^\ell(g,d_A;D_A^{0,*}(\bar g)) + (1-\rho^*(\bar g))V^\ell(g,d_A;D_A^{\infty,*}(\bar g))\big]\Phi^*(\dd\bar g).\label{eq:ResponseFunction2Choice}
		\end{align}
		For any $g\in [\hat g_L,\hat g_H]$, $\rho\in[0,1]$, $d_A^{0}$, $d_A^\infty$, and $\bar d_A\in[0,\hat D_A]$, define
		\begin{align}
			H(g,\rho,d_A^0,d_A^\infty):=\rho h^0(g,d_A^0) + (1-\rho) h^\infty(g,d_A^\infty).\label{eq:ResponseFunctionChoiceAgg}
		\end{align}
		Then, for almost every $g$ under $\Phi^*$, $\big(\rho^*(g),D_A^{0,*}(g),D_A^{\infty,*}(g)\big)$ maximizes $H(g,\rho,d_A^0,d_A^\infty)$ in $(\rho,d_A^0,d_A^\infty)$. Moreover, $H\big(g,\rho^*(g),D_A^{0,*}(g),D_A^{\infty,*}(g)\big)$ is constant in $g\in[\hat g_L,\hat g_H]$. More precisely, let $M$ and $m$ be the essential supremum and infimum of $H\big(g,\rho^*(g),D_A^{0,*}(g),D_A^{\infty,*}(g)\big)$ on $[\hat g_L,\hat g_H]$ under $\Phi^*$, respectively:
		\begin{align*}
			M:=\sup\{a\in\mathbb{R}: \Phi(\{g\in[\hat g_L,\hat g_H]: H(\cdot)>a\})>0\},\\
			m:=\inf\{a\in\mathbb{R}: \Phi(\{g\in[\hat g_L,\hat g_H]: H(\cdot)<a\})>0\}.
		\end{align*}
		Then $M=m$. Moreover, $M\ge 0$, with equality if $\alpha^*<1$.
		\item[(v)] The support of $\Phi^*$ is $[\hat g_L,g_H]$ for some $g_H\in (\hat g_L,\hat g_H]$.
	\end{enumerate}
\end{proposition}

This proposition parallels Proposition \ref{le:RandomizedGasDensity}. Part (i) shows that the trading amount under any transaction reversion setting is positive whenever the arbitrageur chooses that setting with positive probability. It also shows that the first-mover advantage under $\ell=0$ is positive whenever this setting is chosen with positive probability. By Lemma \ref{le:Discontinuity}, the first-mover advantage under the no-revert setting $\ell=\infty$ is positive because the trading amount $D_A^{\ell,*}(g)$ is positive whenever this setting is selected.

In view of Proposition \ref{le:ChooseRandomizedGasDensity}, we assume that $\Phi^*$ admits a density function $\phi^*$ supported on $[\hat g_L,g_H]$ for some $g_H\in (\hat g_L,\hat g_H]$. In the following, we derive the equations satisfied by $(\alpha^*,\phi^*,\rho^*,D_A^{0,*},D_A^{\infty,*})$.

By \eqref{eq:ResponseFunction2Choice}, and since $\Phi^*$ admits a density function $\phi^*$ supported on $[\hat g_L,g_H]$, the reward function can be written as
\begin{align}
	h^\ell(g,d_A) = R_F(g, d_A)-\alpha^* \int_{g}^{g_H}\big[\rho^*(\bar g)V^\ell(g,d_A;D_A^{0,*}(\bar g)) + (1-\rho^*(\bar g))V^\ell(g,d_A;D_A^{\infty,*}(\bar g))\big]\phi^*(\dd\bar g).\label{eq:ResponseFunction3Choice}
\end{align}
By Proposition \ref{le:ChooseRandomizedGasDensity}-(iii), the optimality of $(D_A^{0,*}(g),D_A^{\infty,*}(g))$ implies
\begin{align}
	\frac{\partial H}{\partial d_A^\ell}(g,\rho^*(g),D_A^{0,*}(g),D_A^{\infty,*}(g))=0,\quad \ell\in\{0,\infty\}.\label{eq:OptimalityTradingAmountChoose}
\end{align}

Assume there exist open subsets $B_1$, $B_2$, and $B_3$ of $[\hat g_L,g_H]$ such that $[\hat g_L,g_H]\setminus (B_1\cup B_2\cup B_3)$ contains finitely many elements, with $\rho^*(g)=0$ on $B_1$, $\rho^*(g)=1$ on $B_2$, and $\rho^*(g)\in (0,1)$ on $B_3$. Then, the optimality of $\rho^*(g)$ implies that $\frac{\partial H}{\partial \rho  }(g,\rho^*(g),D_A^{0,*}(g),D_A^{\infty,*}(g))=0$ for $g\in B_3$. Moreover, $\frac{d\rho^*}{dg}(g)=0$ for $g\in B_1\cup B_2$. In consequence, for all $g\in B_1\cup B_2\cup B_3$,
\begin{align}
	\frac{\partial H}{\partial \rho  }(g,\rho^*(g),D_A^{0,*}(g),D_A^{\infty,*}(g))\times \frac{\partial \rho^*}{\partial g}=0.\label{eq:OptimalitySlippageProbChoose}
\end{align}

By Proposition \ref{le:ChooseRandomizedGasDensity}-(iii), $H\big(g,\rho^*(g),D_A^{0,*}(g),D_A^{\infty,*}(g)\big)$ is constant in $g\in [\hat g_L,g_H]$. Thus, its derivative with respect to $g$ is zero. Combining this with \eqref{eq:OptimalityTradingAmountChoose} and \eqref{eq:OptimalitySlippageProbChoose}, we obtain
\begin{align}
	&0 = -1 + \alpha^*\Big[\rho^*(g)V\big(g,\rho^*(g),D_A^{0,*}(g),D_A^{\infty,*}(g);D_A^{0,*}(g)\big)\notag\\
	&\quad + (1-\rho^*(g))V\big(g,\rho^*(g),D_A^{0,*}(g),D_A^{\infty,*}(g);D_A^{\infty,*}(g)\big)\Big]\phi^*(g)\notag\\
	&\quad + \alpha^*(1-r)\rho^*(g)\int_{g}^{g_H}\phi^*(\bar g)\dd\bar g,\label{eq:GasDensityEquationChoice}
\end{align}
where $V(g,\rho,d_A^0,d_A^\infty,\bar d_A):=\rho V^0(g,d_A^0;\bar d_A) + (1-\rho)V^\infty(g,d_A^\infty;\bar d_A)$.

Define
\begin{align}
	&\hat \rho(z):=\rho^*(g_H-L_Bz),\label{eq:TransrhohatChoice}\\
	&\hat x^\ell(z):=D_A^{\ell,*}(g_H-L_B z)/y_A, \quad \ell\in\{0,\infty\},\label{eq:TransxhatChoice}\\
	&\hat \theta(z):=\alpha^*L_B\phi^*(g_H-L_B z),\label{eq:TransthetahatChoice}\\
	&K_1^\ell(z,x,\bar x):=V^\ell(g_H-L_Bz,y_Ax,y_A\bar x)/L_B,\quad \ell\in \{0,\infty\}.\label{eq:FunK1Choice}
\end{align}
Then, \eqref{eq:GasDensityEquationChoice} becomes
\begin{align}
	\hat \theta(z) &= \Big(1-(1-r)\hat \rho(z)\int_{0}^{z}\hat \theta(\bar z)\dd\bar z\Big)\notag\\
	&\times \Big\{\hat \rho(z)\big[\hat \rho(z)K_1^0(z,\hat x^0(z),\hat x^0(z)) + (1-\hat \rho(z))K_1^\infty(z,\hat x^\infty(z),\hat x^0(z))\big]\notag\\
	&\quad + (1-\hat \rho(z))\big[\hat \rho(z)K_1^0(z,\hat x^0(z),\hat x^\infty(z)) + (1-\hat \rho(z))K_1^\infty(z,\hat x^\infty(z),\hat x^\infty(z))\big]\Big\}^{-1}.\label{eq:GasDensityEquationTranChoice}
\end{align}

On the other hand, \eqref{eq:OptimalityTradingAmountChoose} implies
\begin{align}
	&\rho^*(g) \frac{\partial R_F}{\partial d_A}(g,D_A^{0,*}(g))
	- \alpha^*\rho^*(g)\int_{g}^{g_H}\Big[\rho^*(\bar g)\frac{\partial V^0}{\partial d_A}\big(g,D_A^{0,*}(g);D_A^{0,*}(\bar g)\big)\notag\\
	&\qquad + \big(1-\rho^*(\bar g)\big)\frac{\partial V^0}{\partial d_A}\big(g,D_A^{0,*}(g);D_A^{\infty,*}(\bar g)\big)\Big]\phi^*(\bar g)\dd\bar g=0,\label{eq:TradingEquation1Choice}\\
	& (1-\rho^*(g)) \frac{\partial R_F}{\partial d_A}(g,D_A^{\infty,*}(g))
	- \alpha^*(1-\rho^*(g))\int_{g}^{g_H}\Big[\rho^*(\bar g)\frac{\partial V^\infty}{\partial d_A}\big(g,D_A^{\infty,*}(g);D_A^{0,*}(\bar g)\big)\notag\\
	&\qquad + \big(1-\rho^*(\bar g)\big)\frac{\partial V^\infty}{\partial d_A}\big(g,D_A^{\infty,*}(g);D_A^{\infty,*}(\bar g)\big)\Big]\phi^*(\bar g)\dd\bar g=0.\label{eq:TradingEquation2Choice}
\end{align}

Denote
$
	K_0^\ell(z,x,\bar x):=\frac{y_A}{L_B}\frac{\partial V^\ell}{\partial d_A}(g_H-L_B z,y_Ax;y_A\bar x),~ \ell\in\{0,\infty\},
$
and recall $Q$ as defined in \eqref{eq:QFun}. Then, \eqref{eq:TradingEquation1Choice} and \eqref{eq:TradingEquation2Choice} become
\begin{align}
	& \hat \rho(z) Q(\hat x^0(z))
	- \hat \rho(z)\int_{0}^{z}\Big[\hat \rho(\bar z)K_0^0(z,\hat x^0(z),\hat x^0(\bar z))\notag\\
	&\qquad + \big(1-\hat \rho(\bar z)\big)K_0^0\big(z,\hat x^0(z),\hat x^\infty(\bar z)\big)\Big]\hat \theta(\bar z)\dd\bar z=0,\label{eq:TradingEquation1TransChoice}\\
	& \big(1-\hat \rho(z)\big) Q(\hat x^\infty(z))
	- \big(1-\hat \rho(z)\big)\int_{0}^{z}\Big[\hat \rho(\bar z)K_0^\infty(z,\hat x^\infty(z),\hat x^0(\bar z))\notag\\
	&\qquad + \big(1-\hat \rho(\bar z)\big)K_0^\infty\big(z,\hat x^\infty(z),\hat x^\infty(\bar z)\big)\Big]\hat \theta(\bar z)\dd\bar z=0.\label{eq:TradingEquation2TransChoice}
\end{align}
We can rewrite \eqref{eq:TradingEquation1TransChoice} and \eqref{eq:TradingEquation2TransChoice} as
\begin{align}
	& Q(\hat x^0(z))
	- \int_{0}^{z}\Big[\hat \rho(\bar z)K_0^0(z,\hat x^0(z),\hat x^0(\bar z))\notag\\
	&\qquad + \big(1-\hat \rho(\bar z)\big)K_0^0\big(z,\hat x^0(z),\hat x^\infty(\bar z)\big)\Big]\hat \theta(\bar z)\dd\bar z=0,\quad \text{for } z \text{ with } \hat \rho(z)>0,\label{eq:TradingEquation1TransAChoice}\\
	& Q(\hat x^\infty(z))
	- \int_{0}^{z}\Big[\hat \rho(\bar z)K_0^\infty(z,\hat x^\infty(z),\hat x^0(\bar z))\notag\\
	&\qquad + \big(1-\hat \rho(\bar z)\big)K_0^\infty\big(z,\hat x^\infty(z),\hat x^\infty(\bar z)\big)\Big]\hat \theta(\bar z)\dd\bar z=0,\quad \text{for } z \text{ with } \hat \rho(z)<1.\label{eq:TradingEquation2TransAChoice}
\end{align}

The optimality of $\rho^*(g)$ implies
\begin{align}
	&\rho^*(g)\in (0,1)\Longrightarrow h^0(g,D_A^{0,*}(g)) = h^\infty(g,D_A^{\infty,*}(g)),\label{eq:SlippageProbEq1}\\
	&h^0(g,D_A^{0,*}(g)) > h^\infty(g,D_A^{\infty,*}(g))\Longrightarrow \rho^*(g)=1,\label{eq:SlippageProbEq2}\\
	&h^0(g,D_A^{0,*}(g)) < h^\infty(g,D_A^{\infty,*}(g))\Longrightarrow \rho^*(g)=0.\label{eq:SlippageProbEq3}
\end{align}
Denote
% \begin{align}
$
	P(x):= \frac{R_F(g,d_Ax)+g}{L_B} = x\big((1+x)^{-1}-O^{-1}\big).
$
	% \label{eq:FunP}
% \end{align}
Then, the optimality equations \eqref{eq:SlippageProbEq1}--\eqref{eq:SlippageProbEq3} can be written as
\begin{align}
	&\hat \rho(z)\in (0,1)\Longrightarrow P(\hat x^0(z)) - \int_0^z\left[\hat \rho(\bar z)K_1^0(z,\hat x^0(z),\hat x^0(\bar z)) +(1-\hat \rho(\bar z))K_1^0(z,\hat x^0(z),\hat x^\infty(\bar z)\right]\hat \theta(\bar z)\dd \bar z\notag\\
	& =P(\hat x^\infty(z)) - \int_0^z\left[\hat \rho(\bar z)K_1^\infty(z,\hat x^\infty(z),\hat x^0(\bar z)) +(1-\hat \rho(\bar z))K_1^\infty(z,\hat x^\infty(z),\hat x^\infty(\bar z)\right]\hat \theta(\bar z)\dd \bar z,\label{eq:SlippageProbEqTrans1}\\
	&  P(\hat x^0(z)) - \int_0^z\left[\hat \rho(\bar z)K_1^0(z,\hat x^0(z),\hat x^0(\bar z)) +(1-\hat \rho(\bar z))K_1^0(z,\hat x^0(z),\hat x^\infty(\bar z)\right]\hat \theta(\bar z)\dd \bar z\notag\\
	& >P(\hat x^\infty(z)) - \int_0^z\left[\hat \rho(\bar z)K_1^\infty(z,\hat x^\infty(z),\hat x^0(\bar z)) +(1-\hat \rho(\bar z))K_1^\infty(z,\hat x^\infty(z),\hat x^\infty(\bar z)\right]\hat \theta(\bar z)\dd \bar z\Longrightarrow \hat \rho(z)=1,\label{eq:SlippageProbEqTrans2}\\
	&  P(\hat x^0(z)) - \int_0^z\left[\hat \rho(\bar z)K_1^0(z,\hat x^0(z),\hat x^0(\bar z)) +(1-\hat \rho(\bar z))K_1^0(z,\hat x^0(z),\hat x^\infty(\bar z)\right]\hat \theta(\bar z)\dd \bar z\notag\\
	& <P(\hat x^\infty(z)) - \int_0^z\left[\hat \rho(\bar z)K_1^\infty(z,\hat x^\infty(z),\hat x^0(\bar z)) +(1-\hat \rho(\bar z))K_1^\infty(z,\hat x^\infty(z),\hat x^\infty(\bar z)\right]\hat \theta(\bar z)\dd \bar z\Longrightarrow \hat \rho(z)=0.\label{eq:SlippageProbEqTrans3}
\end{align}

To summarize, we derive equations \eqref{eq:GasDensityEquationTranChoice}, \eqref{eq:TradingEquation1TransAChoice}, \eqref{eq:TradingEquation2TransAChoice}, \eqref{eq:SlippageProbEqTrans1}, \eqref{eq:SlippageProbEqTrans2}, and \eqref{eq:SlippageProbEqTrans3}, which must be satisfied by $(\hat \theta,\hat \rho,\hat x^0,\hat x^\infty)$ on $[0,(g_H-\hat g_L)/L_B]$. Finally, we determine $g_H$ and the trading probability $\alpha^*$. Note from \eqref{eq:ResponseFunction3Choice} that $h^\ell(g_H,d_A) = R_F(g_H,d_A)$ for $\ell\in \{0,\infty\}$. Therefore, $(\rho,\hat D_A,\hat D_A)$ maximizes $H(g_H,\rho,d_A^0,d_A^\infty)$ for any $\rho \in[0,1]$. Consequently, the arbitrageur's expected profit is
\begin{align*}
	H(g_H,\rho^*(g_H),D_A^{0,*}(g_H),D_A^{\infty,*}(g_H)) = R_F(g_H, \hat D_A) = \hat g_H-g_H.
\end{align*}
Moreover, by \eqref{eq:TransthetahatChoice} and recalling that $\phi^*$ is a probability density supported on $[\hat g_L,g_H]$, we obtain
\begin{align*}
	\alpha^* = \int_0^{(g_H-\hat g_L)/L_B}\hat \theta(z)\dd z.
\end{align*}
Then, in view of Proposition \ref{le:ChooseRandomizedGasDensity}, if there exists $g_H< \hat g_H$ such that $\int_0^{(g_H-\hat g_L)/L_B}\hat \theta(z)\dd z=1$, then the gas fee is supported on $[\hat g_L,g_H]$, the trading probability satisfies $\alpha^*=1$, and the expected profit is positive. Otherwise, the gas fee is supported on $[\hat g_L,\hat g_H]$, the trading probability is
\begin{align}
	\alpha^* = \int_0^{(\hat g_H-\hat g_L)/L_B}\hat \theta(z)\dd z,
\end{align}
and the expected profit is zero.

\subsection{Comparative Statics in the Selectable-Revert Setting}\label{appx:propertyTCR}

In this subsection, we conduct numerical experiments to study the impact of $O$, $L_B$, and $\hat g_L$ on the probability of trading, trading amount, and gas fee in the selectable-revert setting. We take the values of $L_B$, $\hat g_L$, and $r$ from Table \ref{tab:ParameterValues}, and set $O=1.00109$ and $\gamma=0.4$, as other values of $O$ and $\gamma$ yield qualitatively similar results.

\begin{figure}
	\centering
	\includegraphics[width=0.32\linewidth]{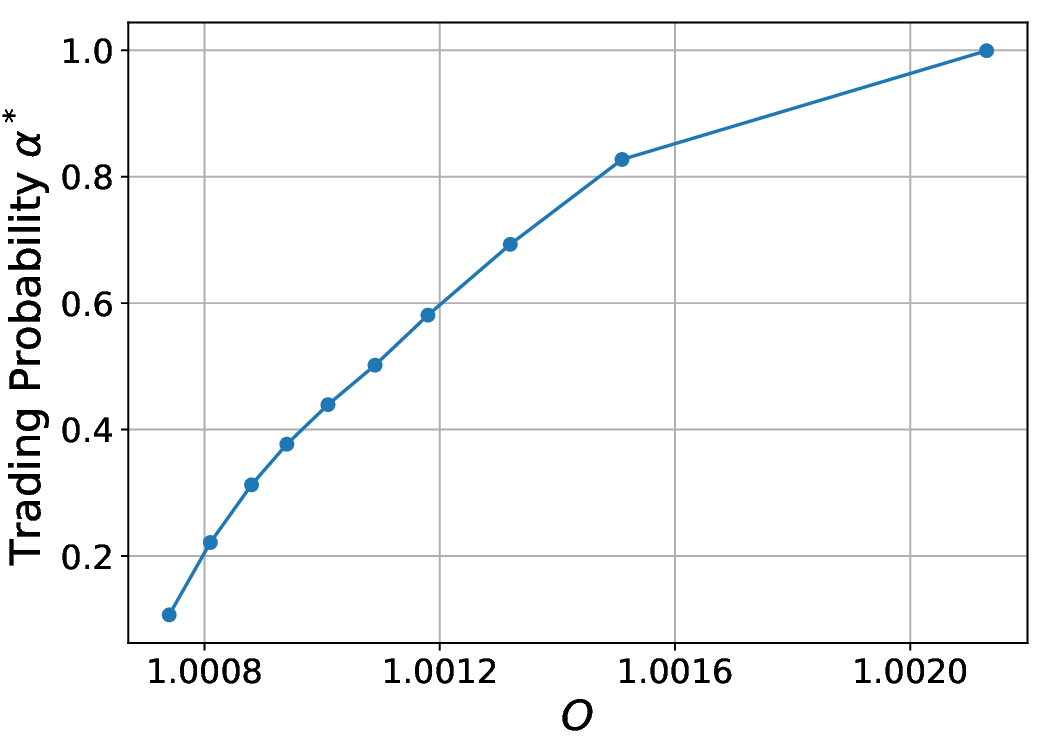}
	\includegraphics[width=0.32\linewidth]{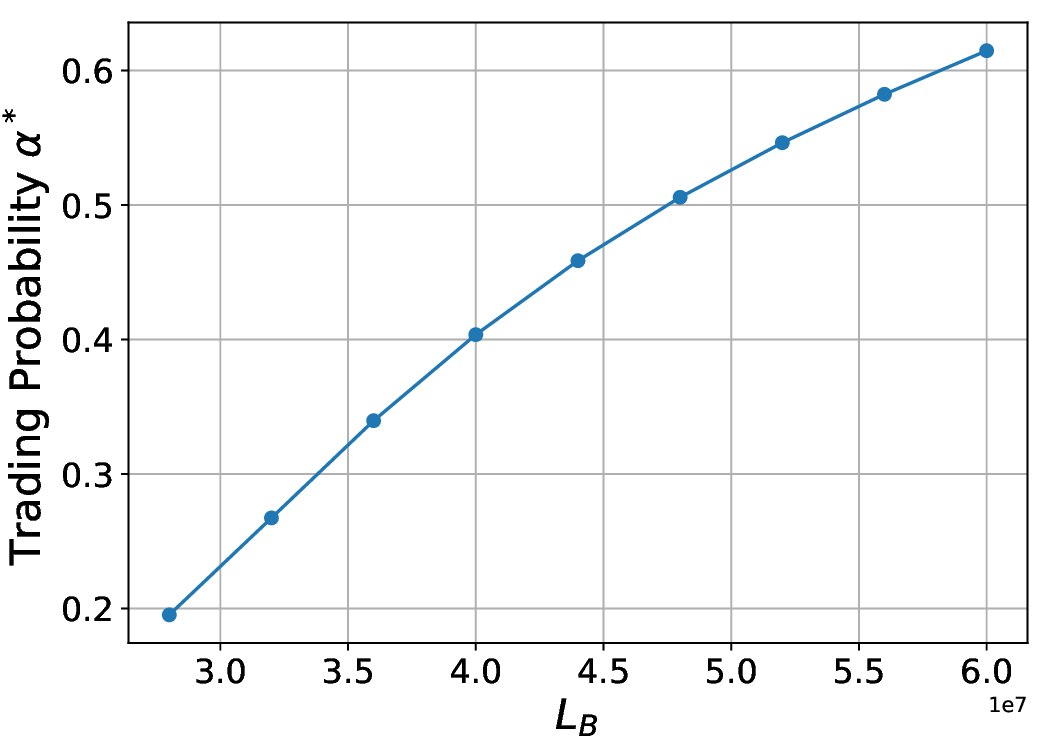}
	\includegraphics[width=0.32\linewidth]{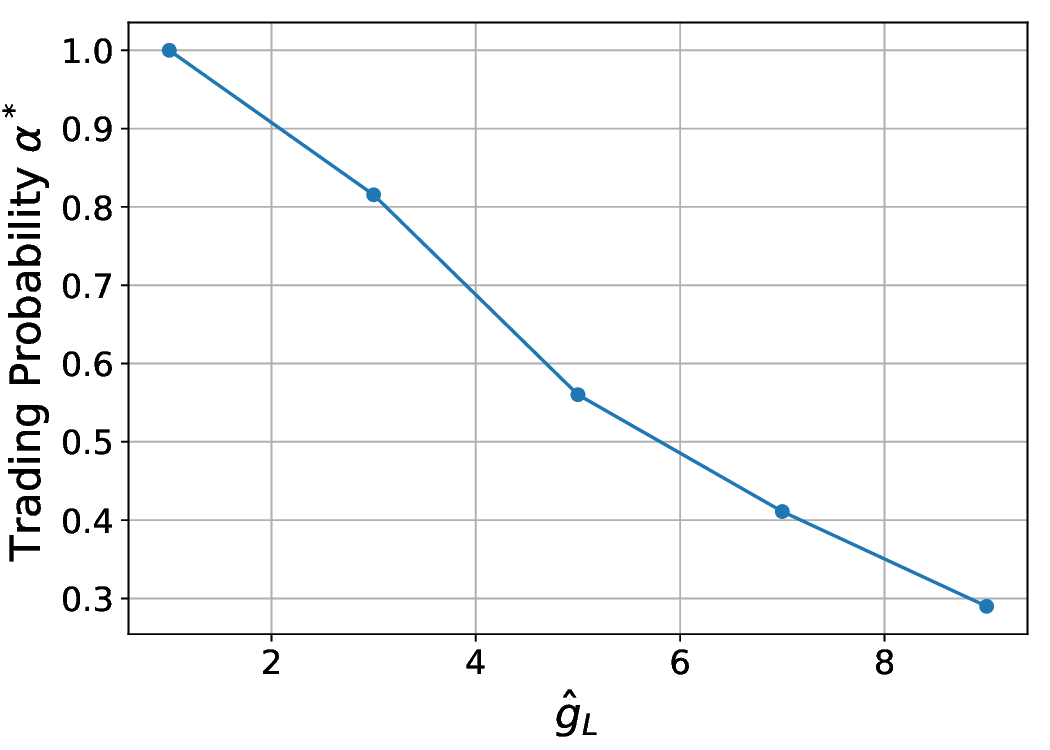}
	\caption{Probability of trading $\alpha^*$ with respect to the price discrepancy $O$, liquidity $L_B$, and base gas fee $\hat g_L$ in the selectable-revert setting. Default parameters are given in Table \ref{tab:ParameterValues} with $\gamma=0.4$. In the middle and right panels, we set $O=1.00109$.}\label{fig:alpha_all}
\end{figure}

\begin{figure}
	\centering
	\includegraphics[width=0.46\linewidth]{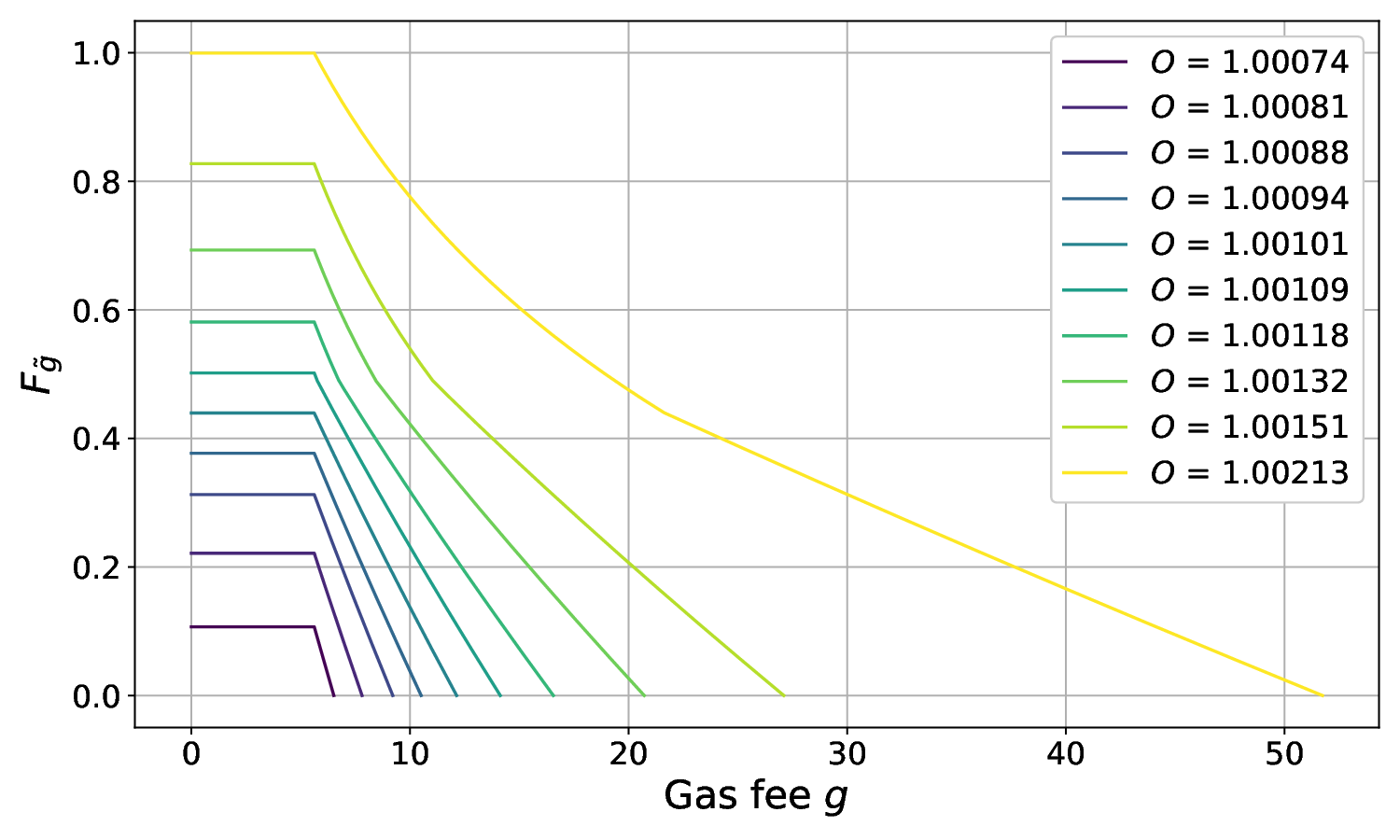}
	\includegraphics[width=0.46\linewidth]{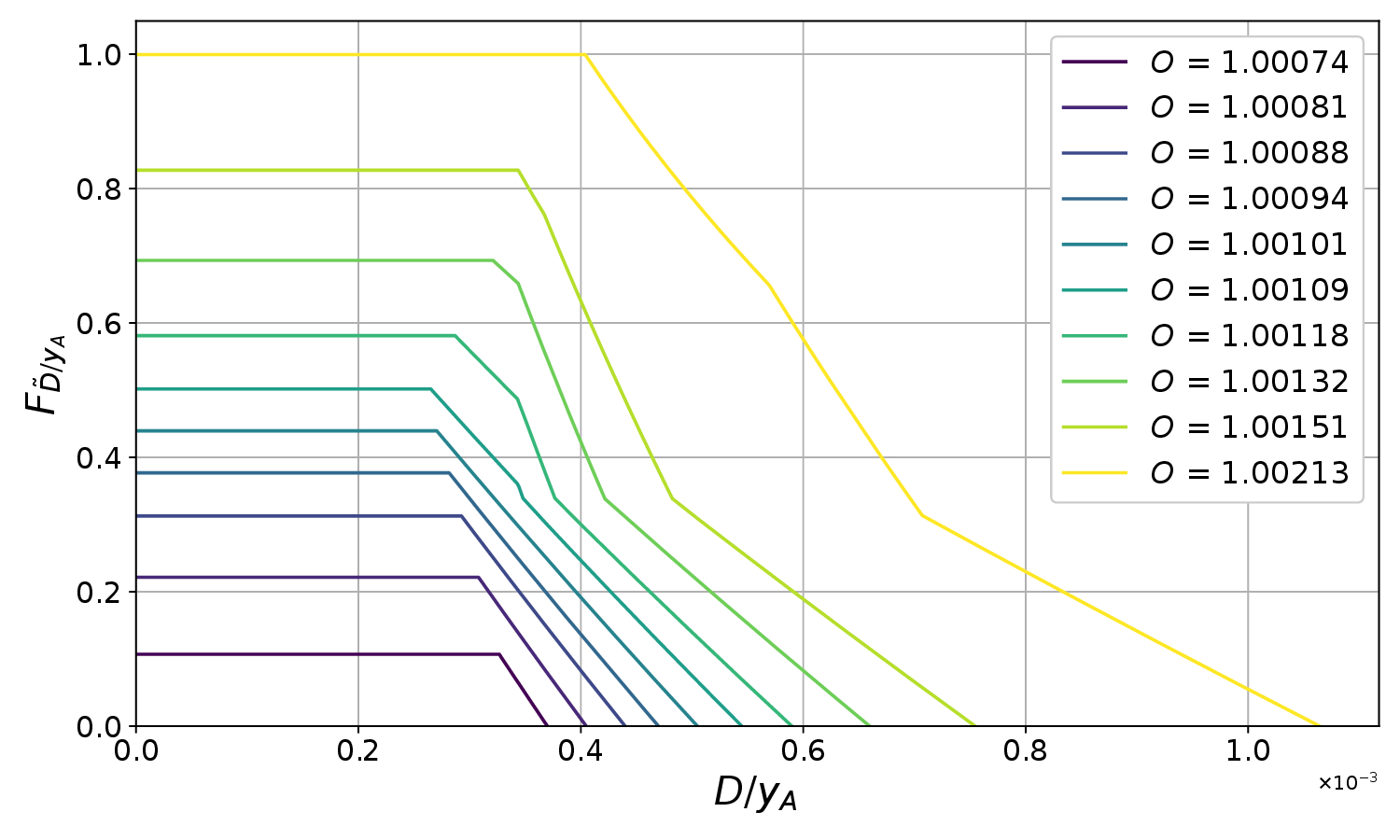}\\
	\includegraphics[width=0.46\linewidth]{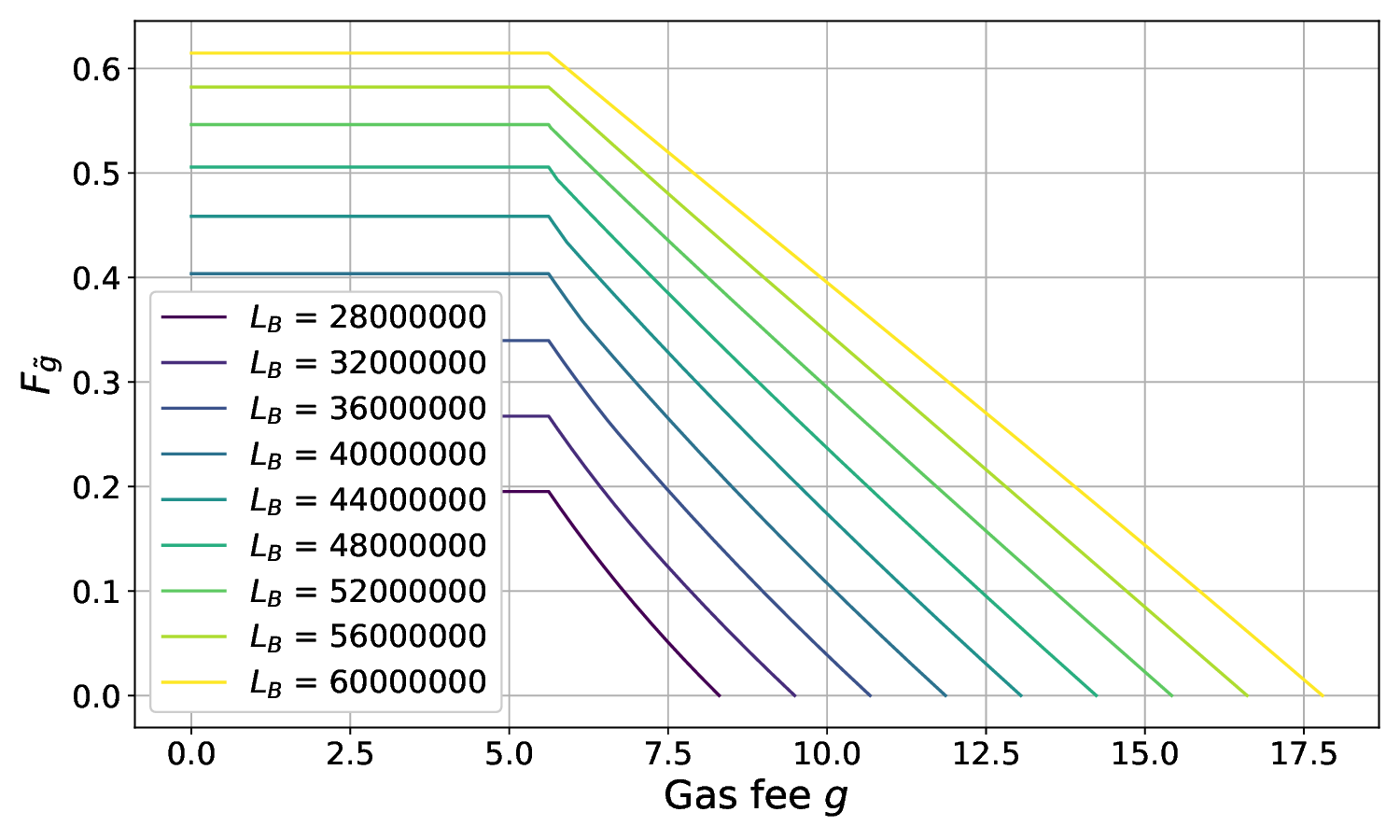}
	\includegraphics[width=0.46\linewidth]{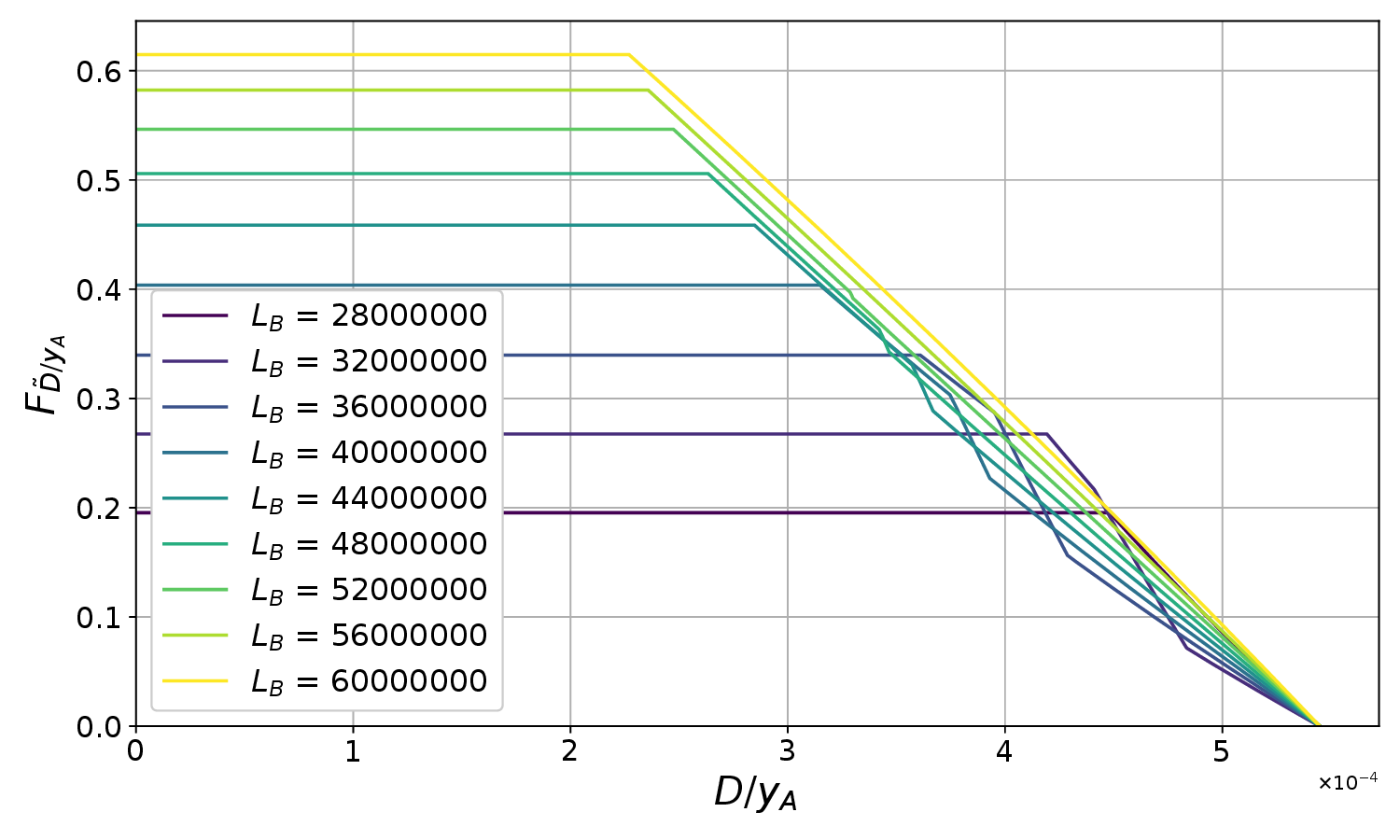}\\
	\includegraphics[width=0.46\linewidth]{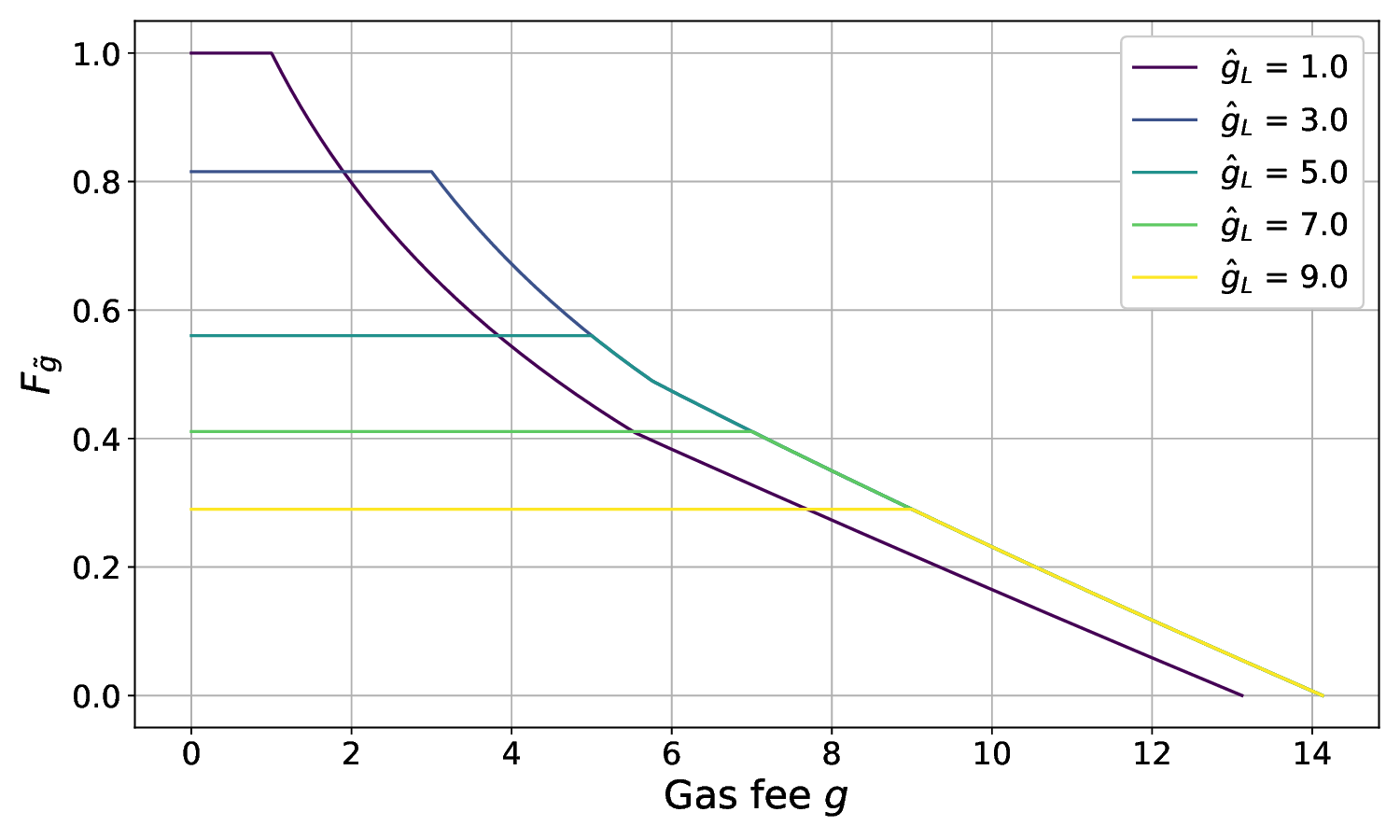}
	\includegraphics[width=0.46\linewidth]{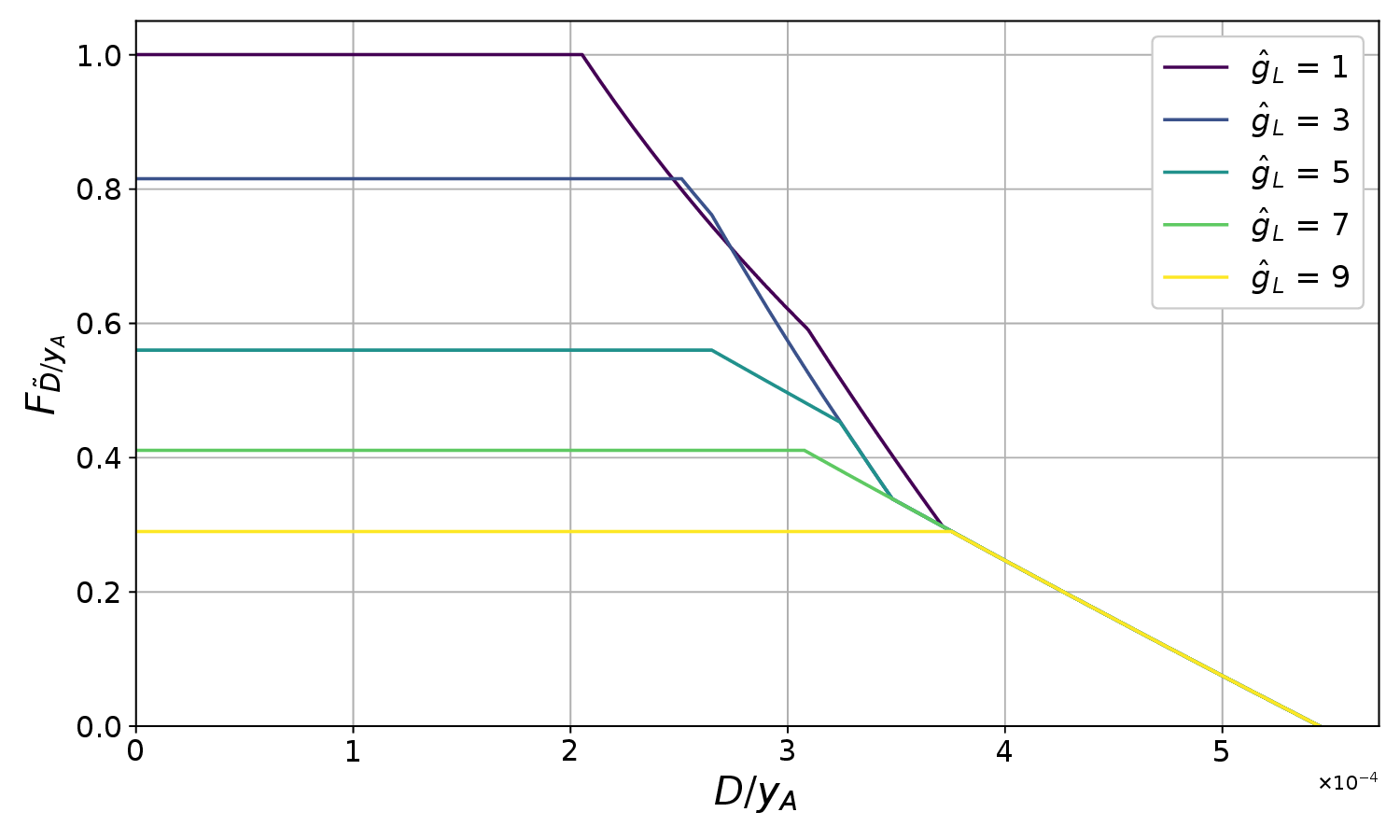}
	\caption{Decumulative distribution function of the gas fee $\tilde g$ (left panels) and the relative trading amount $\tilde D/y_A$ (right panels) with respect to the price discrepancy $O$, liquidity $L_B$, and base gas fee $\hat g_L$ in the selectable-revert setting. Default parameters are given in Table \ref{tab:ParameterValues} with $\gamma=0.4$. In the middle and bottom panels, we set $O=1.00109$.}\label{fig:FD_mix}
\end{figure}

Figure \ref{fig:alpha_all} plots the trading probability $\alpha^*$ with respect to $O$, $L_B$, and $\hat g_L$. Similar to the no-revert and auto-revert settings, we observe that $\alpha^*$ increases with $O$ and $L_B$, and decreases with $\hat g_L$. The right panels of Figure \ref{fig:FD_mix} shows the decumulative distribution of the relative trading amount $\tilde D/y_A$ with respect to $O$, $L_B$, and $\hat g_L$. The relative trading amount increases with $O$, consistent with the no-revert and auto-revert settings. However, its dependence on $L_B$ and $\hat g_L$ is not monotone. The left panels of Figure \ref{fig:FD_mix} presents the decumulative distribution of the gas fee $\tilde g$ with respect to $O$, $L_B$, and $\hat g_L$. The gas fee increases with $O$ and $L_B$, while its dependence on $\hat g_L$ is again non-monotone.

\clearpage

\pagenumbering{arabic}
\renewcommand*{\thepage}{B-\arabic{page}}

\setcounter{equation}{0}
\renewcommand{\theequation}{B.\arabic{equation}}

\section{Proofs} \label{sec:proofs}

\begin{pfof}{Lemma \ref{le:Discontinuity}}
	Lemma \ref{le:Discontinuity} follows directly from Equations \eqref{eq:FirstMoverReward}, \eqref{eq:SecondMoverReward}, \eqref{eq:SecondMoverReward_0}, and \eqref{eq:FMAInfS}. \halmos
\end{pfof}

\begin{pfof}{Proposition \ref{prop:propertyEq}}
	We first prove (i) by contradiction. Suppose $\alpha^*=0$. Consider the strategy $(\alpha,\mu)$ with $\alpha=1$ and $\mu(\{(\hat g_L,\hat D_A)\})=1$. It is straightforward to see that
	\begin{align*}
		\Pi^\ell(\alpha,\mu;\alpha^*,\mu^*)=R_F(\hat g_L, \hat D_A)>0=\Pi^\ell(\alpha^*,\mu^*;\alpha^*,\mu^*),
	\end{align*}
	contradicting the assumption that $(\alpha^*,\mu^*)$ is a mixed Nash equilibrium.
	
	Next, we prove (ii). By the dominated convergence theorem, $h^\ell(g,d_A)$ is continuously differentiable in $d_A$. It is straightforward to see that
	\begin{align*}
		\mu^*(\{g\in[\hat g_L,\hat g_H]:\mu^*((g,\hat g_H]\times [0,\hat D_A])=1\}\times [0,\hat D_A])=0.
	\end{align*}
	Then, for almost every $g$ under $\mu^*$, the arbitrageur is the first mover with positive probability. By Lemma \ref{le:Discontinuity}, we conclude that $h^\ell(g,d_A)$ is strictly concave in $d_A$ and thus admits a unique maximizer in the compact interval $[0,\hat D_A]$.
	
	Suppose, on the contrary, that 
	% \begin{align*}
	$
		\mu^*\big(\{(g,d):d=D_A^*(g),g\in[\hat g_L,\hat g_H]\}\big)<1.
	$
	% \end{align*}
	Then, there exist nonempty intervals $(g_1,g_2)\subseteq [\hat g_L,\hat g_H]$ and $(d_1,d_2)\subseteq[0,\hat D_A]$ such that $d\neq D_A^*(g)$ for every $(g,d_A)\in B:=(g_1,g_2)\times(d_1,d_2)$ and $\mu^*(B)>0$. Consequently, $h^\ell(g,d_A)<h^\ell(g,D_A^*(g))$ for all $(g,d_A)\in B$, and therefore
	\begin{align*}
		\int_{B}h^\ell(g,d_A)\mu^*(\dd g\dd d_A)<\int_{B}h^\ell(g,D_A^*(g))\mu^*(\dd g\dd d_A).
	\end{align*}
	Consider the measure $\tilde \mu$ defined by moving the probability mass of $(g_1,g_2)\times(d_1,d_2)$ under $\mu^*$ to the region $\tilde B:=\{(g,d):d=D_A^*(g),g\in(g_1,g_2)\}$. Specifically, set $\tilde \mu(C):=\mu^*(C)$ for $C\subseteq B^c\cap \tilde B^c$, $\tilde \mu(B)=0$, and
	\begin{align*}
		\tilde \mu(C):= \mu^*(C)+\int \mathbf 1_{(g,D_A^*(g))\in C}\int_{(d_1,d_2)}\mu^*(\dd g\dd d_A)
	\end{align*}
	for every $C\subseteq \tilde B$. Straightforward calculation yields
	\begin{align*}
		& \Pi^\ell(\alpha^*,\tilde \mu;\alpha^*,\mu^*) \\
		&= \alpha^*\int h^\ell(g,d_A)\tilde \mu(\dd g\dd d_A) \\
		&= \alpha^*\left(\int h^\ell(g,d_A)\mu^*(\dd g\dd d_A) + \int_{\tilde B}h^\ell(g,d_A)\tilde \mu(\dd g\dd d_A)-\int_{B}h^\ell(g,d_A)\mu^*(\dd g\dd d_A)\right)\\
		&= \alpha^*\left(\int h^\ell(g,d_A)\mu^*(\dd g\dd d_A) + \int_{(g_1,g_2)}h^\ell(g,D_A^*(g))\int_{(d_1,d_2)}\mu^*(\dd g\dd d_A)-\int_{B}h^\ell(g,d_A)\mu^*(\dd g\dd d_A)\right)\\
		&= \alpha^*\left(\int h^\ell(g,d_A)\mu^*(\dd g\dd d_A) + \int_{B}h^\ell(g,D_A^*(g))\mu^*(\dd g\dd d_A)-\int_{B}h^\ell(g,d_A)\mu^*(\dd g\dd d_A)\right)\\
		&>\alpha^*\int h^\ell(g,d_A)\mu^*(\dd g\dd d_A)\\
		&=\Pi^\ell(\alpha^*,\mu^*;\alpha^*,\mu^*).
	\end{align*}
	This contradicts the assumption that $(\alpha^*,\mu^*)$ is an equilibrium. \halmos
\end{pfof}

\begin{pfof}{Proposition \ref{le:RandomizedGasDensity}}
	We first prove assertion (i) by contradiction. Suppose $\Phi^*(B)>0$, where $B:=\{g\in[\hat g_L,\hat g_H]:D_A^*(g)=0\}$. Recall that $\alpha^*>0$, as shown in Proposition \ref{prop:propertyEq}. For any $g\in B$ and any $\bar g\ge 0$, we have $R_F(g,D_A^*(g);\bar g)=-g$, $R_S^\infty(g,D_A^*(g);\bar g)=-g$, and $R_S^0(g,D_A^*(g);\bar g)=-rg$. Consequently,
	\begin{align*}
		R_F(g,D_A^*(g);\bar g)\le -\hat g_L,~~
		R_S^\ell(g,D_A^*(g);\bar g)\le -r\hat g_L.
	\end{align*}
	
	Define $\tilde \alpha:=\alpha^*(1-\Phi^*(B))$ and $\tilde \Phi$ by $\tilde \Phi(C):=\Phi^*(C\cap B^c)/(1-\Phi^*(B))$ for all $C\subseteq [\hat g_L,\hat g_H]$, where we set $\tilde \Phi$ to be any probability measure on $[\hat g_L,\hat g_H]$ when $\Phi^*(B)=1$. Then,
	\begin{align*}
		& \Pi^\ell(\tilde \alpha,\tilde \Phi,D_A^*;\alpha^*, \Phi^*, D_A^*)\\
		& = \alpha^* (1-\alpha^*)\int_{B^c} R_F(g, D_A^*(g))\Phi^*(\dd g)\notag\\
		&\quad +(\alpha^*)^2 \bigg[ \int_{B^c\times [\hat g_L,\hat g_H]} R_F(g, D_A^*(g))\mathbf 1_{g>\bar g}\Phi^*(\dd g) \Phi^*(\dd\bar g)\notag\\
		&\quad + \int_{B^c\times [\hat g_L,\hat g_H]} \frac{1}{2}\big(R_F(g, D_A^*(g))+R_S^\ell(g, D_A(g); D_A^*(\bar g))\big)\mathbf 1_{g=\bar g}\Phi^*(\dd g)\Phi^*(\dd \bar g)\notag\\
		&\quad + \int_{B^c\times [\hat g_L,\hat g_H]} R_S^\ell(g, D_A^*(g);D_A^*(\bar g))\mathbf 1_{g<\bar g}\Phi^*(\dd g) \Phi^*(\dd\bar g)\bigg]\\
		& = \Pi(\alpha^*, \Phi^*,D_A^*;\alpha^*, \Phi^*, D_A^*)-\alpha^* (1-\alpha^*)\int_{B} R_F(g, D_A^*(g))\Phi^*(\dd g)\notag\\
		&\quad -(\alpha^*)^2 \bigg[ \int_{B\times [\hat g_L,\hat g_H]} R_F(g, D_A^*(g))\mathbf 1_{g>\bar g}\Phi^*(\dd g) \Phi^*(\dd\bar g)\notag\\
		&\quad + \int_{B\times [\hat g_L,\hat g_H]} \frac{1}{2}\big(R_F(g, D_A^*(g))+R_S^\ell(g, D_A(g); D_A^*(\bar g))\big)\mathbf 1_{g=\bar g}\Phi^*(\dd g)\Phi^*(\dd\bar g)\notag\\
		&\quad + \int_{B\times [\hat g_L,\hat g_H]} R_S^\ell(g, D_A^*(g);D_A^*(\bar g))\mathbf 1_{g<\bar g}\Phi^*(\dd g) \Phi^*(\dd\bar g)\bigg]\\
		& \ge  \Pi^\ell(\alpha^*, \Phi^*,D_A^*;\alpha^*, \Phi^*, D_A^*)+\alpha^*rg_L\Phi^*(B)\\
		& >\Pi^\ell(\alpha^*, \Phi^*,D_A^*;\alpha^*, \Phi^*, D_A^*).
	\end{align*}
	This contradicts the assumption that $(\alpha^*,\Phi^*,D_A^*)$ is an equilibrium.
	
	By Lemma \ref{le:Discontinuity}-(ii), we immediately conclude that $\Phi^*\big(\{g\in[\hat g_L,\hat g_H]:V^\infty(g,D_A^*(g);D_A^*(g))\le0\}\big)=0$. Now consider the case $\ell=0$. Suppose, on the contrary, that $\Phi^*\big(\{g\in[\hat g_L,\hat g_H]:V^0(g,D_A^*(g);D_A^*(g))\le0\}\big)>0$. Because
	\begin{align*}
		R_S^0(g,d_A;\bar d_A)\le -rg\le -r\hat g_L,\quad \forall g\in[\hat g_L,\hat g_H],\; d_A\ge 0,\; \bar d_A\ge 0,
	\end{align*}
	we conclude that $R_F(g,D_A^*(g);D_A^*(g))\le R_S^0(g,D_A^*(g);D_A^*(g))\le -r\hat g_L<0$ for all $g\in[\hat g_L,\hat g_H]$ such that $V^0(g,D_A^*(g);D_A^*(g))\le 0$. Thus, $\Phi^*(\bar B)>0$, where
	\begin{align*}
		\bar B:=\{g\in[\hat g_L,\hat g_H]:R_F(g,D_A^*(g);D_A^*(g))\le R_S^0(g,D_A^*(g);D_A^*(g))\le -r\hat g_L\}.
	\end{align*}
	As in the proof for $\Phi^*\big(\{g\in[\hat g_L,\hat g_H]:D_A^*(g)=0\}\big)=0$, we can show that the strategy moving the probability mass of $\bar B$ to the action of no trading yields a strictly higher expected profit than $(\alpha^*,\Phi^*)$, contradicting the assumption that $(\alpha^*,\Phi^*,D_A^*)$ is an equilibrium. Therefore, we must have $\Phi^*\big(\{g\in[\hat g_L,\hat g_H]:V^0(g,D_A^*(g);D_A^*(g))\le0\}\big)=0$.
	
	Next, we prove assertion (ii) by contradiction. Suppose that $\Phi^*(\{g_0\})>0$ for some $g_0\in [\hat g_L,\hat g_H]$. By (i), we have $V^\ell(g_0,D_A^*(g_0);D_A^*(g_0))>0$.
	First, we consider the case $g_0<\hat g_H$. There exists a sequence $\{\epsilon_n\}$ in $(0,\hat g_H-g_0)$ such that $\Phi^*(\{g_0+\epsilon_n\})=0$ and $\lim_{n\rightarrow \infty}\epsilon_n=0$. For each $n$, define $\tilde \Phi_n$ by moving the probability mass of $g_0$ under $\Phi^*$ to $g_0+\epsilon_n$, and define $\tilde D_{A,n}$ by setting $\tilde D_{A,n}(g)=D^*(g)$ for $g\neq g_0+\epsilon_n$ and $\tilde D_{A,n}(g_0+\epsilon_n)=D^*(g_0)$. Then,
	\begin{align*}
		&\Pi^\ell(\alpha^*,\tilde \Phi_n,\tilde D_{A,n};\alpha^*,\Phi^*,D_A^*)\\
		& = \Pi^\ell(\alpha^*,\Phi^*, D_{A}^*;\alpha^*,\Phi^*,D_A^*)\\
		&\quad+ \alpha^* (1-\alpha^*)\big(R_F(g_0+\epsilon_n,D_A^*(g_0))-R_F(g_0,D_A^*(g_0))\big)\Phi^*(\{g_0\})\\
		&\quad +(\alpha^*)^2 \bigg[ \Phi^*(\{g_0\})\int_{[\hat g_L,\hat g_H]} \left[R_F(g_0+\epsilon_n, D_A^*(g_0))\mathbf 1_{g_0+\epsilon_n>\bar g}-R_F(g_0, D_A^*(g_0))\mathbf 1_{g_0>\bar g}\right]\Phi^*(\dd\bar g)\notag\\
		&\quad -  \frac{1}{2}\big(R_F(g_0, D_A^*(g_0))+R_S^\ell(g_0, D_A^*(g_0); D_A^*(g_0))\big)\Phi^*(\{g_0\})^2\notag\\
		&\quad + \Phi^*(\{g_0\})\int_{[\hat g_L,\hat g_H]} \Big[R_S^\ell(g_0+\epsilon_n, D_A^*(g_0);D_A^*(\bar g))\mathbf 1_{g_0+\epsilon_n<\bar g} -R_S\ell(g_0, D_A^*(g_0);D_A^*(\bar g))\mathbf 1_{g_0<\bar g}\Big]\Phi^*(\dd\bar g)\bigg]\\
		& = \Pi^\ell(\alpha^*,\Phi^*, D_{A}^*;\alpha^*,\Phi^*,D_A^*)- \alpha^* (1-\alpha^*)\Phi^*(\{g_0\})\epsilon_n\\
		&\quad +(\alpha^*)^2 \bigg[ \Phi^*(\{g_0\})\left(R_F(g_0, D_A^*(g_0))\Phi^*\big([g_0,g_0+\epsilon_n)\big)-\epsilon_n \Phi^*\big([\hat g_L,g_0+\epsilon_n)\big) \right)\notag\\
		&\quad -  \frac{1}{2}\big(R_F(g_0, D_A^*(g_0))+R_S^\ell(g_0, D_A^*(g_0); D_A^*(g_0))\big)\Phi^*(\{g_0\})^2\notag\\
		&\quad + \Phi^*(\{g_0\})\bigg(-\int_{[\hat g_L,\hat g_H]} R_S^\ell(g_0, D_A^*(g_0);D_A^*(\bar g))\mathbf 1_{g_0<\bar g\le g_0+\epsilon_n}\Phi^*(\dd\bar g)\\
		&\quad -\epsilon_n\int_{[\hat g_L,\hat g_H]} R_S^\ell(g_0+\epsilon_n, D_A^*(g_0);D_A^*(\bar g))\mathbf 1_{g_0+\epsilon_n<\bar g}\Phi^*(\dd\bar g)\bigg].
	\end{align*}
	It is straightforward to see that
	\begin{align*}
		&\alpha^*(1-\alpha^*)\Phi^*(\{g_0\})\epsilon_n + (\alpha^*)^2\Phi^*(\{g_0\})\Phi^*\big([\hat g_L,g_0+\epsilon_n)\big)\\
		&+(\alpha^*)^2\Phi^*(\{g_0\})\left(\int_{[\hat g_L,\hat g_H]} R_S^\ell(g_0+\epsilon_n,D_A^*(g_0);D_A^*(\bar g))\mathbf 1_{g_0+\epsilon_n<\bar g}\Phi^*(\dd\bar g)\right)\epsilon_n
	\end{align*}
	converges to 0 as $n\rightarrow \infty$. Moreover,
	\begin{align*}
		&\Phi^*(\{g_0\})R_F(g_0,D_A^*(g_0))\Phi^*\big([g_0,g_0+\epsilon_n)\big)\\
		&\quad - \tfrac{1}{2}\big(R_F(g_0,D_A^*(g_0))+R_S^\ell(g_0,D_A^*(g_0);D_A^*(g_0))\big)\Phi^*(\{g_0\})^2\\
		&\quad - \Phi^*(\{g_0\})\int_{[\hat g_L,\hat g_H]} R_S^\ell(g_0,D_A^*(g_0);D_A^*(\bar g))\mathbf 1_{g_0<\bar g\le g_0+\epsilon_n}\Phi^*(\dd\bar g)\\
		&= \tfrac{1}{2}\big(R_F(g_0,D_A^*(g_0))-R_S^\ell(g_0,D_A^*(g_0);D_A^*(g_0))\big)\Phi^*(\{g_0\})^2\\
		&\quad + \Phi^*(\{g_0\})\int_{[\hat g_L,\hat g_H]} \big(R_F(g_0,D_A^*(g_0))-R_S^\ell(g_0,D_A^*(g_0);D_A^*(\bar g))\big)\mathbf 1_{g_0<\bar g< g_0+\epsilon_n}\Phi^*(\dd\bar g)\\
		&\ge \tfrac{1}{2}\big(R_F(g_0,D_A^*(g_0))-R_S^\ell(g_0,D_A^*(g_0);D_A^*(g_0))\big)\Phi^*(\{g_0\})^2>0,
	\end{align*}
	where the equality holds because $\Phi^*(\{g_0+\epsilon_n\})=0$, and the inequalities hold because $V^\ell(g_0,D_A^*(g_0);D_A^*(g_0))>0$ and $\Phi^*(\{g_0\})>0$. We therefore conclude that
	\begin{align*}
		\liminf_{n\rightarrow \infty}\big(\Pi^\ell(\alpha^*,\tilde \Phi_n,\tilde D_{A,n};\alpha^*,\Phi^*,D_A^*) - \Pi^\ell(\alpha^*,\Phi^*,D_A^*;\alpha^*,\Phi^*,D_A^*)\big)>0.
	\end{align*}
	This contradicts the assumption that $(\alpha^*,\Phi^*,D_A^*)$ is an equilibrium.
	
	Next, we consider the case $g_0=\hat g_H$. Because $R_F(\hat g_H,d_A)\le R_F(\hat g_H,\hat D_A)=0$ for every $d_A\ge 0$,
	\begin{align*}
		R_S^\ell(\hat g_H,D_A^*(\hat g_H);D_A^*(\hat g_H)) < R_F(\hat g_H,D_A^*(\hat g_H)) \le 0,
	\end{align*}
	where the strict inequality holds because $V^\ell(\hat g_H,D_A^*(\hat g_H);D_A^*(\hat g_H))>0$.
	Define $\tilde \alpha=\alpha^*(1-\Phi^*(\{\hat g_H\}))$ and $\tilde \Phi$ by
	$\tilde \Phi(C):=\frac{\Phi^*(C\cap[\hat g_L,\hat g_H))}{1-\Phi^*(\{\hat g_H\})}$, $\forall C\subseteq [\hat g_L,\hat g_H]$,	where we set $\tilde \Phi$ to be an arbitrary probability measure on $[\hat g_L,\hat g_H)$ when $\Phi^*(\{\hat g_H\})=1$. Straightforward calculation yields
	\begin{align*}
		&\Pi^\ell(\tilde \alpha,\tilde \Phi,D_A^*;\alpha^*,\Phi^*,D_A^*)-\Pi^\ell(\alpha^*,\Phi^*,D_A^*;\alpha^*,\Phi^*,D_A^*) \\
		&= -\alpha^*(1-\alpha^*)R_F(\hat g_H,D_A^*(\hat g_H))\Phi^*(\{\hat g_H\})\\
		&\quad -(\alpha^*)^2 \bigg[ R_F(\hat g_H,D_A^*(\hat g_H))\Phi^*([\hat g_L,\hat g_H))\Phi^*(\{\hat g_H\})\\
		&\quad + \tfrac{1}{2}\big(R_F(\hat g_H,D_A^*(\hat g_H))+R_S^\ell(\hat g_H,D_A^*(\hat g_H);D_A^*(\hat g_H))\big)\Phi^*(\{\hat g_H\})^2\bigg]>0.
	\end{align*}
	This contradicts the assumption that $(\alpha^*,\Phi^*,D_A^*)$ is an equilibrium.
	
	Next, we prove assertion (iii). Because $\Phi^*$ has no atoms, we can immediately derive \eqref{eq:ResponseFunction2} from \eqref{eq:ResponseFunction}. For contradiction, suppose $M>m$. Then there exist $a_1<a_2$ and $B_1,B_2\subseteq [\hat g_L,\hat g_H]$ such that $\Phi^*(B_1)>0$, $\Phi^*(B_2)>0$, $h^\ell(g,D_A^*(g))\le a_1$ for all $g\in B_1$, and $h^\ell(g,D_A^*(g))\ge a_2$ for all $g\in B_2$. Define $\tilde \Phi$ by moving the mass of $B_1$ under $\Phi^*$ to $B_2$, i.e.,
	\begin{align*}
		\tilde \Phi(C):=\Phi^*(C\cap B_1^c\cap B_2^c)+\frac{\Phi^*(B_1)+\Phi^*(B_2)}{\Phi^*(B_2)}\Phi^*(C\cap B_2),\quad \forall C\subseteq [\hat g_L,\hat g_H].
	\end{align*}	
	Then straightforward calculation yields
	\begin{align*}
		&\Pi^\ell(\alpha^*,\tilde \Phi,D_A^*;\alpha^*,\Phi^*,D_A^*)-\Pi^\ell(\alpha^*,\Phi^*,D_A^*;\alpha^*,\Phi^*,D_A^*) \\
		&= \alpha^*\left(\int h^\ell(g,D_A^*(g))\tilde \Phi(\dd g)-\int h^\ell(g,D_A^*(g))\Phi^*(\dd g)\right)\\
		&= \alpha^*\bigg(\frac{\Phi^*(B_1)+\Phi^*(B_2)}{\Phi^*(B_2)}\int_{B_2}h^\ell(g,D_A^*(g))\Phi^*(\dd g)\\
		&\quad -\left(\int_{B_1}h^\ell(g,D_A^*(g))\Phi^*(\dd g)+\int_{B_2}h^\ell(g,D_A^*(g))\Phi^*(\dd g)\right)\bigg)\\
		&= \alpha^*\left(\frac{\Phi^*(B_1)}{\Phi^*(B_2)}\int_{B_2}h^\ell(g,D_A^*(g))\Phi^*(\dd g)-\int_{B_1}h^\ell(g,D_A^*(g))\Phi^*(\dd g)\right)\\
		&\ge \alpha^*\big(\Phi^*(B_1)a_2-\Phi^*(B_1)a_1\big)>0.
	\end{align*}
	This contradicts the assumption that $(\alpha^*,\Phi^*,D_A^*)$ is an equilibrium.
	
	Next, for contradiction, suppose $M<0$. Then,
	\begin{align*}
		\Pi^\ell(0,\Phi^*,D_A^*;\alpha^*,\Phi^*,D_A^*)-\Pi^\ell(\alpha^*,\Phi^*,D_A^*;\alpha^*,\Phi^*,D_A^*)
		&=0-\alpha^*\int h^\ell(g,D_A^*(g))\Phi^*(\dd g)=-M>0.
	\end{align*}
	This contradicts the assumption that $(\alpha^*,\Phi^*,D_A^*)$ is an equilibrium. Thus, we must have $M\ge 0$.
	
	Similarly, for contradiction, suppose $M>0$ and $\alpha^*<1$. Then,
	\begin{align*}
		&\Pi^\ell(1,\Phi^*,D_A^*;\alpha^*,\Phi^*,D_A^*)-\Pi^\ell(\alpha^*,\Phi^*,D_A^*;\alpha^*,\Phi^*,D_A^*)\\
		&=(1-\alpha^*)\int h^\ell(g,D_A^*(g))\Phi^*(\dd g)=(1-\alpha^*)M>0.
	\end{align*}
	This contradicts the assumption that $(\alpha^*,\Phi^*,D_A^*)$ is an equilibrium. Thus, $\alpha^*=1$ if $M>0$.
	
	Finally, we prove assertion (iv). Denote by $\mathrm{supp}(\Phi^*)$ the support of $\Phi^*$. Without loss of generality, assume $\mathrm{supp}(\Phi^*)\neq [\hat g_L,\hat g_H]$. We only need to show that for any $g_0\in [\hat g_L,\hat g_H]\setminus \mathrm{supp}(\Phi^*)$, $\Phi^*([g_0,\hat g_H])=0$. Suppose, on the contrary, that there exists $\hat g_0\in [\hat g_L,\hat g_H]\setminus \mathrm{supp}(\Phi^*)$ such that $\Phi^*([\hat g_0,\hat g_H])>0$. Because $\Phi^*$ has no atoms, we must have $\hat g_0<\hat g_H$. Define
	% \begin{align*}
	$
		\hat g_1:=\sup\{g\ge \hat g_0:\Phi^*([\hat g_0,g))=0\}.
	$
	% \end{align*}
	Since $\hat g_0\in [\hat g_L,\hat g_H)\setminus \mathrm{supp}(\Phi^*)$, which is open, $\Phi^*([\hat g_0,\hat g_H])>0$, and $\Phi^*$ has no atoms, we conclude that $\hat g_1\in (\hat g_0,\hat g_H)$.
	
	By assertion (iii), $\Phi^*(\{g\in[\hat g_L,\hat g_H]: h^\ell(g,D_A^*(g))=M\})=1$ for some constant $M$. Then, by the definition of $\hat g_1$, we can find a decreasing sequence $\{g_n\}$ converging to $\hat g_1$ such that $h^\ell(g_n,D_A^*(g_n))=M$ for all $n$. Let $\hat x_0$ be the maximizer of $h^\ell(\hat g_0,d_A)$ in $d_A\in [0,\hat D_A]$, which uniquely exists due to the continuity and strict concavity of $h^\ell(g,d)$ in $d$. Also recall that $\Phi^*$ has no atoms. Then, for any $n$, we have
	\begin{align*}
		& h^\ell(\hat g_0,\hat x_0)\ge h^\ell(\hat g_0,D_A^*(g_n))\\
		&= (1-\alpha^*)R_F(\hat g_0,D_A^*(g_n))\\
		&\quad + \alpha^*\bigg[R_F(\hat g_0,D_A^*(g_n))\Phi^*([\hat g_L,\hat g_0)) + \int_{(\hat g_0,\hat g_H]} R_S^\ell(\hat g_0,D_A^*(g_n);D_A^*(\bar g))\Phi^*(\dd\bar g)\bigg]\\
		&= (1-\alpha^*)R_F(\hat g_0,D_A^*(g_n))\\
		&\quad + \alpha^*\bigg[R_F(\hat g_0,D_A^*(g_n))\Phi^*([\hat g_L,g_n)) + \int_{(g_n,\hat g_H]} R_S^\ell(\hat g_0,D_A^*(g_n);D_A^*(\bar g))\Phi^*(\dd\bar g)\bigg]\\
		&\quad + \alpha^*\bigg[-R_F(\hat g_0,D_A^*(g_n))\Phi^*([\hat g_1,g_n)) + \int_{(\hat g_1,g_n]} R_S^\ell(\hat g_0,D_A^*(g_n);D_A^*(\bar g))\Phi^*(\dd\bar g)\bigg]\\
		&= h^\ell(g_n,D_A^*(g_n))+(1-\alpha^*)\big(R_F(\hat g_0,D_A^*(g_n))-R_F(g_n,D_A^*(g_n))\big)\\
		&\quad + \alpha^*\bigg[\big(R_F(\hat g_0,D_A^*(g_n))-R_F(g_n,D_A^*(g_n))\big)\Phi^*([\hat g_L,g_n))\\
		&\quad + \int_{(g_n,\hat g_H]} \big(R_S^\ell(\hat g_0,D_A^*(g_n);D_A^*(\bar g))-R_S^\ell(g_n,D_A^*(g_n);D_A^*(\bar g))\big)\Phi^*(\dd\bar g)\bigg]\\
		&\quad + \alpha^*\bigg[-R_F(\hat g_0,D_A^*(g_n))\Phi^*([\hat g_1,g_n)) + \int_{(\hat g_1,g_n]} R_S^\ell(\hat g_0,D_A^*(g_n);D_A^*(\bar g))\Phi^*(\dd\bar g)\bigg]\\
		&\ge h^\ell(g_n,D_A^*(g_n))+r(g_n-\hat g_0)\\
		&\quad + \alpha^*\bigg[-R_F(\hat g_0,D_A^*(g_n))\Phi^*([\hat g_1,g_n)) + \int_{(\hat g_1,g_n]} R_S^\ell(\hat g_0,D_A^*(g_n);D_A^*(\bar g))\Phi^*(\dd\bar g)\bigg],
	\end{align*}
	where the second equality holds because $\Phi^*([\hat g_0,\hat g_1])=0$, and the inequality holds because both $R_F$ and $R_S^\ell$ are linear in the gas fee $g$. The right-hand side converges to $M+r(\hat g_1-\hat g_0)$. Therefore, we conclude that $h^\ell(\hat g_0,\hat x_0)\ge M+r(\hat g_1-\hat g_0)$.
	
	Now, let $\tilde \Phi$ be the probability measure assigning mass 1 to $\hat g_0$, and let $\tilde D_A$ be defined by $\tilde D_A(\hat g_0)=\hat x_0$. Then,
	\begin{align*}
		&\Pi^\ell(\alpha^*,\tilde \Phi,\tilde D_A;\alpha^*,\Phi^*,D_A^*)-\Pi^\ell(\alpha^*,\Phi^*,D_A^*;\alpha^*,\Phi^*,D_A^*)\\
		&= \alpha^*\left(\int h^\ell(g,\tilde D_A(g))\tilde \Phi(\dd g)-\int h^\ell(g,D_A^*(g))\Phi^*(\dd g)\right)\\
		&= \alpha^*\left(h^\ell(\hat g_0,\hat x_0)-\int h^\ell(g,D_A^*(g))\Phi^*(\dd g)\right)\\
		&\ge \alpha^*r(\hat g_1-\hat g_0)>0,
	\end{align*}
	where the inequality holds because $h^\ell(\hat g_0,\hat x_0)\ge M+r(\hat g_1-\hat g_0)$ and $\Phi^*(\{g\in[\hat g_L,\hat g_H]: h^\ell(g,D_A^*(g))=M\})=1$.
	This contradicts the assumption that $(\alpha^*,\Phi^*,D_A^*)$ is an equilibrium. \halmos
\end{pfof}		

\begin{pfof}{Lemma \ref{le:ODE}}
	For any $z\ge 0$ and any positive, continuous function $y$ on $[0,z]$, denote
	\begin{align*}
		F(x,z;y):= \int_0^z K(x,y(\bar z))\dd\bar z - Q(x),\quad x\ge 0.
	\end{align*}
	Then, \eqref{eq:KeyODENormalized} can be written as
	$
		F(\hat x(z),z;\hat x)=0,~ z\ge 0.
	$
	
	First, we prove that the solution to \eqref{eq:KeyODENormalized} must be continuously differentiable and strictly decreasing. To this end, suppose that a positive, continuous function $\hat x$ is a solution to \eqref{eq:KeyODENormalized} on $[0,z_0]$ for some $z_0\in [0,\infty)$. It is straightforward to see that $\hat x(0)=O^{1/2}-1$ and
	\begin{align*}
		\frac{\partial F}{\partial x}(\hat x(0),0;\hat x) = -\frac{\partial Q}{\partial x}(\hat x(0))>0.
	\end{align*}
	For any $z\in (0,z_0]$, we have $Q(\hat x(z))>0$ and $\hat x(z)\in (0,O^{1/2})$ because $K(x,\bar x)>0$ for all $x\ge 0$ and $\bar x>0$, and because $F(\hat x(z),z;\hat x)=0$. Consequently,
	\begin{align}
		\frac{\partial F}{\partial x}(\hat x(z),z;\hat x) &= \int_0^z \frac{\partial K}{\partial x}(\hat x(z),\hat x(\bar z))\dd\bar z - \frac{\partial Q}{\partial x}(\hat x(z))\notag\\
		&= \int_0^z \frac{\partial K}{\partial x}(\hat x(z),\hat x(\bar z))\dd\bar z - \frac{\frac{\partial Q}{\partial x}(\hat x(z))}{Q(\hat x(z))}\int_0^z K(\hat x(z),\hat x(\bar z))\dd\bar z\notag\\
		&= \int_0^z \left[\frac{\frac{\partial K}{\partial x}(\hat x(z),\hat x(\bar z))}{K(\hat x(z),\hat x(\bar z))} - \frac{\frac{\partial Q}{\partial x}(\hat x(z))}{Q(\hat x(z))}\right]K(\hat x(z),\hat x(\bar z))\dd\bar z,\label{eq:pfofODEEq1}
	\end{align}
	where the second equality holds because $F(\hat x(z),z;\hat x)=0$. Straightforward calculation yields
	\begin{align*}
		\frac{\frac{\partial Q}{\partial x}(x)}{Q(x)} &= y_A\frac{\frac{\partial^2 R_F}{\partial d_A^2}(g,y_Ax)}{\frac{\partial R_F}{\partial x}(g,y_Ax)} = \frac{-2(1+x)^{-3}}{(1+x)^{-2}-O^{-1}},\quad x\in [0,O^{1/2}),\\
		\frac{\frac{\partial K}{\partial x}(x,\bar x)}{K(x,\bar x)} &= y_A\frac{\frac{\partial^2 V}{\partial d_A^2}(y_Ax,y_A\bar x)}{\frac{\partial V}{\partial x}(y_Ax,y_A\bar x)} = \frac{-2(1+x)^{-3}+2(1+\bar x+x)^{-3}}{(1+x)^{-2}-(1+\bar x+x)^{-2}},\quad x\ge 0,\;\bar x>0.
	\end{align*}
	Consequently,
	\begin{align}
		&\frac{\frac{\partial K}{\partial x}(x,\bar x)}{K(x,\bar x)} - \frac{\frac{\partial Q}{\partial x}(x)}{Q(x)} \notag \\
		&= \frac{2\bar x(1+x)^{-3}(1+\bar x+x)^{-3}\big(-1+O^{-1}\big((1+\bar x+x)^2+(1+x)^2+(1+x)(1+\bar x+x))\big)}{\big((1+x)^{-2}-O^{-1}\big)\big((1+x)^{-2}-(1+\bar x+x)^{-2}\big)}.\label{eq:ComparativeConcavity}
	\end{align}
	Because $O\le 3$, we have $\frac{\frac{\partial K}{\partial x}(x,\bar x)}{K(x,\bar x)} - \frac{\frac{\partial Q}{\partial x}(x)}{Q(x)}>0$ for all $x\in[0,O^{1/2})$ and $\bar x>0$. We then derive from \eqref{eq:pfofODEEq1} that
	$
		\frac{\partial F}{\partial x}(\hat x(z),z;\hat x)>0.
	$
	By the implicit function theorem, we conclude that $\hat x(z)$ is continuously differentiable in $z\in [0,z_0)$ and
	\begin{align}
		\frac{d\hat x}{dz}(z) &= -\left(\frac{\partial F}{\partial x}(\hat x(z),z;\hat x)\right)^{-1}\frac{\partial F}{\partial z}(\hat x(z),z;\hat x)= -\left(\frac{\partial F}{\partial x}(\hat x(z),z;\hat x)\right)^{-1}K(\hat x(z),\hat x(z))<0.\label{eq:ODESolutionDeri}
	\end{align}
	As a result, $\hat x$ is strictly decreasing on $[0,z_0]$.
	
	Next, we prove the existence and uniqueness of the solution. Recall that \eqref{eq:KeyODENormalized} has a unique solution $\hat x$ at $\{0\}$, where $\hat x(0)=O^{1/2}-1$. Suppose that \eqref{eq:KeyODENormalized} has a unique solution $\hat x$ on $[0,z_0]$ for some $z_0\in [0,\infty)$. We show below that $\hat x$ can be uniquely extended beyond $[0,z_0]$.
	
	We have shown that $\frac{\partial F}{\partial x}(\hat x(z_0),z_0;\hat x)>0$. As a result, there exists $\epsilon_0\in (0,\hat x(z_0))$ such that
	\begin{align}
		\frac{\partial F}{\partial x}(x,z_0;\hat x)\ge \epsilon_0,\quad \forall x\in[\hat x(z_0)-\epsilon_0,\hat x(z_0)].\label{eq:pfofODEEq2}
	\end{align}
	For any $\delta>0$, denote by ${\cal A}_\delta$ the set of continuous functions $y$ on $[z_0,z_0+\delta]$ with $y(z)\in [\hat x(z_0)-\epsilon_0,\hat x(z_0)]$. For every $y\in {\cal A}_\delta$, define
	$
		y_\delta(z):= \hat x(z)\mathbf 1_{z\in[0,z_0]}+y(z)\mathbf 1_{z\in (z_0,z_0+\delta]}.
	$
	Then, for all $z\in [z_0,z_0+\delta]$ and $x\in [\hat x(z_0)-\epsilon_0,\hat x(z_0)]$,
	\begin{align}
		F(x,z;y_\delta) &= \int_0^{z_0}K(x,\hat x(\bar z))\dd\bar z + \int_{z_0}^{z}K(x,y(\bar z))\dd\bar z - Q(x)\notag\\
		&= \int_0^{z_0}\big(K(x,\hat x(\bar z))-K(\hat x(z_0),\hat x(\bar z))\big)\dd\bar z - \big(Q(x)-Q(\hat x(z_0))\big)\notag\\
		&\quad + \int_{z_0}^{z}\big(K(x,y(\bar z))-K(\hat x(z_0),y(\bar z))\big)\dd\bar z + \int_{z_0}^{z}K(\hat x(z_0),y(\bar z))\dd\bar z,\label{eq:pfofODEEq3}
	\end{align}
	where the second equality holds because $F(\hat x(z_0),z_0;\hat x)=0$. Denote
	\begin{align*}
		M_1:=\sup_{x,\bar x\in [\hat x(z_0)-\epsilon_0,\hat x(z_0)]}\left|\frac{\partial K}{\partial x}(x,\bar x)\right|<\infty,\quad
		M_2:=\sup_{\bar x\in [\hat x(z_0)-\epsilon_0,\hat x(z_0)]}K(\hat x(z_0),\bar x)<\infty.
	\end{align*}
	Then, by \eqref{eq:pfofODEEq2}, \eqref{eq:pfofODEEq3}, and the mean value theorem, we derive, for all $z\in [z_0,z_0+\delta]$ and $x\in [\hat x(z_0)-\epsilon_0,\hat x(z_0)]$,
	\begin{align*}
		&\frac{\partial F}{\partial x}(x,z,y_\delta)= \frac{\partial F}{\partial x}(x,z_0;\hat x)+\int_{z_0}^z\frac{\partial K}{\partial x}(x,y(\bar z))\dd\bar z\ge \epsilon_0-M_1\delta, \\
		&F(x,z,y_\delta)\le (\epsilon_0-M_1\delta)(x-\hat x(z_0))+M_2\delta.
	\end{align*}
	Thus, we can find $\delta>0$ such that
	\begin{align}
		\frac{\partial F}{\partial x}(x,z,y_\delta)>0,\quad
		F(\hat x(z_0)-\epsilon_0,z,y_\delta)<0,\quad \forall z\in [z_0,z_0+\delta],\; x\in [\hat x(z_0)-\epsilon_0,\hat x(z_0)].\label{eq:pfofODEEq4}
	\end{align}
	By \eqref{eq:pfofODEEq3}, we have
	\begin{align*}
		F(\hat x(z_0),z,y_\delta)=\int_{z_0}^zK(\hat x(z_0),y(\bar z))\dd\bar z\ge 0,~ \forall z\in [z_0,z_0+\delta].
	\end{align*}
	Therefore, for any $y\in {\cal A}_\delta$ and $z\in [z_0,z_0+\delta]$, there exists a unique solution of $F(x,z,y_\delta)=0$ in $x\in [\hat x(z_0)-\epsilon_0,\hat x(z_0)]$, denoted by $\mathbb{T}y(z)$. Moreover, the implicit function theorem implies that $\mathbb{T}y$ is continuous on $[z_0,z_0+\delta]$. Hence, $\mathbb{T}$ is a mapping from ${\cal A}_\delta$ to itself.
	
	For any $y^i\in {\cal A}_\delta$, denote $x^i=\mathbb{T}y^i$, $i=1,2$. For any $z\in [z_0,z_0+\delta]$ with $x^1(z)\ge x^2(z)$, straightforward calculation yields
	\begin{align*}
		&0=F(x^1(z),z,y_\delta^1)-F(x^2(z),z,y_\delta^2)=F(x^1(z),z_0,\hat x)-F(x^2(z),z_0,\hat x)\\
		&\quad +\int_{z_0}^{z}\big(K(x^1,y^1(\bar z))-K(x^2,y^1(\bar z))\big)\dd\bar z
		+\int_{z_0}^z\big(K(x^2,y^1(\bar z))-K(x^2,y^2(\bar z))\big)\dd\bar z\\
		&\ge (\epsilon_0-M_1\delta)\big(x^1(z)-x^2(z)\big)-M_3\delta |y^1(z)-y^2(z)|,
	\end{align*}
	where
	$
		M_3:=\sup_{x,\bar x\in [\hat x(z_0)-\epsilon_0,\hat x(z_0)]}K(x,\bar x)<\infty,
	$
	and the inequality holds due to \eqref{eq:pfofODEEq2} and the mean value theorem. Similarly, for any $z\in [z_0,z_0+\delta]$ with $x^1(z)\le x^2(z)$, we have
	\begin{align*}
		0=F(x^1(z),z,y_\delta^1)-F(x^2(z),z,y_\delta^2)\le (\epsilon_0-M_1\delta)\big(x^1(z)-x^2(z)\big)+M_3\delta |y^1(z)-y^2(z)|.
	\end{align*}
	Consequently, choosing a sufficiently small $\delta>0$ such that \eqref{eq:pfofODEEq4} holds and $M_3\delta/(\epsilon_0-M_1\delta)\in (0,1)$, $\mathbb{T}$ is a contraction mapping on ${\cal A}_\delta$ endowed with the maximum norm. Thus, $\mathbb{T}$ has a unique fixed point $\hat y$. Consequently, we can extend $\hat x$ from $[0,z_0]$ to $(z_0,z_0+\delta]$ by setting $\hat x=\hat y$ on $(z_0,z_0+\delta]$, and $\hat x$ is a solution to \eqref{eq:KeyODENormalized} on $[0,z_0+\delta]$.
	
	Suppose there is another solution $\tilde x$ to \eqref{eq:KeyODENormalized} extended beyond $[0,z_0]$. Because the solution to \eqref{eq:KeyODENormalized} is unique on $[0,z_0]$ as assumed in the induction step, we conclude that $\tilde x(z_0)=\hat x(z_0)$. We have already proved that $\tilde x$ is strictly decreasing. Then, by the continuity of $\tilde x$, there exists $\tilde \delta\in (0,\delta)$ such that $\tilde x(z)\in [\hat x(z_0)-\epsilon_0,\hat x(z_0)]$ for all $z\in [z_0,z_0+\tilde \delta]$. Thus, $\tilde x(z)$ for $z\in [z_0,z_0+\tilde \delta]$ is a fixed point of $\mathbb{T}$ as defined previously on ${\cal A}_{\tilde \delta}$. Because $\mathbb{T}$ is a contraction mapping, its fixed point is unique, and therefore $\tilde x$ must equal $\hat x$ as constructed above on $[z_0,z_0+\tilde \delta]$. Hence, we prove that \eqref{eq:KeyODENormalized} admits a unique solution on $[0,z_0+\tilde \delta]$.
	
	Next, we show that the unique solution $\hat x$ to \eqref{eq:KeyODENormalized} goes to 0 in finite time. The set of $z\ge 0$ such that $\hat x(z)>0$ must take the form $[0,z_\infty)$ for some $z_\infty\in(0,\infty]$. Because $\frac{\partial F}{\partial x}(\hat x(z),z;\hat x)$ is continuous in $z\in \mathrm{dom}(\hat x)$ and because $\frac{\partial F}{\partial x}(\hat x(z),z;\hat x)>0$ for any $z\in[ z_0,z_\infty)$, we can find $z_0\in (0,z_\infty)$ and positive constants $k_1$ and $k_2$ such that $\frac{\partial F}{\partial x}(\hat x(z),z;\hat x)\in [k_1,k_2]$ for all $z\in[0,z_0]$. Because $\hat x$ is strictly decreasing, $\hat x(z)\le \hat x(z_0)<\hat x(0)=O^{1/2}-1$ for any $z\in[ z_0,z_\infty)$. Consequently, by \eqref{eq:pfofODEEq1} and \eqref{eq:ComparativeConcavity}, there exist positive constants $k_3$ and $k_4$ such that
	\begin{align*}
		k_3\int_0^zK(\hat x(z),\hat x(\bar z))\dd \bar z \le \frac{\partial F}{\partial x}(\hat x(z),z;\hat x)\le k_4\int_0^zK(\hat x(z),\hat x(\bar z))\dd\bar z,\quad \forall z\ge z_0.
	\end{align*}
	Because $\int_0^zK(\hat x(z),\hat x(\bar z))\dd\bar z=Q(\hat x(z))$ and because $0<\hat x(z)\le \hat x(z_0)<O^{1/2}-1$ for all $z\in [z_0,z_\infty)$, we conclude that $\frac{\partial F}{\partial x}(\hat x(z),z;\hat x)$ is bounded above and below uniformly in $z\in[z_0,z_\infty)$. Then, by \eqref{eq:ODESolutionDeri} and the formula for $K$, and recalling that $\hat x(z)\in (0,O^{1/2}-1]$, we conclude that there exist positive constants $C_1$ and $C_2$ such that
	\begin{align}
		-\frac{C_1}{\hat x(z)}\le \frac{d\hat x}{dz}(z)\le -\frac{C_2}{\hat x(z)}\le -\frac{C_2}{O^{1/2}-1},\quad \forall z\in [0,z_\infty).\label{eq:ODESolutionDeriBound}
	\end{align}
	Consequently,
	$
		\hat x(z)-\hat x(0)\le -\frac{C_2}{O^{1/2}-1}z,~ \forall z\in [0,z_\infty),
	$
	implying that $z_\infty$ must be finite and $\lim_{z\uparrow z_\infty}\hat x(z)=0$.
	
	Next, by \eqref{eq:ODESolutionDeriBound}, we conclude that
	\begin{align*}
		\int_0^{z_\infty} (\hat x(\bar z))^{-2}\dd\bar z&=\lim_{z\uparrow z_\infty}\int_0^{z} (\hat x(\bar z))^{-2}\dd\bar z\ge -C_1^{-1}\lim_{z\uparrow z_\infty}\int_0^{z} (\hat x(\bar z))^{-1}\frac{\dd\hat x}{\dd z}(\bar z)\dd\bar z\\
		&= -C_1^{-1}\lim_{z\uparrow z_\infty}\left[\ln \hat x(z)-\ln \hat x(0)\right]=+\infty.
	\end{align*}
	It follows that there exists a unique $\hat z$ such that \eqref{eq:Z0ell} holds. \halmos
\end{pfof}

\begin{pfof}{Theorem \ref{thm:Main}}
	Recall the response function \eqref{eq:ResponseFunction2}. For every $g\in[\hat g_L,g_H]$, because $D_A^*(g)$ is the maximizer of $h(g,d_A)$ in $d_A\in[0,\hat D_A]$ and because $D_A^*(g)>0$ as shown in Proposition \ref{le:RandomizedGasDensity}-(i), we must have $\frac{\partial h}{\partial d_A}(g,D_A^*(g))\ge 0$, with equality when $D_A^*(g)<\hat D_A$. Moreover,
	\begin{align*}
		\frac{\partial h}{\partial d_A}(g,\hat D_A) &= \frac{\partial R_F}{\partial d_A}(g,\hat D_A)-\alpha^* \int_{g}^{\hat g_H}\frac{\partial V^\infty}{\partial d_A}(\hat D_A,D_A^*(\bar g))\Phi^*(\dd\bar g)\le 0,
	\end{align*}
	where the inequality holds because $\frac{\partial R_F}{\partial d_A}(g,\hat D_A)=0$ and
	\begin{align*}
		\frac{\partial V^\infty}{\partial d_A}(\hat D_A,\bar d_A) = \frac{\partial R_F}{\partial d_A}(g,\hat D_A)-\frac{\partial R_S}{\partial d_A}(g,\hat D_A,\bar d_A)\ge 0.
	\end{align*}
	Consequently, $\frac{\partial h}{\partial d_A}(g,D_A^*(g))=0$ for all $g\in[\hat g_L,g_H]$.
	
	By Proposition \ref{le:RandomizedGasDensity}-(iii), $h(g,D_A^*(g))$ is constant on $[\hat g_L,g_H]$. Differentiating $h(g,D_A^*(g))$ with respect to $g$, applying the chain rule, and recalling that $\frac{\partial h}{\partial d_A}(g,D_A^*(g))=0$, we obtain
	\begin{align*}
		\frac{\partial R_F}{\partial g}(g,D_A^*(g))+\alpha^* V^\infty(D_A^*(g),D_A^*(g))\phi^*(g)=0,\quad g\in[\hat g_L,g_H],
	\end{align*}
	where $\phi^*$ denotes the density of $\Phi^*$. Because $\frac{\partial R_F}{\partial g}(g,D_A^*(g))=-1$, we immediately derive
	\begin{align*}
		\phi^*(g)=\frac{1}{\alpha^* V^\infty(D_A^*(g),D_A^*(g))},\quad g\in[\hat g_L,g_H].
	\end{align*}
	Substituting this into $\frac{\partial h}{\partial d_A}(g,D_A^*(g))=0$, we obtain
	\begin{align}
		\frac{\partial R_F}{\partial d_A}(g,D_A^*(g))-\int_{g}^{g_H}\frac{\partial V^\infty}{\partial d_A}(D_A^*(g),D_A^*(\bar g))\frac{1}{V(D_A^*(\bar g),D_A^*(\bar g))}\dd\bar g=0,\quad g\in[\hat g_L,g_H].\label{eq:DAEquilibriuEquation}
	\end{align}
	It is straightforward to see that \eqref{eq:DAEquilibriuEquation} holds if and only if
	$\hat x(z):=x(L_B z)=D_A^*(g_H-L_B z)/y_A$ is a solution to \eqref{eq:KeyODENormalized}.
	Moreover, since $\phi^*$ is a density function on $[\hat g_L,g_H]$, we have
	\begin{align}
		1=\int_{\hat g_L}^{g_H}\phi^*(g)\dd g=\int_{\hat g_L}^{g_H}\frac{1}{\alpha^* V^\infty(D_A^*(g),D_A^*(g))}\dd g=\frac{1}{\alpha^*}\int_0^{(g_H-\hat g_L)/L_B}v(\hat x(\bar z))\dd\bar z,\label{eq:DensityIntegralCond}
	\end{align}
	where the last equality holds because $\hat x(z)=D_A^*(g_H-L_B z)/y_A$ and by the definition of $v$.
	The values of $\alpha^*$ and $g_H$ are determined by condition \eqref{eq:DensityIntegralCond}.
	
	It is straightforward to see that
	\begin{align}
		h(g,D_A^*(g))=h(g_H,D_A^*(g_H))=R_F(g_H,\hat D_A)-\alpha^*\int_{g_H}^{\hat g_H}V^\infty(d_A,D_A^*(\bar g))\Phi^*(\dd\bar g)\notag\\
		=R_F(g_H,\hat D_A)=\hat g_H-g_H,\quad \forall g\in[\hat g_L,g_H],\label{eq:EquiResponseFun}
	\end{align}
	where the first equality follows from Proposition \ref{le:RandomizedGasDensity}-(iii) and the last equality from $R_F(\hat g_H,\hat D_A)=0$.
	
	We first consider the case $L_B\hat z\le \hat g_H-\hat g_L$. We claim that $\alpha^*=1$. Suppose, on the contrary, that $\alpha^*<1$. Then, by \eqref{eq:DensityIntegralCond}, the definition of $\hat z$, and the positivity of $v$, we immediately have $(g_H-\hat g_L)/L_B<\hat z$. Combining this with the assumption $L_B\hat z\le \hat g_H-\hat g_L$, we conclude that $g_H<\hat g_H$. Consequently, from \eqref{eq:EquiResponseFun} we derive
	\begin{align}
		h(g,D_A^*(g))>0,\quad \forall g\in[\hat g_L,g_H],\label{eq:pfofmainTheoremEq1}
	\end{align}
	where the last equality holds due to the definition of $\hat g_H$ and $\hat D_A$. This contradicts the assumption that $\alpha^*<1$; see Proposition \ref{le:RandomizedGasDensity}-(iii). Therefore, we must have $\alpha^*=1$.
	
	Next, consider the case $L_B\hat z>\hat g_H-\hat g_L$. We claim that $g_H=\hat g_H$. Suppose, on the contrary, that $g_H<\hat g_H$. Then, we have \eqref{eq:pfofmainTheoremEq1}. On the other hand, by \eqref{eq:DensityIntegralCond} and the assumption $L_B\hat z>\hat g_H-\hat g_L$, we conclude that $\alpha^*<1$. This contradicts \eqref{eq:pfofmainTheoremEq1}; see Proposition \ref{le:RandomizedGasDensity}-(iii). Therefore, we must have $g_H=\hat g_H$. Consequently, we derive \eqref{eq:EquiAlpha} from \eqref{eq:DensityIntegralCond}.
	
	Finally, by \eqref{eq:EquiResponseFun}, we obtain the arbitrageur's expected profit as
	$
		\alpha^*\int h(g,D_A^*(g))\Phi^*(\dd g)=\alpha^*(\hat g_H-g_H).
	$
	This completes the proof. \halmos
\end{pfof}

\begin{pfof}{Proposition \ref{prop:EquilibriumMonotonocity}}
	Because $D_A^*(g)=y_A\hat x\big((g_H-g)/L_B\big)$ and $\hat x(z)$ is strictly decreasing in $z$, we immediately conclude that $D_A^*(g)$ is strictly increasing in $g\in[\hat g_L,g_H]$. Moreover, $D_A^*(g_H)=y_A\hat x(0)=\hat D_A$.
	
	Straightforward calculation shows that
	\begin{align*}
		\phi^*(g) = \frac{1}{\alpha^* L_B} v\big(\hat x\big((g_H-g)/L_B\big)\big),\quad g\in [\hat g_L,g_H],
	\end{align*}
	where $v$ is given by \eqref{eq:FMAInvNormalized}. We have shown in the proof of Lemma \ref{le:ODE} that $v(x)$ is strictly decreasing in $x$. Because $\hat x(z)$ is strictly decreasing in $z$, we conclude that $\phi^*(g)$ is strictly decreasing in $g\in[\hat g_L,g_H]$. \halmos
\end{pfof}

\begin{pfof}{Proposition \ref{prop:BaseGasFee}}
	First, it is straightforward to see that $\alpha^*=1$ when $\hat g_L\le \hat g_H-L_B\hat z$ and $\alpha^*<1$ when $\hat g_L>\hat g_H-L_B\hat z$. In the latter case, $\alpha^*$ is given by \eqref{eq:EquiAlpha}, which is strictly decreasing in $\hat g_H$.
	
	Next, the arbitrageur's expected profit is 0 when $\hat g_L\ge \hat g_H-L_B\hat z$ and positive when $\hat g_L<\hat g_H-L_B\hat z$. In the latter case, the expected profit is $\hat g_H-g_H=\hat g_H-\hat g_L-L_B\hat z$, which is strictly decreasing in $\hat g_L$.
	
	Next, we prove (iii). When $\hat g_L\le \hat g_H-L_B\hat z$, we have $\alpha^*=1$ and $g_H=\hat g_L+L_B\hat z$. Consequently,
	\begin{align*}
		F_{\tilde g}(g)&=\left(\int_{\max(g,\hat g_L)}^{g_H}\phi^*(\bar g)\dd\bar g\right)\mathbf 1_{g< g_H}=\frac{1}{L_B}\left(\int_{\max(g,\hat g_L)}^{g_H} v\big(\hat x\big((g_H-\bar g)/L_B\big)\big)\dd\bar g\right)\mathbf 1_{g< g_H}\\
		&=\left(\int_0^{\min((g_H-g)/L_B,(g_H-\hat g_L)/L_B)} v(\hat x(\bar z))\dd\bar z\right)\mathbf 1_{g< g_H}\\
		&=\left(\int_0^{\min((\hat g_L+L_B\hat z-g)/L_B,\hat z)} v(\hat x(\bar z))\dd\bar z\right)\mathbf 1_{g< \hat g_L+L_B\hat z},\quad g>0.
	\end{align*}
	It is then straightforward to see that $F_{\tilde g}(g)$ is increasing in $\hat g_L$. When $\hat g_L>\hat g_H-L_B\hat z$, we have $\alpha^*<1$ and $g_H=\hat g_H$. Consequently, $F_{\tilde g}(g)=\alpha^*$ for $g\in [0,\hat g_L)$, $F_{\tilde g}(g)=0$ for $g\ge \hat g_H$, and
	\begin{align*}
		F_{\tilde g}(g)&=\alpha^*\mathbf 1_{g\in(0,\hat g_L)}+\alpha^*\left(\int_g^{\hat g_H}\phi^*(\bar g)\dd\bar g\right)\mathbf 1_{g\in [\hat g_L,g_H)}\\
		&=\alpha^*\mathbf 1_{g\in(0,\hat g_L)}+\frac{1}{L_B}\left(\int_g^{\hat g_H} v\big(\hat x\big((\hat g_H-\bar g)/L_B\big)\big)\dd\bar g\right)\mathbf 1_{g\in [\hat g_L,\hat g_H)}\\
		&=\min\left(\alpha^*,\int_0^{(\hat g_H-g)/L_B} v(\hat x(\bar z))\dd\bar z\right)\mathbf 1_{g\le \hat g_H},\quad g>0.
	\end{align*}
	Because $\alpha^*$ is decreasing in $\hat g_L$, we conclude that $F_{\tilde g}(g)$ is decreasing in $\hat g_L$ for all $g$.
	
	Finally, when $\hat g_L\le \hat g_H-L_B\hat z$, we have $\alpha^*=1$ and $g_H=\hat g_L+L_B\hat z$. Consequently,
	\begin{align*}
		F_{\tilde D/y_A}(d)=F_{\tilde g}\left(\hat g_L+L_B\hat z-L_B\hat x^{-1}(d)\right)\mathbf 1_{d<\hat x(0)}=\left(\int_0^{\max(\hat x^{-1}(d),0)} v(\hat x(\bar z))\dd\bar z\right)\mathbf 1_{d<\hat x(0)},~d>0
	\end{align*}
	which is independent of $\hat g_L$. When $\hat g_L>\hat g_H-L_B\hat z$, we have $\alpha^*<1$ and $g_H=\hat g_H$. Consequently,
	\begin{align*}
		F_{\tilde D/y_A}(d)=F_{\tilde g}\left(\hat g_H-L_B\hat x^{-1}(d)\right)\mathbf 1_{d<\hat x(0)}=\min\left(\alpha^*,\int_0^{\hat x^{-1}(d)} v(\hat x(\bar z))\dd\bar z\right)\mathbf 1_{d<\hat x(0)},~d>0.
	\end{align*}
	Because $\alpha^*$ is decreasing in $\hat g_L$, we conclude that $F_{\tilde D/y_A}(d)$ is decreasing in $\hat g_L$. \halmos
\end{pfof}

\begin{pfof}{Proposition \ref{prop:Liquidity}}
	First, it is straightforward to see that $\alpha^*=1$ when $\hat g_L\le \big(O^{-1}(O^{1/2}-1)^2-\hat z\big)L_B$ and $\alpha^*<1$ when $\hat g_L>\big(O^{-1}(O^{1/2}-1)^2-\hat z\big)L_B$. Because $O^{-1}(O^{1/2}-1)^2>\hat z$, we conclude that $\alpha^*=1$ when $L_B\ge \hat g_L/\big(O^{-1}(O^{1/2}-1)^2-\hat z\big)$ and $\alpha^*<1$ when $L_B<\hat g_L/\big(O^{-1}(O^{1/2}-1)^2-\hat z\big)$. In the latter case, $\alpha^*$ is given by \eqref{eq:EquiAlpha}, which is strictly increasing in $L_B$.
	
	Next, when $L_B\hat z \ge \hat g_H-\hat g_L$, i.e., when $L_B\le \hat g_L/\big(O^{-1}(O^{1/2}-1)^2-\hat z\big)$, the expected profit is 0. When $L_B\hat z< \hat g_H-\hat g_L$, i.e., when $L_B> \hat g_L/\big(O^{-1}(O^{1/2}-1)^2-\hat z\big)$, the expected profit is
	\begin{align*}
		\hat g_H-g_H &= L_BO^{-1}(O^{1/2}-1)^2 - (\hat g_L+L_B\hat z) = \big(O^{-1}(O^{1/2}-1)^2-\hat z\big)L_B-\hat g_L,
	\end{align*}
	which is positive and strictly increasing in $L_B$.
	
	When $L_B\ge \hat g_L/\big(O^{-1}(O^{1/2}-1)^2-\hat z\big)$, we have $\alpha^*=1$ and $g_H=\hat g_L+L_B\hat z$. Consequently,
	\begin{align*}
		F_{\tilde g}(g) &= \left(\int_{\max(g,\hat g_L)}^{g_H}\phi^*(\bar g)\dd\bar g\right)\mathbf 1_{g< g_H}= \frac{1}{L_B}\left(\int_{\max(g,\hat g_L)}^{g_H} v\big(\hat x\big((g_H-\bar g)/L_B\big)\big)\dd\bar g\right)\mathbf 1_{g< g_H}\\
		&= \left(\int_0^{\min((g_H-g)/L_B,(g_H-\hat g_L)/L_B)} v(\hat x(\bar z))\dd\bar z\right)\mathbf 1_{g< g_H}\\
		&= \mathbf 1_{g\le \hat g_L}+\left(\int_0^{(\hat g_L-g)/L_B+\hat z} v(\hat x(\bar z))\dd\bar z\right)\mathbf 1_{\hat g_L<g< \hat g_L+L_B\hat z},\quad g>0.
	\end{align*}
	It is then straightforward to see that $F_{\tilde g}(g)$ is increasing in $\hat g_L$. When $L_B< \hat g_L/\big(O^{-1}(O^{1/2}-1)^2-\hat z\big)$, we have $\alpha^*<1$ and $g_H=\hat g_H=O^{-1}(O^{1/2}-1)^2L_B$. Consequently, $F_{\tilde g}(g)=\alpha^*$ for $g\in [0,\hat g_L)$, $F_{\tilde g}(g)=0$ for $g\ge O^{-1}(O^{1/2}-1)^2L_B$, and
	\begin{align*}
		F_{\tilde g}(g) &= \alpha^*\mathbf 1_{g\in(0,\hat g_L)} + \alpha^*\left(\int_g^{O^{-1}(O^{1/2}-1)^2L_B}\phi^*(\bar g)\dd\bar g\right)\mathbf 1_{g\in [\hat g_L,O^{-1}(O^{1/2}-1)^2L_B)}\\
		&= \alpha^*\mathbf 1_{g\in(0,\hat g_L)} + \frac{1}{L_B}\left(\int_g^{O^{-1}(O^{1/2}-1)^2L_B} v\big(\hat x\big((O^{-1}(O^{1/2}-1)^2L_B-\bar g)/L_B\big)\big)\dd\bar g\right)\mathbf 1_{g\in [\hat g_L,O^{-1}(O^{1/2}-1)^2L_B)}\\
		&= \min\left(\alpha^*,\int_0^{O^{-1}(O^{1/2}-1)^2-g/L_B} v(\hat x(\bar z))\dd\bar z\right)\mathbf 1_{g< O^{-1}(O^{1/2}-1)^2L_B},\quad g>0.
	\end{align*}
	Because $\alpha^*$ is increasing in $L_B$, we conclude that $F_{\tilde g}(g)$ is increasing in $L_B$ for every $g$.
	
	Finally, when $L_B\ge \hat g_L/\big(O^{-1}(O^{1/2}-1)^2-\hat z\big)$, we have $\alpha^*=1$ and $g_H=\hat g_L+L_B\hat z$. In this case,
	\begin{align*}
		F_{\tilde D/y_A}(d) &= F_{\tilde g}\left(g_H-L_B\hat x^{-1}(d)\right)\mathbf 1_{d<\hat x(0)} = \left(\int_0^{\min(\hat x^{-1}(d),\hat z)} v(\hat x(\bar z))\dd\bar z\right)\mathbf 1_{d<\hat x(0)},~ d>0,
	\end{align*}
	which is independent of $\hat g_L$. When $L_B< \hat g_L/\big(O^{-1}(O^{1/2}-1)^2-\hat z\big)$, we have $\alpha^*<1$ and $g_H=\hat g_H$. Consequently,
	\begin{align*}
		F_{\tilde D/y_A}(d) &= F_{\tilde g}\left(\hat g_H-L_B\hat x^{-1}(d)\right)\mathbf 1_{d<\hat x(0)} = \min\left(\alpha^*,\int_0^{\hat x^{-1}(d)} v(\hat x(\bar z))\dd\bar z\right)\mathbf 1_{d<\hat x(0)},~ d>0.
	\end{align*}
	Because $\alpha^*$ is increasing in $L_B$, we conclude that $F_{\tilde D/y_A}(d)$ is increasing in $L_B$. \halmos
\end{pfof}

\begin{pfof}{Proposition \ref{prop:Slippage0}}
	For contradiction, suppose $\alpha^*=1$. Then, from \eqref{eq:ResponseFunction2Slippage0} we derive
	\begin{align*}
		h^0(\hat g_L,D_A^*(\hat g_L)) = R_F(\hat g_L,D_A^*(\hat g_L)) - V^0(\hat g_L,D_A^*(\hat g_L)) = R_S^0(\hat g_L,D_A^*(\hat g_L)) \le -r\hat g_L < 0,
	\end{align*}
	where $R_S^0(g,d_A)$ denotes the second mover's payoff \eqref{eq:SecondMoverReward_0}. This contradicts Proposition \ref{le:RandomizedGasDensity}-(iii). Thus, we must have $\alpha^*\in(0,1)$.
	
	On the other hand, from \eqref{eq:ResponseFunction2Slippage0} we obtain $h^0(g_H,d_A)=R_F(g_H,d_A)$. By Proposition \ref{prop:propertyEq}-(ii), we conclude
	\begin{align*}
		D_A^*(g_H) &= \argmax_{d_A\in[0,\hat D_A]} h^0(g_H,d_A) = \argmax_{d_A\in[0,\hat D_A]} R_F(g_H,d_A) = \hat D_A.
	\end{align*}
	Consequently,
	$
		h^0(g_H,D_A^*(g_H)) = R_F(g_H,\hat D_A) = \hat g_H-g_H.
	$
	By Proposition \ref{le:RandomizedGasDensity}-(iii) and recalling that $\alpha^*<1$, we must have $\hat g_H-g_H=0$, and the expected profit is zero. \halmos
\end{pfof}

\begin{pfof}{Lemma \ref{le:ODESlippage0}}
	We first prove (i). In this case, $K^0_0=Q$. From \eqref{eq:KeyODESlippage0} it is straightforward to see that $\hat x(0)=O^{1/2}-1$ and $\hat \theta(0)=(r\hat g_H/L_B)^{-1}>0$. For any $z>0$ such that $\hat \theta$ remains positive on $[0,z)$, because $K^0_0=Q$ and $\int_0^z \hat \theta(\bar z)\dd\bar z>0$, we immediately conclude $Q(\hat x(z))=0$, which implies $\hat x(z)=O^{1/2}-1$. Consequently, $\hat \theta$ satisfies
	\begin{align}
		\hat \theta(z) = \frac{1-(1-r)\int_0^{z}\hat \theta(\bar z)\dd\bar z}{\hat g_H/L_B-(1-r)(\hat g_H/L_B-z)}.\label{eq:ODESolutionSlippage0Gamma0phiInt}
	\end{align}
	
	When $r=1$, from \eqref{eq:ODESolutionSlippage0Gamma0phiInt} we conclude $\hat\theta^*(z)=(\hat g_H/L_B)^{-1},\;z\ge 0$. When $r<1$, straightforward rearrangement yields the differential form of \eqref{eq:ODESolutionSlippage0Gamma0phiInt}:
	\begin{align*}
		\hat \theta'(z) = -\frac{2(1-r)\hat \theta(z)}{r\hat g_H/L_B+(1-r)z},\quad \hat \theta(0)=(r\hat g_H/L_B)^{-1}.
	\end{align*}
	Solving this gives $\hat \theta$ as in \eqref{eq:ODESolutionSlippage0Gamma0phi}. Moreover, straightforward calculation yields \eqref{eq:ODESolutionSlippage0Gamma0phiIntSol}.
	
	Next, we prove (ii). The proof is similar to that of Lemma \ref{le:ODE}. For every fixed $T>0$ and $y\in\mathcal{C}([0,T])$, it is straightforward to see that the equation
	\begin{align}
		\hat \theta(z) = \frac{1-(1-r)\int_0^{z}\hat \theta(\bar z)\dd\bar z}{K^0_1(z,y(z))}\label{eq:Slippage0ODEthetay}
	\end{align}
	admits a unique continuous solution on $[0,T]$. Denote
	\begin{align*}
		F(x,z;y):=K_0^0(x)\int_0^z \hat \theta(\bar z)\dd\bar z-Q(x),
	\end{align*}
	where $\hat \theta$ is given by \eqref{eq:Slippage0ODEthetay}.
	
	It is straightforward to see that \eqref{eq:KeyODESlippage0} admits a unique solution at $\{0\}$ with
	\begin{align*}
		\hat x(0)=O^{1/2}-1,\quad \hat \theta(0)=\frac{1}{K_1^0(0,\hat x(0))}=\left[r\hat g_H/L_B+\Gamma\big(y_A(O^{1/2}-1)\big)/L_B\right]^{-1}>0.
	\end{align*}
	Moreover,
	$
		\frac{\partial F}{\partial x}(\hat x(0),0;\hat x)=-\frac{dQ}{dx}(\hat x(0))>0.
	$
	Then,
	\begin{align*}
		\frac{\partial F}{\partial x}(\hat x(z_0),z_0;\hat x) &= \frac{dK_0^0}{dx}(\hat x(z_0))\int_0^{z_0}\hat \theta(\bar z)\dd\bar z-\frac{dQ}{dx}(\hat x(z_0))=Q(\hat x(z_0))\left(\frac{\frac{dK_0^0}{dx}(\hat x(z_0))}{K_0^0(\hat x(z_0))}-\frac{\frac{dQ}{dx}(\hat x(z_0))}{Q(\hat x(z_0))}\right)>0,
	\end{align*}
	where the second equality holds because $F(\hat x(z_0),z_0;\hat x(z_0))=0$, and the inequality holds due to \eqref{eq:K0FunSlippage0}, the observation that
	\begin{align*}
		\frac{Q'(x)}{Q(x)}<0\le \frac{y_A\Gamma''(x)}{\Gamma'(x)},\quad \forall x\ge 0,
	\end{align*}
	and the assumption $\hat x(z_0)<O^{1/2}-1$.
	
	Now, suppose \eqref{eq:KeyODESlippage0} has a unique solution $(\hat x,\hat \theta)$ on $[0,z_0]$ for some $z_0\ge 0$, such that $\hat x\in\mathcal{C}([0,z_0])$ and satisfies $\frac{\partial F}{\partial x}(\hat x(z_0),z_0;\hat x)>0$, and $\hat \theta$ is positive on $[0,z_0]$. Following the same proof as for Lemma \ref{le:ODE}, we can extend the solution to $[0,z_0+\delta]$ for small $\delta>0$, with $\hat x\in\mathcal{C}([0,z_0+\delta])$ satisfying $\hat x(z)<\hat x(z_0)$ for $z\in(z_0,z_0+\delta]$, and with $\hat \theta$ positive on $[0,z_0+\delta]$. Therefore, we can derive a solution $(\hat x,\hat \theta)$ to \eqref{eq:KeyODESlippage0} up to
	\begin{align}
		\bar z_\infty:=\sup\{z\ge 0:\forall\ 0<u<z,\ \hat x(u)>\varphi((1-r)(\hat g_H/L_B-u)),\ \hat \theta(u)>0\},\label{eq:Slippage0Explosion2}
	\end{align}
	and $\hat x$ must be strictly decreasing. Again, following the same proof as for Lemma \ref{le:ODE} shows that the solution is unique.
	
	By definition, $K_0^0(x)=\frac{\partial K_1^0(z,x)}{\partial x}(z,x)$.
	Straightforward calculation then yields the differential form of \eqref{eq:KeyODESlippage0}:
	\begin{align*}
		\hat \theta'(z) = -\frac{\big(2(1-r)+K_0^0(\hat x(z))\hat x'(z)\big)\hat \theta(z)}{K_1^0(z,\hat x(z))},\quad z\ge 0,\quad
		\hat \theta(0)=\left[r\hat g_H/L_B+\Gamma\big(y_A(O^{1/2}-1)\big)/L_B\right]^{-1}.
	\end{align*}
	As a result, for any interval $[0,z_0]$ with $\hat x(z)>\varphi((1-r)(\hat g_H/L_B-z))$ for all $z\in[0,z_0]$,
	since $K_0^0(x)$ is bounded on $[0,O^{1/2}-1]$ and $K_1^0(z,\hat x(z))$ is bounded below by a positive number on $[0,z_0]$,
	we conclude that $\hat \theta$ uniquely exists and is positive on $[0,z_0]$.
	Thus, $\bar z_\infty$ as given by \eqref{eq:Slippage0Explosion2} must coincide with $z_\infty$ as defined in \eqref{eq:Slippage0Explosion}.\halmos
\end{pfof}

\begin{pfof}{Theorem \ref{thm:MainSlippage0}}
	By Proposition \ref{prop:Slippage0}, the subsequent discussion, and the discussion in Appendix \ref{subse:KeyODEEll0},
	$(\alpha^*,\Phi^*,D_A^*)\in{\cal A}$ is a mixed symmetric equilibrium strategy if and only if
	(i) $g_H=\hat g_H$,
	(ii) $(\hat x,\hat \theta)$ with $\hat x(z):=D_A^*(\hat g_H-L_B z)/y_A$ and $\hat \theta(z):=\alpha^*L_B\phi^*(\hat g_H-L_B z)$ is a solution to \eqref{eq:KeyODESlippage0} on $[0,(\hat g_H-\hat g_L)/L_B]$, with $\hat x$, $\hat \theta$, and $K_1^0(z,\hat x(z))$ positive on $[0,(\hat g_H-\hat g_L)/L_B]$, and
	(iii)
	\begin{align*}
		\int_0^{(\hat g_H-\hat g_L)/L_B}\hat\theta(z)\dd z
		&= \alpha^*\int_0^{(\hat g_H-\hat g_L)/L_B}L_B\phi^*(\hat g_H-L_B z)\dd z = \alpha^*\int_{\hat g_L}^{\hat g_H}\phi^*(g)\dd g = \alpha^*\in(0,1),
	\end{align*}
	where the last equality holds because $\int_{\hat g_L}^{\hat g_H}\phi^*(g)\dd g=1$.
	
	We first prove (i). In this case, by Lemma \ref{le:ODESlippage0}-(i), \eqref{eq:KeyODESlippage0} admits a unique solution $(\hat x,\hat \theta)$ on $[0,\infty)$, given by \eqref{eq:ODESolutionSlippage0Gamma0phi}, and $\hat x(z)$, $\hat \theta(z)$, and $K_1^0(z,\hat x(z))$ are positive for $z\in[0,\infty)$. Moreover, by \eqref{eq:ODESolutionSlippage0Gamma0phiIntSol}, we immediately conclude that $\int_0^{(\hat g_H-\hat g_L)/L_B}\hat\theta(z)\dd z\in(0,1)$. Therefore, the mixed symmetric equilibrium strategy uniquely exists and is given by \eqref{eq:Slippage0TradingProb}--\eqref{eq:Slippage0TradingAmount}.
	
	Next, we prove (ii). In this case, by Lemma \ref{le:ODESlippage0}-(ii), \eqref{eq:KeyODESlippage0} admits a unique solution $(\hat x,\hat \theta)$ on $[0,\infty)$, and $\hat x(z)$, $\hat \theta(z)$, and $K_1^0(z,\hat x(z))$ are positive for $z\in[0,z_\infty)$. Because $z_\infty\ge (1-O^{-1/2})^2=\hat g_H/L_B$, we conclude that $(\hat x,\hat \theta)$ exist on $[0,(\hat g_H-\hat g_L)/L_B]$, and $\hat x$, $\hat \theta$, and $K_1^0(z,\hat x(z))$ are positive on this interval. Moreover, because $z_\infty\ge \hat g_H/L_B>(\hat g_H-\hat g_L)/L_B$ and by the definition of $z_\infty$, we conclude that
	\begin{align*}
		\hat x\big((\hat g_H-\hat g_L)/L_B\big)
		> \varphi\Big((1-r)\big(\hat g_H/L_B-(\hat g_H-\hat g_L)/L_B\big)\Big)
		= \varphi\big((1-r)\hat g_L/L_B\big)\ge \varphi(0)=0.
	\end{align*}
	Consequently,
	\begin{align*}
		K_0^0\Big(\hat x\big((\hat g_H-\hat g_L)/L_B\big)\Big)
		&= Q\Big(\hat x\big((\hat g_H-\hat g_L)/L_B\big)\Big)
		+ \frac{y_A}{L_B}\Gamma'\Big(y_A\hat x\big((\hat g_H-\hat g_L)/L_B\big)\Big)> Q\Big(\hat x\big((\hat g_H-\hat g_L)/L_B\big)\Big),
	\end{align*}
	where the inequality holds because $\Gamma'(d_A)>0$ for every $d_A>0$. We then derive from \eqref{eq:KeyODESlippage0} that $\int_0^{(\hat g_H-\hat g_L)/L_B}\hat\theta(z)\dd z\in(0,1)$. Therefore, the mixed symmetric equilibrium strategy uniquely exists and is given by \eqref{eq:Slippage0rTradingProb}--\eqref{eq:Slippage0rTradingAmount}. \halmos
\end{pfof}

\begin{pfof}{Proposition \ref{prop:CSNoSlippageO}}
	It is straightforward to see from \eqref{eq:Slippage0TradingProb} that $\alpha^*$ is strictly increasing in $O$. It is also straightforward to verify \eqref{eq:TradingProbInfOSetting0}.
	
	Recall that $\hat \theta(z)=\alpha^*L_B\phi^*(\hat g_H-L_Bz)$, where $\hat \theta$ is given in Lemma \ref{le:ODESlippage0}-(i). Then, for $r\le 1, g\in[\hat g_L,\hat g_H)$,
	\begin{align*}
		F_{\tilde g}(g) &= \alpha^*\int_g^{\hat g_H}\phi^*(\bar g)\dd\bar g
		= \int_0^{(\hat g_H-g)/L_B}\hat \theta(\bar z)\dd\bar z= (\hat g_H/L_B-g/L_B)\big(\hat g_H/L_B-(1-r)g/L_B\big)^{-1}.
	\end{align*}
	It is straightforward to see that $F_{\tilde g}(g)$ is strictly increasing in $\hat g_H/L_B=(1-O^{-1/2})^2$, so $F_{\tilde g}(g)$ is strictly increasing in $O$ for $g\in[\hat g_L,\hat g_H)$. For $g\in[0,\hat g_L)$, $F_{\tilde g}(g)=\alpha^*$, so by part (i) of the lemma, we conclude that $F_{\tilde g}(g)$ is strictly increasing in $O$. Finally, note that $\hat g_H=(1-O^{-1/2})^2L_B$ is strictly increasing in $O$ and that $F_{\tilde g}(g)=0$ for $g\ge \hat g_H$. The proof of part (ii) is then complete.
	
	Finally, because $D_A^*(g)\equiv \hat D$, we derive
	$
		F_{\tilde D/y_A}(d)=\alpha^*,~ \forall d\in[0,\hat D).
	$
	Moreover, $\hat D$ is strictly increasing in $O$ and $F_{\tilde D/y_A}(d)=0$ for $d\ge \hat D$. Part (iii) of the lemma then follows immediately from part (i). \halmos
\end{pfof}

\begin{pfof}{Propositions \ref{prop:CSgLSlippageO} and \ref{prop:CSLBSlippageO}}
	The proof is similar to that of Proposition \ref{prop:CSNoSlippageO}. \halmos
\end{pfof}

\begin{pfof}{Proposition \ref{prop:propertyEqChooseSlippage}}
	The proof is similar to that of Proposition \ref{prop:propertyEq}. \halmos
\end{pfof}

\begin{pfof}{Proposition \ref{le:ChooseRandomizedGasDensity}}
	Denote by $\Pi(\alpha,\Phi,\rho,D_A^{0},D_A^{\infty})$ the arbitrageur's payoff under strategy $(\alpha,\Phi,\rho,D_A^{0},D_A^{\infty})$ when the competitor adopts $(\alpha^*,\Phi^*,\rho^*,D_A^{0,*},D_A^{\infty,*})$. Then, it is straightforward to see that
	\begin{align}
		&\Pi(\alpha,\Phi,\rho,D_A^{0},D_A^{\infty};\alpha^*,\Phi^*,\rho^*,D_A^{0,*},D_A^{\infty,*})\notag\\
		&= \alpha \int_{[\hat g_L,\hat g_H]}\big[\rho(g)h^0(g,D_A^0(g))+(1-\rho(g))h^\infty(g,D_A^\infty(g))\big]\Phi^*(\dd g),\label{eq:ExpectedProfitChoice2}
	\end{align}
	where $h^\ell$ is given by \eqref{eq:ResponseFunctionChoice} and can be reformulated as
	\begin{align}
		h^\ell(g,d_A) &= (1-\alpha^*)R_F(g,d_A)+\alpha^*\bigg[\int R_F(g,d_A)\mathbf 1_{g>\bar g}\Phi^*(\dd \bar g)\notag\\
		&\quad+\int \tfrac{1}{2}\Big(R_F(g,d_A)+\rho^*(\bar g)R_S^\ell(g,d_A;D^{0,*}(\bar g))+(1-\rho^*(\bar g))R_S^\ell(g,d_A;D^{\infty,*}(\bar g))\Big)\mathbf 1_{g=\bar g}\Phi^*(\dd \bar g)\notag\\
		&\quad+\int \Big(\rho^*(\bar g)R_S^\ell(g,d_A;D^{0,*}(\bar g))+(1-\rho^*(\bar g))R_S^\ell(g,d_A;D^{\infty,*}(\bar g))\Big)\mathbf 1_{g<\bar g}\Phi^*(\dd \bar g)\bigg].\label{eq:ResponseFunctionChoice1}
	\end{align}
	
	Suppose, for contradiction, that
	$
		\Phi^*\big(\{g\in[\hat g_L,\hat g_H]:\rho^*(g)\ge \epsilon,\;D_A^{0,*}(g)=0\}\big)>0
	$
	for some $\epsilon>0$. Denote $B:=\{g\in[\hat g_L,\hat g_H]:\rho^*(g)\ge \epsilon,\;D_A^{0,*}(g)=0\}$. Then,
	\begin{align*}
		h^0(g,D_A^0(g))=h^0(g,0)\le -rg,\quad \forall g\in B.
	\end{align*}
	Define $B_1:=\{g\in B:h^\infty(g,D_A^\infty(g))\le 0\}$. If $\Phi^*(B_1)=0$, consider $\tilde \rho(g)$ defined as $\rho^*(g)$ for $g\notin B$ and $\rho^*(g)-\epsilon$ for $g\in B$. It is straightforward to see that
	\begin{align*}
		&\Pi(\alpha^*,\Phi^*,\tilde \rho,D_A^{0,*},D_A^{\infty,*};\alpha^*,\Phi^*,\rho^*,D_A^{0,*},D_A^{\infty,*})-\Pi(\alpha^*,\Phi^*,\rho^*,D_A^{0,*},D_A^{\infty,*};\alpha^*,\Phi^*,\rho^*,D_A^{0,*},D_A^{\infty,*})\\
		&= \alpha^*\int_B \epsilon\big(h^\infty(g,D_A^\infty(g))-h^0(g,D_A^\infty(g))\big)\Phi^*(\dd g)\ge \alpha^*\epsilon r\hat g_L\Phi^*(B)>0,
	\end{align*}
	contradicting the optimality of $(\alpha^*,\Phi^*,\rho^*,D_A^{0,*},D_A^{\infty,*})$.	If $\Phi^*(B_1)>0$, consider
	\begin{align*}
		\tilde \alpha:=\alpha^*(1-\Phi^*(B_1)),\quad
		\tilde \Phi(\dd g):=(1-\Phi^*(B_1))^{-1}\mathbf 1_{B_1}(g)\Phi^*(\dd g).
	\end{align*}
	Straightforward calculation yields
	\begin{align*}
		&\Pi(\tilde \alpha,\tilde \Phi,\rho^*,D_A^{0,*},D_A^{\infty,*};\alpha^*,\Phi^*,\rho^*,D_A^{0,*},D_A^{\infty,*})-\Pi(\alpha^*,\Phi^*,\rho^*,D_A^{0,*},D_A^{\infty,*};\alpha^*,\Phi^*,\rho^*,D_A^{0,*},D_A^{\infty,*})\\
		&= -\alpha^*\int_{B_1}\big[\rho(g)h^0(g,D_A^0(g))+(1-\rho(g))h^\infty(g,D_A^\infty(g))\big]\Phi^*(\dd g)\ge \alpha^*\epsilon r\hat g_L\Phi^*(B_1)>0,
	\end{align*}
	where the inequality holds because $h^0(g,D_A^0(g))\le -r\hat g_L$, $\rho(g)\ge \epsilon$, and $h^\infty(g,D_A^\infty(g))\le 0$ on $B_1$. This again contradicts the optimality of $(\alpha^*,\Phi^*,\rho^*,D_A^{0,*},D_A^{\infty,*})$. Therefore, we must have $\Phi^*(\{g\in[\hat g_L,\hat g_H]:\rho^*(g)>0,\;D_A^{0,*}(g)=0\})=0$. Similarly, we can prove $\Phi^*(\{g\in[\hat g_L,\hat g_H]:\rho^*(g)<1,\;D_A^{\infty,*}(g)=0\})=0$.
	
	Suppose, for contradiction, that
	$
		\Phi^*(\{g\in[\hat g_L,\hat g_H]:\rho^*(g)\ge \epsilon,\;V^{0}(g,D_A^{0,*}(g))\le 0\})>0
	$
	for some $\epsilon>0$. Denote $C:=\{g\in[\hat g_L,\hat g_H]:\rho^*(g)\ge \epsilon,\;V^{0}(g,D_A^{0,*}(g))\le 0\}$. Then,
	\begin{align*}
		R_F(g,D_A^{0,*}(g))=R_S^0(g,D_A^{0,*}(g))+V^0(g,D_A^{0,*}(g))\le -rg\le -r\hat g_L,\quad \forall g\in C.
	\end{align*}
	By \eqref{eq:ResponseFunctionChoice}, we derive
	\begin{align*}
		h^0(g,D_A^{0,*}(g))\le -r\hat g_L,~ \forall g\in C.
	\end{align*}
	Using the same construction as above, we can find a strategy that yields a strictly higher expected payoff than $(\alpha^*,\Phi^*,\rho^*,D_A^{0,*},D_A^{\infty,*})$. This is a contradiction, so we must have $\Phi^*(\{g\in[\hat g_L,\hat g_H]:\rho^*(g)>0,\;V^{0}(g,D_A^{0,*}(g))\le 0\})=0$.
	
	The proof of (ii) is similar to that of Proposition \ref{le:RandomizedGasDensity}.
	
	Next, it is straightforward to derive \eqref{eq:ResponseFunction2Choice} from \eqref{eq:ResponseFunctionChoice1}. Moreover, by \eqref{eq:ExpectedProfitChoice2}, we have
	\begin{align*}
		\Pi(\alpha,\Phi^*,\rho,D_A^{0},D_A^{\infty};\alpha^*,\Phi^*,\rho^*,D_A^{0,*},D_A^{\infty,*})
		= \alpha \int_{[\hat g_L,\hat g_H]} H\big(g,\rho(g),D_A^{0}(g),D_A^{\infty}(g)\big)\Phi^*(\dd g).
	\end{align*}
	By the optimality of $(\alpha^*,\Phi^*,\rho^*,D_A^{0,*},D_A^{\infty,*})$, we conclude that $\big(\rho^*(g),D_A^{0,*}(g),D_A^{\infty,*}(g)\big)$ maximizes $H(g,\rho,d_A^0,d_A^\infty)$ in $(\rho,d_A^0,d_A^\infty)$ for almost every $g$ under $\Phi^*$. The remaining part of (iii) can be proved similarly to Proposition \ref{le:RandomizedGasDensity}.
	
	Finally, the proof of (iv) is similar to that of Proposition \ref{le:RandomizedGasDensity}. \halmos
\end{pfof}

\begin{pfof}{Proposition \ref{prop:MixedIC0}}
	It is straightforward to see that
	\begin{align*}
		R_S^\infty(g,d_A;\hat D_A)
		&= d_A\left(\frac{\tfrac{y_Ay_B}{y_A+\hat D_A}}{y_A+\hat D_A+d_A}p_B-(1+f)p_A\right)-g <-g \le -rg = \inf_{d_A\ge 0}R_S^0(g,d_A;\hat D_A)
	\end{align*}
	for every $d_A>0$, where the strict inequality follows from \eqref{eq:MaximumTradingAmount} and the last equality holds because $\Gamma(d_A)\equiv 0$. Consequently, $V^\infty(g,d_A;\hat D_A)>\inf_{d_A\ge 0}V^0(g,d_A;\hat D_A)$.
	
	Now, consider the competitor's strategy $(\alpha^*,\Phi^*,\rho^*,D_A^{0,*},D_A^{\infty,*})$ with $\rho^*(g)\equiv 1$ and $(\alpha^*,\Phi^*,D_A^{0,*})$ as given in Theorem \ref{thm:MainSlippage0}-(i). We have $D_A^{0,*}(g)\equiv \hat D_A$. Then, by \eqref{eq:ResponseFunction2Choice}, we obtain
	\begin{align}
		h^0(g,d_A)\ge h^\infty(g,d_A),\quad \forall g\in[\hat g_L,\hat g_H],\; d_A\in[0,\hat D_A].\label{eq:pfofnoslippageeq1}
	\end{align}
	Denote by $\Pi(\alpha,\Phi,\rho,D_A^{0},D_A^{\infty};\alpha^*,\Phi^*,\rho^*,D_A^{0,*},D_A^{\infty,*})$ the arbitrageur's payoff under strategy $(\alpha,\Phi,\rho,D_A^0,D_A^\infty)$. The arbitrageur's expected payoff, given the competitor takes $(\alpha^*,\Phi^*,\rho^*,D_A^{0,*},D_A^{\infty,*})$, is
	\begin{align*}
		&\Pi(\alpha,\Phi,\rho,D_A^{0},D_A^{\infty};\alpha^*,\Phi^*,\rho^*,D_A^{0,*},D_A^{\infty,*})\\
		&= \alpha \int_{[\hat g_L,\hat g_H]}\big[\rho(g)h^0(g,D_A^0(g))+(1-\rho(g))h^\infty(g,D_A^\infty(g))\big]\Phi(\dd g)\\
		&\le \alpha \int_{[\hat g_L,\hat g_H]}\big[\rho(g)h\infty(g,D_A^0(g))+(1-\rho(g))h^\infty(g,D_A^\infty(g))\big]\Phi(\dd g)\\
		&\le \alpha^* \int_{[\hat g_L,\hat g_H]}h^0(g,D_A^{0,*}(g))\Phi^*(\dd g)\\
		&= \Pi(\alpha^*,\Phi^*,\rho^*,D_A^{0,*},D_A^{\infty,*};\alpha^*,\Phi^*,\rho^*,D_A^{0,*},D_A^{\infty,*}),
	\end{align*}
	where the first inequality holds because of \eqref{eq:pfofnoslippageeq1}, and the second inequality holds because $(\alpha^*,\Phi^*,D_A^{0,*})$ is an equilibrium strategy when $\ell$ is fixed to be 0. Therefore, $(\alpha^*,\Phi^*,\rho^*,D_A^{0,*},D_A^{\infty,*})$ is an equilibrium strategy. \halmos
\end{pfof}

\begin{pfof}{Proposition \ref{prop:MixedICLarge}}
	We first prove that
	\begin{align}
		R_S^\infty(g,d_A;\bar d_A)-R_S^0(g,d_A;\bar d_A)>0,\quad
		\forall d_A\in(0,\hat D_A],\;\bar d_A\in[0,\hat D_A].\label{eq:InventoryCostLowerBoundPf}
	\end{align}
	In view of \eqref{eq:InventoryCostLowerBound}, we assume without loss of generality that $\Gamma(d_A)=\tfrac{1}{2}\gamma d_A^2$ with
	\begin{align*}
		\gamma:=2(1+f)p_Ay_A^{-1}O^{-1/2}.
	\end{align*}
	
	For any $d_A\ge 0$ and $\bar d_A\in[0,\hat D_A]$, we have
	\begin{align*}
		R_S^\infty(g,d_A;\bar d_A)\ge R_S^\infty(g,d_A;\hat D_A)
		&= d_A\left(\frac{\tfrac{y_Ay_B}{y_A+\hat D_A}}{y_A+\hat D_A+d_A}p_B-(1+f)p_A\right)-g\\
		&= (1+f)p_Ad_A\Big((1+y_A^{-1}O^{-1/2}d_A)^{-1}-1\Big)-g.
	\end{align*}
	Consequently,
	\begin{align*}
		R_S^\infty(g,d_A;\bar d_A)-R_S^0(g,d_A;\bar d_A)
		\ge (1+f)p_Ad_A\Big((1+y_A^{-1}O^{-1/2}d_A)^{-1}-1\Big)+\tfrac{1}{2}\gamma d_A^2 =:\zeta(d_A).
	\end{align*}
	To prove \eqref{eq:InventoryCostLowerBoundPf}, it suffices to show that $\zeta(d_A)>0$ for all $d_A\in(0,\hat D_A]$.
	
	It is straightforward to see that $\zeta(0)=0$. Moreover,
	\begin{align*}
		\zeta'(d_A) &= -(1+f)p_Ay_A^{-1}O^{-1/2}d_A(2+y_A^{-1}O^{-1/2}d_A)(1+y_A^{-1}O^{-1/2}d_A)^{-2}+\gamma d_A,\\
		\zeta''(d_A) &= -2y_A^{-1}O^{-1/2}(1+f)p_A(1+y_A^{-1}O^{-1/2}d_A)^{-3}+\gamma.
	\end{align*}
	It is straightforward to verify that $\zeta''(0)=0$ by the definition of $\gamma$, and that $\zeta''$ is strictly increasing. This implies that $\zeta''$ is positive on $(0,+\infty)$. Noting that $\zeta'(0)=0$, we conclude that $\zeta'$ is positive on $(0,\infty)$. As a result, $\zeta$ is strictly increasing on $(0,\infty)$. Since $\zeta(0)=0$, to show $\zeta(d_A)>0$ for all $d_A\in(0,\hat D_A]$, it suffices to note that $\zeta(\hat D_A)>0$.
	
	Next, we show that if $(\alpha^*,\Phi^*,\rho^*,D_A^{0,*},D_A^{\infty,*})$ is a mixed symmetric equilibrium strategy, then $\rho^*\equiv 0$. By \eqref{eq:ResponseFunction2Choice} and \eqref{eq:InventoryCostLowerBoundPf}, for almost all $g$ under $\Phi^*$, we have
	\begin{align}
		h^\infty(g,d_A)>h^0(g,d_A),\quad \forall d_A\in(0,\hat D_A].\label{eq:InventoryCostLowerBoundPfEq2}
	\end{align}
	By Proposition \ref{le:ChooseRandomizedGasDensity}-(iii), for almost every $g$, $(\rho^*(g),D_A^{0,*}(g),D_A^{\infty,*}(g))$ maximizes $H(g,\rho,d_A^0,d_A^\infty)$ as defined in \eqref{eq:ResponseFunctionChoiceAgg}. We claim that $\rho^*(g)=0$ for almost every $g$. Otherwise, $B:=\{g\in[\hat g_L,\hat g_H]:\rho^*(g)>0\}$ has positive measure under $\Phi^*$. By Proposition \ref{le:ChooseRandomizedGasDensity}-(i), $D_A^{0,*}(g)>0$ on $B$. From \eqref{eq:InventoryCostLowerBoundPfEq2}, we then have $h^\infty(g,D_A^{0,*}(g))>h^0(g,D_A^{0,*}(g))$ for $g\in B$. Consider $\tilde \rho$ and $\tilde D_A^{\infty}$ with $\tilde \rho(g)=\rho^*(g)$ and $\tilde D_A^\infty(g)=D_A^{\infty,*}(g)$ for $g\notin B$, and $\tilde \rho(g)=0$ and $\tilde D_A^\infty(g)=D_A^{0,*}(g)$ for $g\in B$. Recall the payoff function \eqref{eq:ExpectedProfitChoice2}. Then, straightforward calculation yields
	\begin{align*}
		&\Pi(\alpha^*,\Phi^*,\tilde \rho,D_A^{0,*},\tilde D_A^{\infty};\alpha^*,\Phi^*,\rho^*,D_A^{0,*},D_A^{\infty,*})-\Pi(\alpha^*,\Phi^*,\rho^*,D_A^{0,*},D_A^{\infty,*};\alpha^*,\Phi^*,\rho^*,D_A^{0,*},D_A^{\infty,*})\\
		&= \alpha^*\int_B \rho^*(g)\big(h^\infty(g,D_A^{0,*}(g))-h^0(g,D_A^{0,*}(g))\big)\Phi^*(\dd g)>0,
	\end{align*}
	where the inequality holds because $\alpha^*>0$ (Proposition \ref{prop:propertyEqChooseSlippage}-(i)), by \eqref{eq:InventoryCostLowerBoundPfEq2}, since $\rho^*(g)>0$ and $D_A^{0,*}(g)>0$ on $B$, and because $\Phi^*(B)>0$. This contradicts the optimality of $(\alpha^*,\Phi^*,\rho^*,D_A^{0,*},D_A^{\infty,*})$, so we must have $\rho^*\equiv 0$.
	
	Finally, we show that $(\alpha^*,\Phi^*,\rho^*,D_A^{0,*},D_A^{\infty,*})$ with $\rho^*\equiv 0$ and $(\alpha^*,\Phi^*,D_A^{\infty,*})$ as given in Theorem \ref{thm:Main} is an equilibrium strategy. To this end, consider any strategy $(\alpha,\Phi,\rho,D_A^0,D_A^\infty)$. Straightforward calculation yields
	\begin{align*}
		&\Pi(\alpha,\Phi,\rho,D_A^{0},D_A^{\infty};\alpha^*,\Phi^*,\rho^*,D_A^{0,*},D_A^{\infty,*})\\
		&= \alpha \int_{[\hat g_L,\hat g_H]}\big[\rho(g)h^0(g,D_A^0(g))+(1-\rho(g))h^\infty(g,D_A^\infty(g))\big]\Phi^*(\dd g)\\
		&\le \alpha \int_{[\hat g_L,\hat g_H]}h^\infty(g,D_A^\infty(g))\Phi^*(\dd g)\\
		&\le \alpha^* \int_{[\hat g_L,\hat g_H]}h^\infty(g,D_A^{\infty,*}(g))\Phi^*(\dd g)\\
		&= \Pi(\alpha^*,\Phi^*,\rho^*,D_A^{0,*},D_A^{\infty,*};\alpha^*,\Phi^*,\rho^*,D_A^{0,*},D_A^{\infty,*}),
	\end{align*}
	where the first inequality holds due to \eqref{eq:InventoryCostLowerBoundPfEq2}, and the second inequality holds because $(\alpha^*,\Phi^*,D_A^{0,*})$ is an equilibrium strategy when $\ell$ is fixed to be $\infty$. Therefore,  $(\alpha^*,\Phi^*,\rho^*,D_A^{0,*},D_A^{\infty,*})$ is an equilibrium strategy.\halmos
\end{pfof}

\end{appendices}
\end{document}